%% file: main.tex
\documentclass[english]{article}
\RequirePackage{etex}
\usepackage{babel}
\usepackage[utf8]{inputenc}
\usepackage{a4wide}
\usepackage{geometry}
\usepackage{authblk}
\usepackage{hyperref}
\usepackage{float}
\usepackage{graphicx}
\usepackage{color}
\usepackage{caption}
\usepackage{amsfonts}
\usepackage{amsmath}
\usepackage{caption}
\usepackage{subcaption}
\usepackage{standalone}
\usepackage{lineno}
\usepackage{multirow}
\usepackage{bbm}
\usepackage{csquotes}

\newlength\fheight 
\newlength\fwidth 
\usepackage{xr}

\usepackage{tikz-network}
\usepackage{pgfplots}
\pgfplotsset{compat=newest}
\usepackage{hhline}
\definecolor{mycolor1}{rgb}{0.00000,0.80000,0.80000}
\definecolor{mycolor2}{rgb}{0.00000,0.50000,0.50000}
\definecolor{mycolor3}{rgb}{0.00000,0.30000,0.30000}
\definecolor{mycolor4}{rgb}{0.00000,0.10000,0.10000}
\definecolor{mycolor5}{rgb}{0.00000,1,1}

\newcommand{\review}[1]{{#1}}
\newcommand{\revsec}[1]{{#1}}

\newcommand{\E}[1]{{\mathbb E}\left[#1\right]}
\renewcommand{\P}{\mathbb P}

\newcommand{\eqn}[1]{\begin{equation}#1\end{equation}}

\newtheorem{proposition}{Proposition}

\renewcommand{\deg}{{\rm deg}}

\usepackage[style=ieee]{biblatex}
\bibliography{references}

\title{
Learning the mechanisms of network growth
}

\author[a,1]{Lourens Touwen}
\author[b,2]{Doina Bucur}
\author[a,3]{Remco van der Hofstad}
\author[a,4]{Alessandro Garavaglia}
\author[a,5,*]{Nelly Litvak}
\affil[a]{\footnotesize Department of Mathematics and
  Computer Science, Eindhoven University of Technology, Groene Loper 3, 5612 AE Eindhoven, The Netherlands}
\affil[b]{\footnotesize Faculty of Electrical Engineering, Mathematics and Computer Science,
				 University of Twente, Drienerlolaan 5, 7522 NB, Enschede, The Netherlands}
\affil[$ $]{ {\itshape Email address}: $^1$touwenlourens@gmail.com, $^2$d.bucur@utwente.nl, $^3$rhofstad@win.tue.nl, $^4$ale.garavaglia@gmail.com,  $^{5}$n.v.litvak@tue.nl, $^*$ corresponding author}

\date{\today}

\begin{document}

\maketitle

\begin{abstract} 
  We propose a novel model-selection method for \revsec{dynamic networks}. Our approach involves training a classifier on a large body of synthetic network data. The data is generated by simulating nine state-of-the-art random graph models for dynamic networks, with parameter range chosen to ensure exponential growth of the network size in time. We design a conceptually novel type of {\em dynamic} features that count new links received by a group of vertices in a particular time interval. The proposed features are easy to compute, analytically tractable, and interpretable. Our approach achieves a near-perfect classification of synthetic networks, exceeding the state-of-the-art by a large margin. Applying our classification method to real-world citation networks gives credibility to the claims in the literature that models with preferential attachment, fitness and aging fit real-world citation networks best, although sometimes, the predicted model does not involve vertex fitness. 
\end{abstract}

\section{Introduction}
In this paper, we investigate model selection for dynamic growing networks using machine learning. By model selection we mean choosing a generative mathematical model that describes the dynamic network in the best possible way. We apply our methods to citation networks, as these are prominent examples of dynamic networks where vertices and edges remain in the network forever.

\subsection{Model selection for complex networks}

In dynamic networks, model selection identifies generative {\em mechanisms} that drive the formation of new connections. There are numerous ways in which model selection can help solve network problems: (a) The specific topology of a graph affects the performance of algorithms for common tasks such as computing the shortest paths~\cite{DialATrees, MadduriAnInstances} and finding strongly connected components~\cite{NiewenhuisVerbanescu2022efficient}. Therefore, the classification of generative network-growth mechanisms of real-life networks is useful for selecting suitable network algorithms and forecasting their performance as well as their requirements for computing power and storage. (b) Scale-free networks with the same exponent but different degree cut-off values can be small- or ultra-small~worlds~\cite{HofstadKomjathy2017scale}. As such, the right model is necessary for accurate prediction of the shortest paths in networks. (c) Exact assumptions on the scale-free model may lead to the opposite conclusions on the epidemic threshold being positive~\cite{Pastor2001epidemic} (the network has initial resistance to an epidemic) or zero~\cite{Chatterjee2009contact} (even the smallest infection spreads widely). (d) Predictions of the network robustness depend on the fine details of the model. Indeed, a scale-free preferential attachment model of the Internet predicts an alarming sensitivity to targeted attacks~\cite{Albert2000error}, while more realistic models suggest a much higher robustness~\cite{Doyle2005robust}. (e) Different models assign different importance to the vertices. For example, PageRank follows a power-law distribution with the same exponent as the in-degree in a scale-free directed configuration model~\cite{Chen2017PageRank}, but, surprisingly, has a smaller exponent, thus higher variability, in directed preferential attachment networks~\cite{BanerjeeOlvera2022pagerank}. (f) Last but not least, realistic models enable us to predict the growth and evolution of networks in the future. 

The common approach for model selection is to fit the parameters of a generative model to the target network, generate synthetic networks with these fitted parameters, and compare their features (e.g., degree distributions, shortest paths, etc.) to the target network~\cite{BarAlb99,Mahadevan2006systematic,Krioukov2010hyperbolic, GarHofWoe17,Attar2017ClassificationFeatures, Langendorf2020EmpiricallyMechanisms}. 
However, fitting the model parameters is a difficult problem in itself, so only simple models admit explicit analytical results~\cite{Gao2021statistical, Gomez2013likelihood}. Numerical methods for parameter fitting are available for a wider range of models, for instance, \cite{Overgoor2019NetworkFormationCitations} fits a discrete choice model, where the probability of an edge from $u$ to $v$ depends on the features (e.g., degrees) of $u$ and $v$. The parameters are fitted using logit regression, where a large regression coefficient implies that the feature is important. The downside is that the model must be fitted for each network separately, and the parameter estimates are analytically intractable, thus conclusions do not easily generalize to other networks. Interestingly, experiments in \cite{Overgoor2019NetworkFormationCitations} find all features statistically significant, even when coefficients are small. This illustrates a fundamental challenge in model selection for complex networks, in that statistical tests are likely to reject any generative model.

Here we propose a machine learning approach to model selection that works as follows: First, many synthetic networks are generated from a particular set of generative models. Then a classifier is trained based on the network features. As a result, the classifier labels any network as being produced by one of the generative models.
Recent work on machine learning for model selection usually compares fundamentally different models, such as Erdős–Rényi (ER) \review{(random connections)} vs.\ Barabási–Albert (BA) \review{(connection probability to a vertex is proportional to its current degree)}~\cite{Dehmamy2019}, or ER vs. BA vs. Chung-Lu \review{(connection probability to a vertex is proportional to its fixed weight)} vs. Hyperbolic Random Graphs \review{(connection probability to a vertex is defined by its position in a hyperbolic space)}~\cite{Blasius2018TowardsModels}. \review{Some works also compare distinct models to} real-life networks from different domains~\cite{CanningNetworkCategorization, Rossi2019ComplexDomain}. In this work, we view a generative model as a combination of growth mechanisms. More specifically, we apply our methods for selecting the best combinations of growth mechanisms used in the literature to describe citation networks: fitness, aging, and preferential attachment~\cite{WanSonBar13,A.Garavaglia2019PreferentialNetworks,Overgoor2019NetworkFormationCitations,Yasui2022stochastic,Chang2021generative,Zhou2023}.

An important contribution of this work is the novel feature design for dynamic networks. The literature mostly uses a set of features that in fact are metrics of the network's final snapshot, for instance, the degree distribution, the PageRank distribution, the number of triangles, etc.~\cite{CanningNetworkCategorization,Blasius2018TowardsModels, Langendorf2020EmpiricallyMechanisms,Attar2017ClassificationFeatures,Bonner2016} (see more details in Section~\ref{ssec:ML-methods}). There are also deep learning methods that take the entire adjacency matrix as input~\cite{Dehmamy2019, Hegde2018NetworkImage}. We call these input features {\it static} because they do not explicitly include the network evolution. Instead, we propose new {\it dynamic} features inspired by the state-of-the-art network models \cite{Wang2013QuantifyingImpact, GarHofWoe17}\review{, and the growing data availability of dynamic networks\footnote{https://networkrepository.com/dynamic.php} \cite{dynamicnetworks}.} Our features are aggregated statistics of the network growth in time. These statistics are interpretable, easy to compute, and have explicit analytical expressions suitable for mathematical derivations of their properties. Our method achieves almost perfect classification of synthetic data obtained from the state-of-the-art network models \cite{Wang2013QuantifyingImpact, GarHofWoe17}. Our results on real-world citation data \revsec{mostly} support the current models \cite{WanSonBar13,A.Garavaglia2019PreferentialNetworks,Yasui2022stochastic,Chang2021generative}  with fitness, aging and preferential attachment as discussed next, but also \revsec{show that the selected model may depend on a specific feature design, thus} telling a cautionary tale for blanket application of these models as well as the usage of machine learning for model selection.

\subsection{Dynamic network mechanisms} We propose network models where edges do not disappear. Citation networks are a natural real-world example of such networks. Motivated by the properties of citation networks, our models include three generative mechanisms: fitness, aging and preferential attachment.

  \noindent {\bf Fitness.} Some papers generate tremendous follow-up work, while others may not. One may say that in terms of attracting citations, not all papers are equally {\em fit}. Fitness is the mechanism that allows relatively young papers to be cited a lot. \textcite{WanSonBar13} estimate the fitness of papers based on citations they receive. \review{\textcite{Zhou2023} propose a model of link formation based on fitness and degrees. They  find that, for citation networks like most non-social networks, the evolution based on fitness describes the degree distribution and the degree ratio of adjacent nodes better than the evolution based on degrees.} Moreover, in models with fitness but without preferential attachment, power-law distribution of fitness results in power-law degree distribution (see Proposition~A.\ref{prop:fitness-only} in the Supplementary Material).

  \noindent {\bf Aging.} Citations to a paper depend on the (calendar) time since its publication (see \cite{WanYuYu09} and Figure~\ref{fig-empirical-properties-main}(b) in the Supplementary Material). The literature suggests that aging occurs according to a log-normal distribution in calendar time (see \cite{WanSonBar13} and Figure~\ref{fig-empirical-properties-main}(c) in the Supplementary Material).

  \noindent {\bf Preferential Attachment.} Highly-cited papers can be expected to get even more citations, thus giving rise to a preferential attachment effect. This effect was already observed in the context of citation networks by De Solla Price \cite{Pri65}, and was put forward as a main driver for the occurrence of power-law degrees in \cite{BarAlb99}. See e.g., \cite{WanSonBar13}, as well as \cite{WanYuYu08} for an attempt to measure this effect. Contrary to fitness, preferential attachment means that papers receive extra citations {\em because} they have already been cited, and not because they are fit. This effect may be due to scientists reading paper A that cites paper B, and thus citing B as well.

\subsection{Modeling networks in calendar time}
Most models of dynamic networks take the {\it network growth time} approach, that is, vertices arrive one by one at time $t=1,2,\ldots$. We instead propose a continuous-time approach that mimics the networks' exponential growth in {\it calendar time} (illustrated in Figure \ref{fig-empirical-properties-main}(a) of the Supplementary Material). Our approach relies on the powerful concept of {\em continuous-time branching processes} (CTBP) \cite{Jage75, JagNer96, GarHofWoe17}. We use CTBPs to produce random trees, and then transition from trees to graphs with random out-degrees using a {\em collapsing} procedure that identifies several vertices in the CTBP to one vertex in the random graph. We refer to Section \ref{sec-CTBP-networks} for details. 

\section{Results}

\subsection{Dynamic network models}
\label{sec-dyn-network-models}

\begin{figure*}[!htb]
  \centering
  \includegraphics[width=\textwidth]{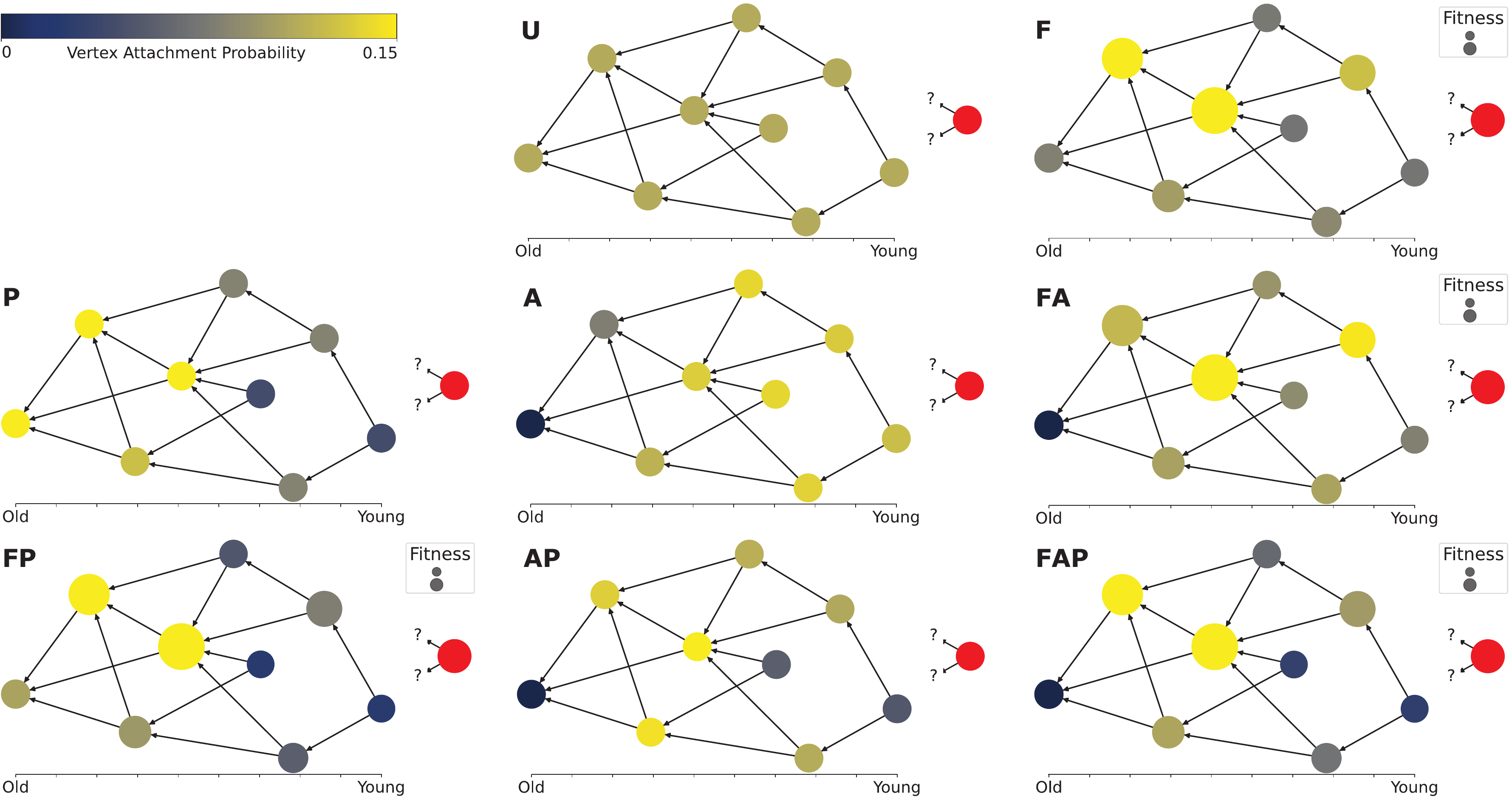}
\caption{An example of the eight combinations of growth mechanisms. \textsc{F} stands for fitness, \textsc{A} for aging, and \textsc{P} for Preferential Attachment. \textsc{U} stands for uniform attachment, when a new vertex connects to existing ones uniformly at random. The circles are the vertices, arranged horizontally from left to right in the order of their arrival. The vertical positioning is chosen to make all directed edges visible, it has no further meaning. In the models with fitness, the size of the circles is proportional to their fitness. The red circle on the right is the new vertex that will connect to one of the existing vertices. The color of the other vertices from dark blue to yellow corresponds to the connection probability, from low to high, of the red vertex to an older vertex. }
\label{fig:8mechanismsnew}
\end{figure*}

We model a network as a collapsed CTBP. By switching each of the generative mechanisms -- fitness, aging, and preferential attachment -- on or off, we obtain a flexible set of models that we wish to compare to real-world network data sets. Figure~\ref{fig:8mechanismsnew} illustrates the eight possible combinations of mechanisms. We use acronyms \textsc{F} (fitness), \textsc{A} (aging) and \textsc{P} (preferential attachment). The illustrations demonstrate how in the models with aging, the probability to connect to the older vertex is low; in the models with preferential attachment, the connection probability increases with degree; and in the models with fitness, the fit vertices have a higher connection probability. In addition to the eight different combinations of mechanisms, in models \textsc{F} and \textsc{FA}, we consider two distributions of fitness: power-law and exponential. We denote this by \textsc{F\textsubscript{pl}}  and \textsc{F\textsubscript{exp}}, respectively. Model \textsc{FP} requires bounded fitness~\cite{Borgs2007FirstFitness}, therefore we assume that the fitness has a uniform distribution and denote this by \textsc{F\textsubscript{unif}}. Finally, we disregard model A (only aging) because it lacks the property of exponential growth in calendar time (see Section~\ref{sup:supercritical} of the Supplementary Material). Altogether, we obtain a set of nine models from which we generate synthetic data to train our classifier for machine learned model selection. In the classification problem, each generative model corresponds to one category, or class.

The models proposed in this paper come with many parameters. We choose these parameters in such a way that the network characteristics obey asymptotic laws that are close to those observed in real-world networks. In particular, to mimic citation networks, we aim for models whose sizes grows exponentially in time. Mathematically, it means that the collapsed CTBP must satisfy a certain {\em supercriticality condition}, explained in Section~\ref{sup:supercritical} of the Supplementary Material. Table \ref{tab:dynamic_models} in Section \ref{sec-methods} gives an overview of the models and the parameters used to generate the synthetic data set.

\subsection{Classification of dynamic networks}
\paragraph{\bf Generating the training data.}
For training, we have generated $6733$ synthetic networks, about $750$ for each of the nine generative models (categories), of size $20,000$. The numbers and size of synthetic networks are chosen so that the learning curves of machine learning models flatten at these data and network sizes (see Section~\ref{secsup:learning-curves} in the Supplementary Material). Synthetic networks are generated by the following procedure. For a given generative model, the parameters are chosen uniformly at random from the range in Table~\ref{tab:dynamic_models}. Then we let the network grow until it reaches $20,000$ vertices. If the CTBP dies out, we repeat the simulations with the same parameter combination at most 1000 times, until one of the runs reaches 20,000 vertices before dying out. The resulting network of $20,000$ vertices is included in the dataset. A few parameter combinations result is a very low survival probability, so we do not succeed in 1000 attempts; then we have slightly less than $750$ synthetic networks of the corresponding category. 

\paragraph{\bf Static features.}
Static features use only the final snapshot of the graph. We use 36 static features from the recent literature~\cite{Ikehara2017CharacterizingDomains, Attar2017ClassificationFeatures, Blasius2018TowardsModels, Rossi2019ComplexDomain,Langendorf2020EmpiricallyMechanisms}, that are feasible to compute on many medium-to-large size networks. They include assortativity, transitivity, and the distributions of numerical values associated with each vertex: degree, coreness, local clustering coefficient, and the number of triangles. See Section~\ref{ssec:ML-methods} for details. 

\paragraph{\bf Dynamic features.} Our novel {\it dynamic} features significantly depart from the standard static features: they explicitly capture the network's aggregated dynamics. The literature either learns the model from the final snapshots of many networks~\cite{Ikehara2017CharacterizingDomains, Attar2017ClassificationFeatures, Blasius2018TowardsModels, Rossi2019ComplexDomain,Langendorf2020EmpiricallyMechanisms} thus ignoring the dynamics, or considers each edge as a data point and learns the model from the set of edges of one network~\cite{Overgoor2019NetworkFormationCitations} thus allowing no generalization of the learned model to other networks. Our approach strikes the balance between viewing the network as a whole and at the same time including its dynamics. Specifically, denote the network graph at time $t\in [0,T]$ by $G(t)=(V(t), E(t))$, where $V(t)$ is the set of vertices, and $E(t)$ is the set of edges. We subdivide vertices in $V(T)$ in equal groups $V_j$, $j\in \{1,2,\ldots, r\}$, by their degrees in the final graph $G(T)$ from large to small. Furthermore, we subdivide the time interval $T$ in $s$ sub-intervals. We do this in two ways: using \textit{size}-cohorts (where each sub-interval contains the same number of arrivals, $|V(T)|/s$), and using \textit{time}-cohorts (where each sub-interval is of the same length, $T/s$, in calendar time). The dynamic features $D_{ij}$ are the average numbers of edges per vertex, received by group $V_j$ in the $i$-th sub-interval (see Section~\ref{ssec:ML-methods} for the formal definition). This way the features $(D_{ij})_{1\le i\le s,1\le j\le r}$ track the network's dynamics. We call $D=(D_{ij})$ the Dynamic Feature Matrix (DFM).

\subsection{Near-perfect classification of synthetic dynamic networks}
\label{ssec:classifying-synthetic}

\begin{figure*}[th]
\centering
\begin{subfigure}{0.3\textwidth}
\centering
\includegraphics[height=\textwidth]{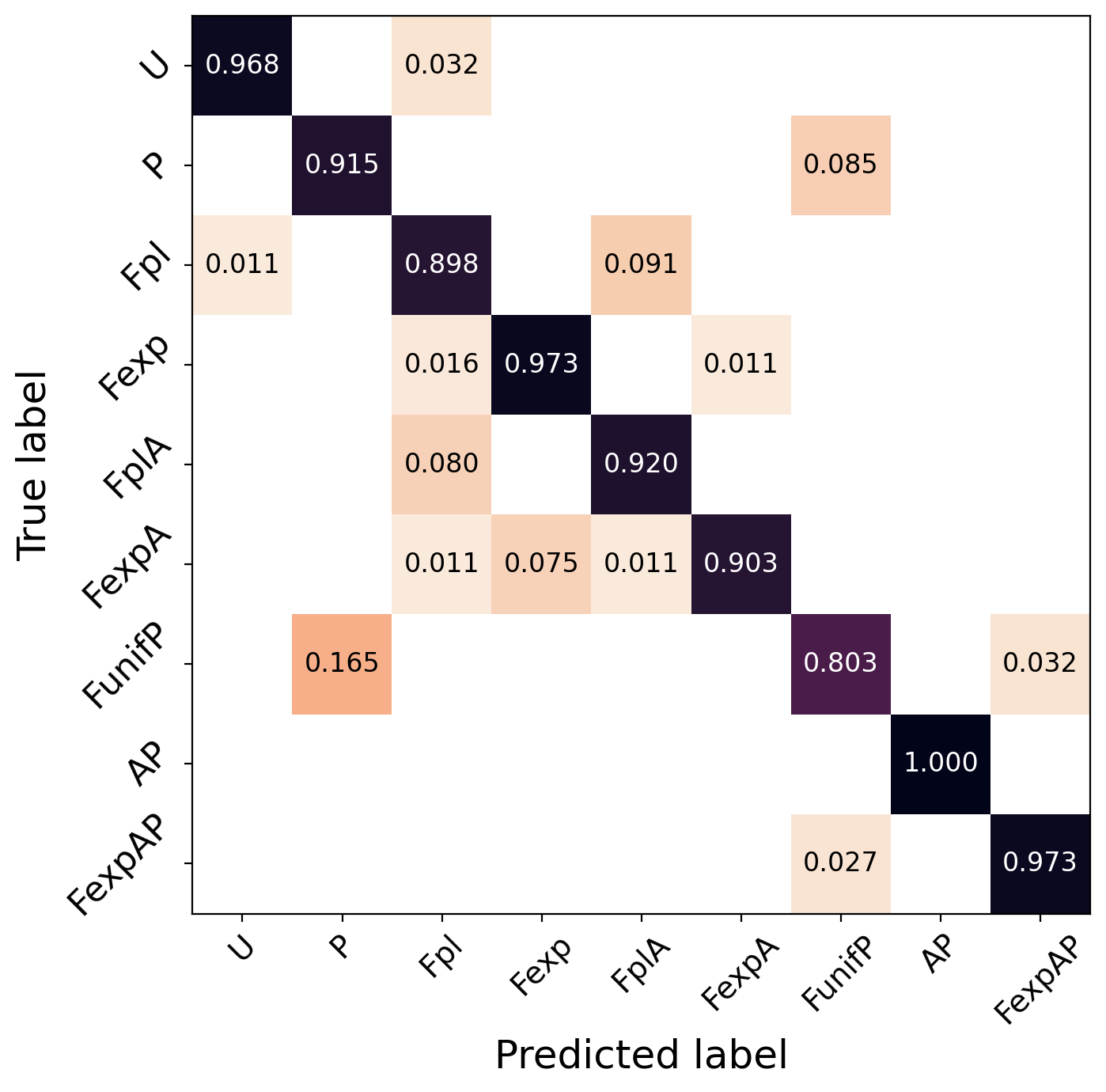}
\caption{}
\end{subfigure}
\begin{subfigure}{0.3\textwidth}
\centering
\includegraphics[height=\textwidth]{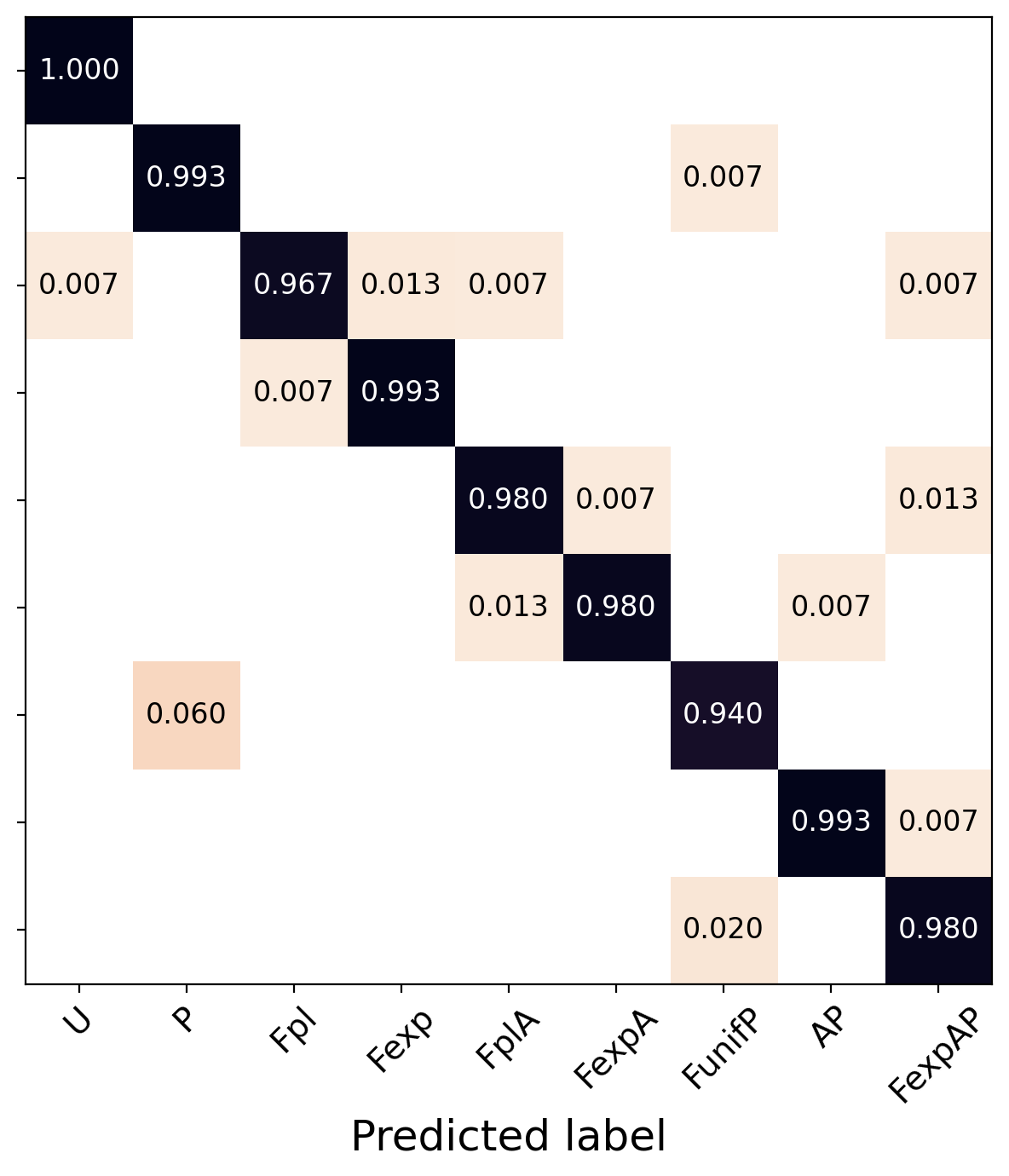}
\caption{}
\end{subfigure}
\begin{subfigure}{0.3\textwidth}
\centering
\includegraphics[height=\textwidth]{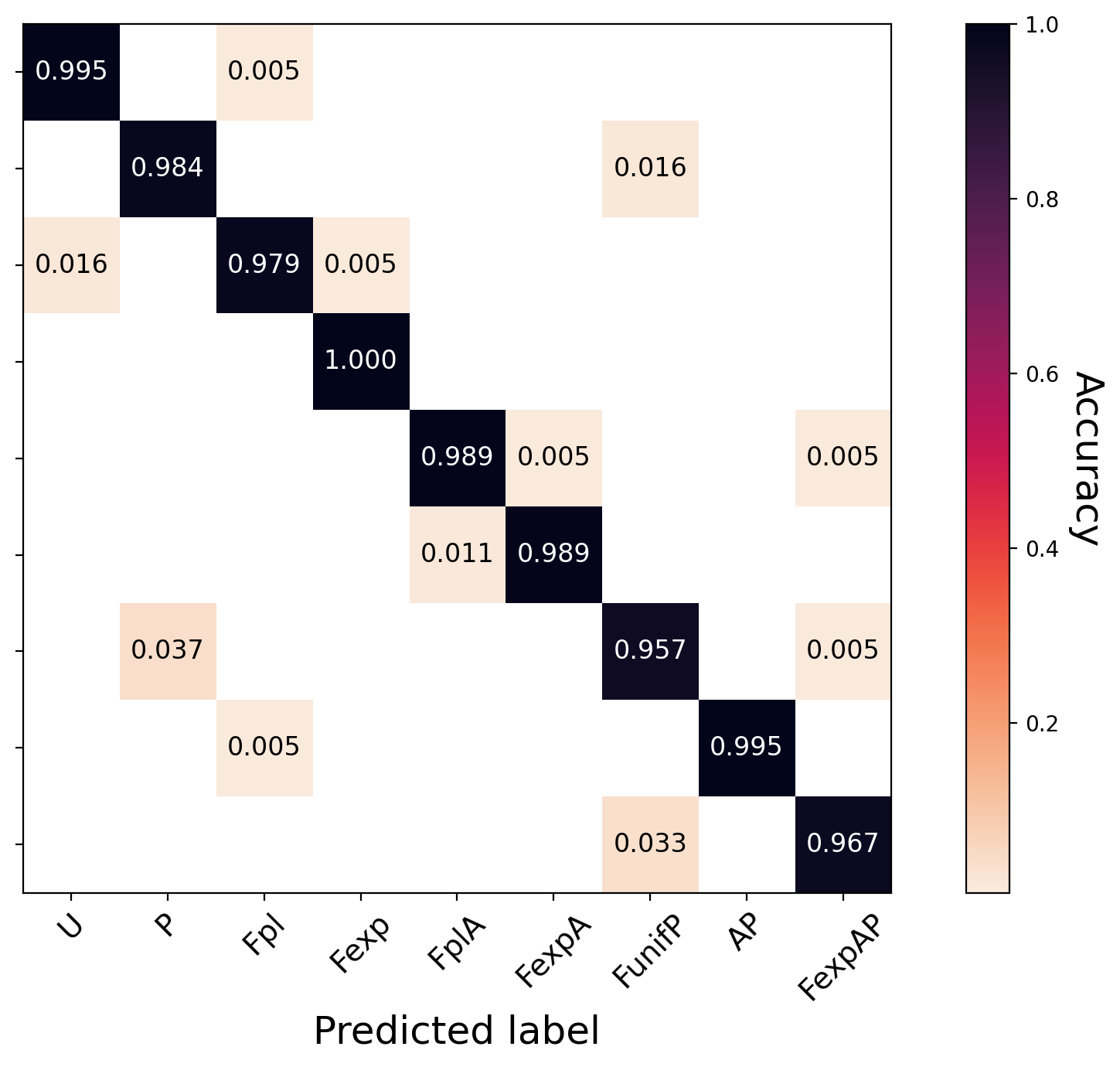}
\caption{}
\end{subfigure}
\caption{Confusion matrix using: (a) static features, (b) dynamic features with time cohorts, (c) both static features and dynamic features with time cohorts.} 
\label{fig:ML-results}
\end{figure*}

\paragraph{Classification results with static features.} Static features already yield high accuracy of $92.81\pm 1\%$, where $92.81\%$ is the accuracy in the reported train-test split and $1$\% is the maximal observed absolute difference with other train-test splits in our experiments. The confusion matrix in Figure~\ref{fig:ML-results}(a) predictably shows that the most common confusion is between \textsc{P} and \textsc{F\textsubscript{unif}}P. For discussion of feature importance, see Section~\ref{sup:feature-importance} and Figure~\ref{fig:feature-importance}(a) in the Supplementary Material. 

\paragraph{Classification results with dynamic features.} Our dynamic features convincingly outperform static features, yielding near-perfect accuracy $98.06\pm 0.5$\% with time-cohorts, and $97.62\pm 0.5\%$ with size-cohorts. 
The confusion matrix with time-cohorts in Figure~\ref{fig:ML-results}(b) shows a significant improvement. 

\begin{figure*}[th]
  \centering
  \includegraphics[width=\textwidth]{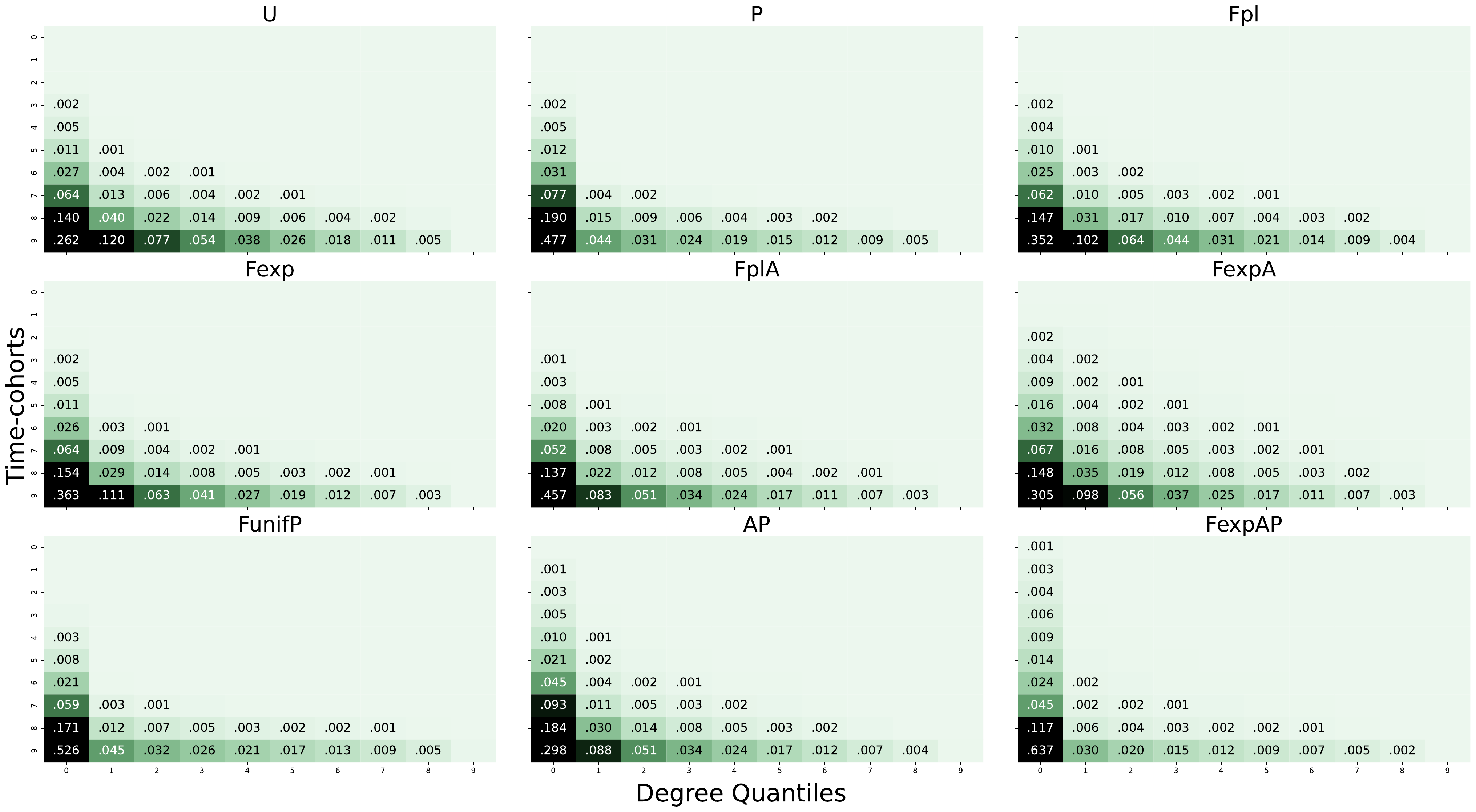}
  \caption{The values of the dynamic features with time-cohorts for the nine models. The shape of the matrix is similar for all models but the values reduce with their distance to $D_{10,1}$ in different ways.}
  \label{fig:feature-matrix}
\end{figure*}

\begin{figure*}
  \centering
  \includegraphics[width=\textwidth]{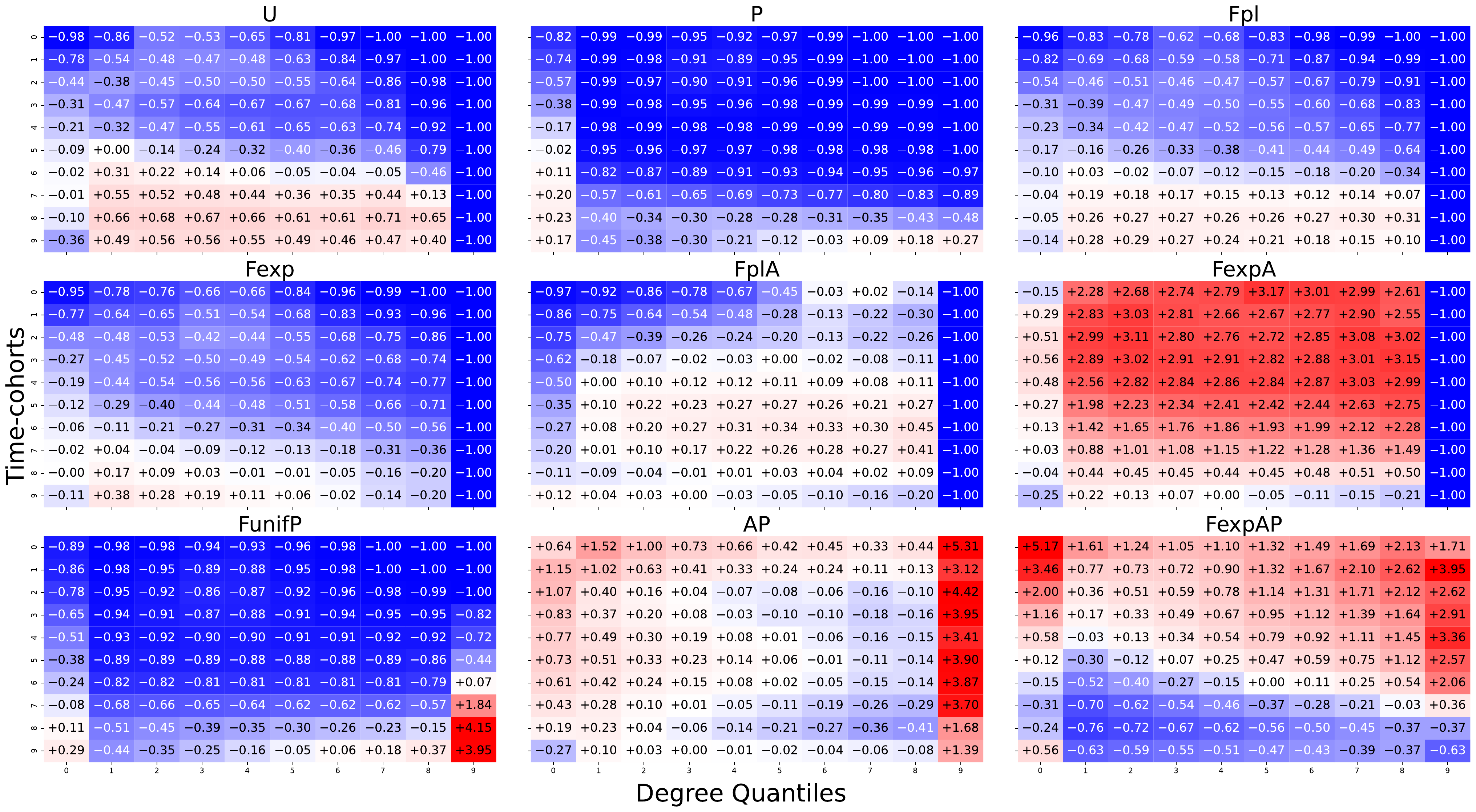}
  \caption{The relative difference, $\delta_{ij}=(D_{ij}-\bar{D}_{ij})/\bar{D}_{ij}$ with time-cohorts. The sum of elements $i,j$ over all classes equals zero. The value $-1$ appears often because it corresponds to $D_{ij}=0$.}
  \label{fig:matrixvisualization}
\end{figure*}

The next natural question is, can we interpret the difference between generative  models using the dynamic features? The standard permutation importance approach (see Section~\ref{sup:feature-importance} and Figure~\ref{fig:feature-importance}(b) in the Supplementary Material)  as well as extensive computations of correlations between features (omitted in this article, see Section~\ref{ssec:correlations} in the Supplementary Material), did not offer any intelligible way to identify most informative features. 
Therefore, we propose a new approach to interpretation of the dynamic features. To begin with, Figure~\ref{fig:feature-matrix} shows the values of the dynamic features for the nine models.
We see that all feature matrices have a similar shape. The largest feature is $D_{10,1}$ in the bottom left corner, as expected, because it is the average degree increment of the top-10\% degree vertices obtained from the last, by far the largest (due to the exponential growth of the network in calendar time), time-cohort. The values of the other features gradually reduce with their distance to $D_{10,1}$. The top-right corner is always empty, conform the old-get-richer phenomenon: low-degree vertices did not get any link early on. Surprisingly, this is true even for the Uniform model and for the models with aging. While all feature matrices have a similar shape, it turns out that the classifier discriminates the models by the differences between the numbers in the cells. We visualize this in Figure~\ref{fig:matrixvisualization} as follows. For each feature $D_{ij}$ we compute its average over all synthetic networks, $\bar{D}_{ij}$. Then for each model, we plot the relative difference $\delta_{ij}=(D_{ij}-\bar{D}_{ij})/\bar{D}_{ij}$. Note that for fixed $i,j$, the sum of $\delta_{ij}$ over all models equals zero because $\bar{D}_{ij}$ is the average over the nine models. However, the elements of the $\delta$ matrix for a given model are not comparable because they are normalized by different values of $\bar{D}_{ij}$.
The matrices in Figure~\ref{fig:matrixvisualization} are strikingly different. This explains the success of the classifier. Moreover, Figure~\ref{fig:matrixvisualization} yields several interesting insights. First, note that the largest negative value of $\delta_{ij}$ is $-1$, when $D_{ij}=0$. In particular, in the Uniform model, and all models with {\it fitness} but without {\it preferential attachment}, the last column has only $-1$'s, so at least 10\%  vertices in these models have in-degree zero. As expected, the models with {\it aging} smooth out the degree increments. Indeed, we see that $\delta$ has values closer to zero, in other words, if $D_{ij}$'s in the models with aging are relatively close to $\bar{D}_{ij}$. Next, we now see that in Figure~\ref{fig:matrixvisualization}, the `old-get-richer' effect is most prominent in the pure {\it preferential attachment} model, P. Indeed, low-quantile vertices are the ones that are born late in the process because this is when they start gaining positive degree increments. 

\paragraph{Classification results with static and dynamic features.} With static and dynamic features together, the accuracy increases to $98.40\pm 0.3$\%.  It is on average higher than with the dynamic features alone, but the error margins do overlap. The confusion matrix is shown in Figure~\ref{fig:ML-results}(c). We conclude that adding static features does not significantly improve the accuracy. 

\subsection{Classification applied to citation networks} \label{sec-classifying-citation}
\review{We use citation network data from the Web of Science. Our dataset contains publications between 1980 and 2015. Citations are included only to the papers contained in the dataset. In order to obtain somewhat homogeneous real-life networks, we use citation networks separately for different scientific fields. See the brief summary of the data in Table~\ref{tab:wos_eda}.
\begin{table*}[t]
  \centering
  \begin{tabular}{c|l c c}
    Acronym & Field & Original size & Size of connected component \\
    \hline
    AP & Astrophysics & 477113 & 451730 \\
    BT & Biotechnology & 537867 & 421643\\
    PS & Probability and statistics & 185167 & 157500\\
    GE & Geology & 56692 & 46672\\
    NP & Nuclear Physics & 223321 & 198162\\
    OC & Organic Chemistry & 567146 & 535945\\
    OP & Optics & 501817 & 416893\\
    SO & Sociology & 222416 & 91736\\
  \end{tabular}
  \caption{The summary of the Web of Science citation networks for several scientific fields. In the experiments, only the connected component was used.}
  \label{tab:wos_eda}
\end{table*}
Section~\ref{sec:citation} in the Supplementary Material contains detailed discussion of typical empirical properties of citation networks reported in the literature.}

The classifier takes the dynamic citation network as input and returns the probability distribution over the nine categories. Table \ref{tab:overview_wos_results} shows the most likely category for each research field using static features, dynamic features and the combination thereof. For the complete distribution over classes, see Tables~\ref{tab:static_features_class_probabilities}--\ref{tab:time_cohort_both_features_class_probabilities} in the Supplementary Material. 

With static features only, citation networks are classified \textsc{F\textsubscript{exp}A} or AP. This is surprising because neither of these growth mechanisms gives rise to power-law in-degrees. With the size-cohorts dynamic features or the combination of the static features and the size-cohorts dynamic features, most citation networks are classified as \textsc{F\textsubscript{exp}AP}, the model that has been suggested in the literature \cite{WanSonBar13}. Interestingly, precisely {\it this} combination of mechanisms makes the in-degrees have  dynamic power-laws, where the exponent decreases over time to some limit \cite{GarHofWoe17}, as observed in the data (see Figure~\ref{fig-empirical-properties-main}(a) in the Supplementary Material). SO is clearly different, it is the only citation network never being classified as \textsc{F\textsubscript{exp}AP}. Furthermore, it is classified as \textsc{P} by dynamic features with size-cohorts, which is the only classification that does not involve aging. 
\revsec{Notably, the results with time-cohorts dynamic features differ from those with size-cohorts dynamic features; the citation networks are often classified as \textsc{AP} and sometimes \textsc{F\textsubscript{exp}A}.
Altogether, with the best classifiers, we find it likely that aging is part of the real-world mechanisms, but whether this is combined with preferential attachment, exponential fitness, or both, is inconclusive. It is also clear that \textsc{F\textsubscript{exp}} appears statistically most likely out of the considered fitness options.}

{\small
\begin{table*}[th]
  \centering
  \begin{tabular}{l|l|c c c c c }
    \multicolumn{2}{l|}{Static Features} & Yes & No & Yes & No & Yes \\
    \hline
    \multicolumn{2}{l|}{Dynamic Features} & No & Size-cohorts & Size-cohorts & Time-cohorts & Time-cohorts \\
    \hline
\multicolumn{2}{l|}{Accuracy on synthetic data} & {92.81\%} & {97.33\%} & {97.62\%} & {98.06\%} & {98.40\%} \\
    \hline
    AP & Astrophysics  & \textsc{F\textsubscript{exp}A} & \textsc{F\textsubscript{exp}AP} & \textsc{F\textsubscript{exp}AP} & \textsc{AP} & \textsc{F\textsubscript{exp}A} \\
    BT & Biotechnology & \textsc{AP} & \textsc{F\textsubscript{exp}AP} & \textsc{F\textsubscript{exp}AP} & \textsc{AP} & \textsc{F\textsubscript{exp}AP}\\
    GE & Geology & \textsc{AP} & \textsc{F\textsubscript{exp}AP} & \textsc{AP} & \textsc{AP} & \textsc{AP} \\
    NP & Nuclear Physics & \textsc{F\textsubscript{exp}A} & \textsc{F\textsubscript{exp}AP} & \textsc{F\textsubscript{exp}AP} & \textsc{AP} & \textsc{F\textsubscript{exp}A}\\
    OC & Organic Chemistry & \textsc{F\textsubscript{exp}A} & \textsc{F\textsubscript{exp}AP} & \textsc{F\textsubscript{exp}AP} & \textsc{AP} & \textsc{F\textsubscript{exp}A}\\
    OP& Optics & \textsc{AP} & \textsc{F\textsubscript{exp}AP} & \textsc{F\textsubscript{exp}AP} & \textsc{AP} & \textsc{F\textsubscript{exp}AP} \\
    PS & Probability and Statistics & \textsc{AP} & \textsc{F\textsubscript{exp}AP} & \textsc{F\textsubscript{exp}AP} & \textsc{AP} & \textsc{F\textsubscript{exp}AP}\\
    SO & Sociology & \textsc{AP} & \textsc{P} & \textsc{AP} & \textsc{AP} & \textsc{AP} \\
  \end{tabular}
  \caption{Classification results on WoS citation networks.}
  \label{tab:overview_wos_results}
\end{table*}
}

\section{Discussion}

\paragraph{Excellent classification of dynamic network models.}
The classification of the synthetic data based on networks dynamics has excellent accuracy higher than 98\%, greatly superseding the accuracy with state-of-the-art static features (92.8\%). The accuracy may improve only slightly when adding static features to the dynamic ones. Our dynamic features are interpretable in that we identify clear differences between distinct models. This is a highly desirable property for machine learned model selection.
\paragraph{Application to citation networks.}
Using dynamic features, \textsc{F\textsubscript{exp}AP} is often the most likely classification for citation networks. SO is an outlier in that it is never classified \textsc{F\textsubscript{exp}AP}. However, in some cases, other models are also quite plausible in terms of the likelihood of their classification. For instance, in Table~\ref{tab:size_cohort_both_features_class_probabilities} in the Supplementary Material we see that the classifier (with both static features and dynamic features with size-cohorts) estimates the likelihood of uniform attachment (U) for OC as 47.484\%, which is only slightly smaller than the likelihood 50.6094\% of \textsc{F\textsubscript{exp}AP}. This demonstrates that the most likely classification of real-life networks is not always robust, the entire distribution over the nine models is more informative. \revsec{Furthermore, in practice, we recommend to use both dynamic features designs -- with size-cohorts and time-cohorts -- for more reliable conclusions. In particular, in our application to citation networks, all feature choices have clear preferences for just a few classes (\textsc{F\textsubscript{exp}AP}, \textsc{AP}, sometimes \textsc{F\textsubscript{exp}A} and \textsc{P}). At the same time, many other classes are clearly ruled out, e.g., the quite plausible \textsc{F\textsubscript{pl}A} and most models without aging, see Tables~\ref{tab:static_features_class_probabilities} -- \ref{tab:size_cohort_both_features_class_probabilities}).}   

In search for a \revsec{simple} explanation why different features result in a different classification on real-life networks, \revsec{we have performed many additional experiments, for instance, computing distances between the synthetic and real-life feature vectors. These attempts however were not successful in the sense that simple measures like distances did not explain the classification. As an example, and for completeness, we report the results of our best attempt, the} two-dimensional UMAP (Uniform Manifold Approximation and Projection) embeddings of the features, see Section~\ref{sec:umap} of the Supplementary Material. \revsec{We see that the two-dimensional embeddings of different synthetic classes have significant overlap.} \revsec{This makes classification of citation networks with UMAP highly unreliable, and} exemplifies the complexity of learning the right model, even for {\em synthetic} networks. \revsec{Similarly, in our experiments with other known methods, the complex structure of our features could not be captured in spaces of much lower dimensions.}

\paragraph{Power laws in networks.}
Interestingly, by \cite{GarHofWoe17}, the \textsc{F\textsubscript{exp}AP} model, that is often selected by our classifier for citation networks, has {dynamical} power-law in-degrees with exponent decreasing over time. This is a new argument in favor of using power laws as a mathematical model for degree distributions. There is a lively debate about whether power-law in-degrees are rare or ubiquitous (see \cite{BroCla19, VoiHooHofKri19, Holm19} and the references therein). One of the problems with statistical evaluation of power laws is that, fundamentally, a power law holds only in the infinite network size limit~\cite{VoiHooHofKri19}. Model selection offers an alternative, albeit arguably qualitative, approach to infer the presence of power laws: first, identify the model based on a finite network; then analyze this model when the network size grows to infinity. 

\paragraph{A cautionary tale on ML techniques for model selection.}
Our machine learning methods work really well to distinguish between synthetic networks having different growth mechanisms. Moreover, these methods are general in that they identify the combination of mechanisms regardless the exact value of parameters. However, applying these methods to real-world networks with the aim to distinguish network growth mechanisms may lead to non-robust results because models are never perfect representations of the real-world networks, and machine-learning methods pick up differences \revsec{easily. We believe that the true reason for this sensitivity is that} in reality, citation networks are not accurately described \revsec{at a detailed level} by any of the models considered here. Therefore, depending on the features, the model selection yields the result that is {\it most consistent} with the chosen features. Hence, the best modeling choice depends on which features we want to capture.  This calls for systematic \review{approaches to feature design depending on a specific research question and application, as well as} validation methods of machine learned model selection. 

\paragraph{Further research.} 
It would be of interest to study {\em other} possible generative models. Particularly for citation networks, models with {\em copying}~\cite{kumar2000stochastic} or high probability of triangles~\cite{Overgoor2019NetworkFormationCitations} might be appropriate. This will account for the situation when authors cite both a paper and some references from that paper. One may also want to include geometry and make closer vertices more likely to connect~\cite{flaxman2006geometric} mimicking research topics and communities. \revsec{It would be highly interesting to find a set of mechanisms that will be consistently selected by different classifiers for all citation networks, if such a combination of mechanisms exists.} Furthermore, \revsec{we may consider networks where} links between fixed vertices change over time as many related real-world networks, such as social networks and the World Wide Web. 

Our dynamic features are computationally light, this enables further investigation of machine learning methods. For instance, one could include {varying network sizes}, explore other approaches to feature importance (e.g. Shapley values), and use regression to fit model parameters for real-life networks. \review{Moreover,} our features have explicit analytical expressions, thus we may hope to analytically derive the accuracy of machine learning methods and identify most informative features using the theory of continuous-time branching processes. 

\review{In terms of model selection, our cautionary tale requires further investigation. Our results show that model selection is sensitive to the choice of features. This calls for new methods of systematic feature design that is suitable for the problem at hand. New experiments can be done to go further in details in choosing the best classifier (e.g. training binary classifiers which predict the presence or absence of a dynamic network mechanism such as aging, broader parameter range for simulations, larger tuning hyperspace).}

\section{Methods}
\label{sec-methods}

\subsection{Collapsed CTBP models for dynamic networks}
\label{sec-CTBP-networks}

\paragraph{Modeling growth mechanisms with CTBPs.}
 Our models build on a continuous-time branching process (CTBP) \review{representation of preferential attachment models. The equivalence of these processes was established, for instance, in ~\cite[Theorem 3.3]{Athreya2007PreferentialFunction} and \cite[Theorem 2.1]{Athreya2008GrowthProcesses}. Furthermore, collapsed CTBPs represent the local weak limit of directed PAMs as established in \cite[Theorem 1]{rudas2007random} and \cite[Proposition 6.10]{garavaglia2020local}, see also a unified treatment of this result in \cite[Section 4.2]{BanerjeeOlvera2022pagerank}.} The CTBP's  are random growing trees. By defining the rates, at which vertices of a CTBP produce their offspring, one can model desired mechanisms of network's growth. In this work, the rates explicitly depend on the {\em fitness} of the vertex, its {\em age}, and its {\em degree}  (preferential attachment). Fix a vertex $u$ of CTBP, and let $t_u$ denote its birth time. Then, the rate at which vertex $u$ produces offspring at time $t$ is given by
  \eqn{
  \label{eq:CTBP-rate}
  \review{\eta_u h(t-t_u) f(\mbox{in-degree}_u(t)),\quad t>t_u,}
  }
where $h\colon [0,\infty)\to [0,\infty)$ denotes the aging function, $\eta_u$ the fitness of vertex $u$, $f$ the preferential attachment function, and $\mbox{in-degree}_u(t)$ is the in-degree, that is, number of offspring of vertex $u$ at time $t$. This gives a flexible class of CTBPs models for aging, fitness, and preferential attachment, moreover, we can switch off any of these mechanisms by taking $h$, $\eta_u$ and/or $f$ equal to 1. 
Motivated by empirical properties of citation networks (see Section \ref{sec:citation} in the Supplementary Material), we make the following choices in \eqref{eq:CTBP-rate}.

The {\it preferential attachment function} $f$, when present, is {\em affine}, a key example being the linear preferential attachment function used in the Barab\'asi-Albert model \cite{BarAlb99}. We take $f(k)=ak+b$, $a,b>0$.

 The {\it fitness variables} $(\eta_v)_{v\in V(T)}$ are positive i.i.d.\ random variables. In the FP models, the distribution of $\eta_v$'s must have finite support, since otherwise condensation occurs that is absent in citation networks~\cite{BiaBar01a,Borgs2007FirstFitness}. Therefore, we assume the uniform distribution on the interval $[c,d]$. In FAP models, in order to obtain sufficient variability in the network's degree distribution, $\eta_v$'s must have unbounded support, with at least an exponential tail~\cite{GarHofWoe17}, therefore we assume the exponential distribution with parameter $\lambda$. Finally, in F and FA models, we consider $\eta_v$'s to have either the exponential distribution or a power law, that is pure Pareto with density $(\tau-1)x_{\rm min}^{\tau-1}x^{-\tau}$, $x>x_{\rm min}$.

We let the {\it aging function} $h$, when present, be the density of a log-normal distribution conform to the log-normal aging as in Figure \ref{fig-empirical-properties-main}, 
\begin{equation*}
h(t)=\frac{1}{x\sigma\sqrt{2\pi}}\,e^{-\frac{(\ln(t)-\mu)^2}{2\sigma^2}}, \qquad t>0.
\end{equation*}

\paragraph{From trees to graphs with collapsed CTBP's.}
A CTBP produces a tree, which we convert to a graph $G(t)$ with out-degrees greater than $1$, using the following collapsing procedure. 

First, we generate random out-degrees $M_1,M_2,\ldots$ as independent copies of a random variable $M$ sampled from the empirical distribution of the number of references in a paper in the WoS OC network, with $\mathbb{E}[M] \approx 10.29$. The
OC network was chosen because it is sufficiently representative for the out-degree distributions in the other citation networks as well. See Section~\ref{sup:ecdf} in the Supplementary Material for details. 

Next, we start generating the CTBP, simultaneously assigning its vertices to  {\it batches} of sizes $M_1,M_2,\ldots$ in the order of their arrival. At time  $t$, graph $G(t)$ is produced by collapsing every batch into one vertex $v$ of $G(t)$. When generating the CTBP, all vertices in the batch of $v$ receive the same fitness $\eta_v$ and the same birth time $t_v$, equal to the birth time of the {\it oldest} CTBP vertex in the batch.  Furthermore, we make the total rate, at which a collapsed vertex $v$ produces offspring, independent of its out-degree $M_v$, by distributing the original rate \eqref{eq:CTBP-rate} equally among the $M_v$ collapsed vertices. Summarising, in our collapsed model, every vertex $u$ of the CTBP, that is to be collapsed into vertex $v$ of $G(t)$, produces offspring at rate
\begin{equation}
    \label{eq:rate-collapsed-CTBP}
\review{\frac{1}{M_v}\eta_v h(t-t_v) f(\mbox{in-degree}_u(t)),\quad t>t_v.}
\end{equation}
In application to citation networks, each collapsed vertex $v$ models one paper; this paper cites $M_v$ other papers; the publication date is the birth time of the oldest collapsed vertex; the citations to this paper are the offspring  produced by the $M_v$ collapsed vertices together.

\paragraph{Parameters range of the collapsed CTBP.} 
Table \ref{tab:dynamic_models} summarises the generative models, on which we have trained our machine-learning methods. The last column ({\bf `Power law'}) state which models have power-law in-degrees, based on the literature and Proposition~A.\ref{prop:fitness-only} in Section~\ref{sup:power-law-proof} of the Supplementary Material. Column {\bf `Range'} displays the parameter ranges that we used for different models. These ranges are derived from the following considerations. There are only three possible scenarios of the CTBP's growth: 1) CTBP is {\it subcritical}, it dies out after producing a random finite tree, 2) CTBP is {\it supercritical}, it grows exponentially in time  with positive probability, or 3) CTBP is {\it explosive}, it produces an infinite tree in a finite time. Since real-life citation networks grow exponentially in time (as in Figure \ref{fig-numberpublic-tailtogether}(a) in the Supplementary Material), we choose the range of parameters $a,b,c,d,\lambda, x_{\rm min},\tau, \mu,\sigma$ inside the region where the CTBP with the original rate \eqref{eq:CTBP-rate} is supercritical. We derive these conditions in Section~\ref{sup:supercritical}, and discuss the supercriticality of the collapsed process in Section~\ref{sup:supercritical-collapsed} of the Supplementary Material. In the models with aging, a supercritical branching process dies out with positive probability. In the experiments, for each parameter combination, we remove networks that die out, and keep the first network that achieves a size of $20,000$ within at most $1000$ attempts. For example, Figure~\ref{fig:class4_tau} in the Supplementary Material shows the number of \textsc{F\textsubscript{pl}A} networks that died out for different values of $\tau$. 
For simplicity of interpretation, we choose the parameter range so that the average rate of producing offspring is similar across different models. Finally, we choose the models such that the CTBP cannot be explosive. Thus, we exclude FP models with exponential or power law fitness, as such models do explode in finite time \cite{Borgs2007FirstFitness}.

\begin{table*}[th]
  \centerline{
  \begin{tabular}{l l|p{35mm} p{20mm} p{30mm} l}
    \textbf{Code} & \textbf{Name} & \textbf{Growth Mechanism} & \textbf{Parameters} & \textbf{Range} & \textbf{Power-law}\\
    \hline
    0 & \textsc{U} & Uniform Attachment & None & - & No~\cite[Exercise 8.17]{Hofstad2016random} \\
    \hline
    1 & \textsc{P} & Affine PA & $a,b$ & $a \in (1, 4), b \in (1, 4)$ & Yes~\cite[Theorem 8.3]{Hofstad2016random} \\
    \hline
    2 & \textsc{F\textsubscript{pl}} &Power-law Fitness & $(x_{\rm min}, \tau)$ & $x_{\rm min} \in (0.5, 1), \tau \in (2, 4)$ & Yes (Proposition~A.\ref{prop:fitness-only})\\
    \hline
    3 & \textsc{F\textsubscript{exp}} & Exponential Fitness & $\lambda$ & $\lambda 
    \in (0.1, 3)$ & No (Proposition~A.\ref{prop:fitness-only})\\
    \hline
    4 & \textsc{F\textsubscript{pl}A} & Power-law Fitness \newline Lognormal Aging & $(x_{\rm min}, \tau)$ \newline $(\mu, \sigma)$ & $x_{\rm min} \in (0.5, 1), \tau \in (2, 2.7)$ \newline $\mu \in (0.1, 3), \sigma = 1$ & Yes (Proposition~A.\ref{prop:fitness-only})\\
    \hline
    5 & \textsc{F\textsubscript{exp}A} & Exponential Fitness \newline Lognormal Aging & $\lambda$ \newline $(\mu, \sigma)$ & $\lambda 
    \in (0.1, 1)$ \newline $\mu \in (0.1, 3), \sigma = 1$ & No (Proposition~A.\ref{prop:fitness-only})\\
    \hline
    6 & \textsc{F\textsubscript{unif}P} & Affine PA \newline Uniform Fitness & $(a,b)$ 
    \newline $(c,d)$ & $a \in (1, 4), b \in (1, 4)$ \newline $ c \in (0.1, 1), d \in (1, 5)$ & Yes \cite[Theorem 3]{Borgs2007FirstFitness}\\
    \hline
    7 & \textsc{AP} & Affine PA \newline Lognormal Aging & $(a,b)$ 
    \newline $(\mu, \sigma)$ & $a \in (3.3, 7), b \in (1, 4)$ \newline $\mu \in (0.1, 3), \sigma = 1$ & No \cite[Section 5.1]{GarHofWoe17}\\
    \hline
    8 & \textsc{F\textsubscript{exp}AP} & Affine PA \newline Exponential Fitness \newline Lognormal Aging & $(a,b)$ \newline $\lambda$ \newline $(\mu, \sigma)$ & $a \in (1, 4), b \in (1, 4)$ \newline $\lambda 
    \in (0.1, a + \frac{b}{\E{M}})$ \newline $\mu \in (0.1, 3), \sigma = 1$ & Yes \cite[Section 5.2]{GarHofWoe17}\\
    \hline
  \end{tabular}
  }
  \caption{Dynamic network models with parameters. The parameter range is chosen so that the CTBP is supercritical, i.e., it grows exponentially in time (see Section~\ref{sup:supercritical} for details).}
  \label{tab:dynamic_models}
\end{table*}

\subsection{Machine learning methodology}
\label{ssec:ML-methods}
\paragraph{Static features.} The 36 static features are selected from recent literature. We compute static features on the {\it undirected} version of our networks because most of these features are designed for undirected graphs. Two static features are global characteristics of the graph: the assortativity coefficient~\cite{Ikehara2017CharacterizingDomains, Attar2017ClassificationFeatures, Rossi2019ComplexDomain} is the Pearson's correlation coefficient between degrees of two vertices connected by an edge; the global clustering coefficient (or, transitivity)~\cite{Ikehara2017CharacterizingDomains, Attar2017ClassificationFeatures, Langendorf2020EmpiricallyMechanisms} is the fraction of triangles among connected triplets of vertices. Other features describe the distributions of numerical {\it values} of individual vertices. For each value we compute the minimum, the maximum, the mean, the standard deviation, and 3 to 5 quantiles that proved descriptive for different networks. Two values are vertex centralities: in-degrees~\cite{Attar2017ClassificationFeatures,Blasius2018TowardsModels, Rossi2019ComplexDomain, Langendorf2020EmpiricallyMechanisms} (quantiles 0.125, .25, .5, .75, .875), and coreness~\cite{Blasius2018TowardsModels,Rossi2019ComplexDomain} (quantiles .25, .5, .75). The coreness of a vertex equals the largest number $k$ such that the vertex is in the $k$-core -- the maximal induced subgraph with all vertices having degree $k$ or higher. Another type of values describes the vertex neighborhoods. Of these, we use the number of triangles that contain the vertex~\cite{Rossi2019ComplexDomain,Langendorf2020EmpiricallyMechanisms} (quantiles .80, .90, .95, .97, .99, which have sufficient variance to be used as features), and the local clustering coefficient~\cite{Rossi2019ComplexDomain,Blasius2018TowardsModels} (quantiles .5, .6, .7, .8, .9). The latter is the fraction of pairs of neighbors of the vertex that are connected to each other. 
Our features do not need to be independent of the network size~\cite{Ikehara2017CharacterizingDomains}, because 
all synthetic networks in our training set have the same size. 

We do not include edge density~\cite{Blasius2018TowardsModels}, because our networks are of comparable density defined by the average in/out-degree $\E{M}$.  We do not use PageRank~\cite{Blasius2018TowardsModels,Langendorf2020EmpiricallyMechanisms} and Katz index~\cite{Blasius2018TowardsModels}, because of their similarity to in-degree in our models. Diameter~\cite{Blasius2018TowardsModels}, closeness and betweenness centrality~\cite{Blasius2018TowardsModels} and motif counts~\cite{Ikehara2017CharacterizingDomains,Langendorf2020EmpiricallyMechanisms} are excluded because their infeasible computational costs on many networks of size 20,000. Finally, our synthetic networks are not expected to contain communities, thus we omit the number of communities~\cite{Langendorf2020EmpiricallyMechanisms}. 

\paragraph{Novel dynamic features.}

A core contribution of this paper is the \textit{dynamic feature design} that directly summarizes the {\em evolution} of the dynamic graph $G(t)=(V(t), E(t))$, $t\in (0,T]$. Our dynamic features form a Dynamic Feature Matrix (DFM) $D$ of size $s \times r$, defined as follows. 
The {\it columns} $j\in [r]$ of the matrix $D$ correspond to groups of vertices $V_j$ between $(r-j)/r$-th and $(r-j+1)/r$-th quantiles of the empirical degree distribution at time $T$. Formally, 
\begin{align*}
V_{j} = \Big\{ & v \in V(T) \colon 
\frac{r-j}{r} < \frac{1}{|V(T)|} \sum_{w \in V(T)} \\
&\mathbbm{1}\{\deg_{w}(T)\le \deg_{v}(T)\} \leq \frac{r-j+1}{r} 
\Big\}, \; j \in [r].
\end{align*}
For instance, $V_1$ contains the vertices whose degree at time $T$ is in the highest $(100/r)$\% of $V(T)$. 

The {\it rows} $i\in[s]$ of the matrix $D$ stand for cohorts of vertices in the order of their arrival. More specifically, we define time instants $0=\tau_0<\tau_1<\cdots<\tau_s=T$, and let cohort $i$ consist of those vertices that have arrived in the time interval $(\tau_{i-1},\tau_i]$. To define the $\tau_i$'s, we consider {\it size}-cohorts and {\it time}-cohorts. Size-cohorts  all contain the same number of vertices, and $\tau_i$ is the time when the $\lfloor |V(T)| i/s \rfloor$-th vertex arrives. Time-cohorts correspond to the calendar time, they consist of vertices that have arrived in equal time intervals, and $\tau_i = Ti/s$. Note that in the CTBP, the number of vertices in time-cohort $i$ increases exponentially in $i$, so the highest-degree vertices in $V_1$ attract many edges from the last, the largest, time-cohort. 

The {\it entries} $ij$ of the matrix $D$ are the normalized average number of edges received by vertices in $V_j$ from cohort $i$: 
\begin{equation}
\label{eq:Dij}
  D_{ij} = C\Delta_{ij}, \quad i\in [s], j\in [r],
\end{equation}
where 
\begin{equation}
\label{eq:Delta}
\Delta_{ij}= \frac{1}{|V_j|}\sum_{v \in V_j}(\deg_{v}(\tau_{i}) - \deg_{v}(\tau_{i-1})),
\end{equation}
and 
$C$ is the normalization constant so that $\sum_{i=1}^s\sum_{j=1}^rD_{ij}=1$. This normalization is a common practice to prevent overfitting. In our case, $\Delta_{ij}$ increases with the graph size, and, by construction, decreases in $s$ and $r$. The normalization ensures that our learned classes are robust with respect to the choice of $|V(T)|$, $s$ and $r$. Intuitively, the dynamic features are meaningful for model selection because different models prescribe different likelihood of vertices to attract new edges. This is exactly what we are able to track with dynamic features. 

\paragraph{Choice of classifier} is explained in Section~\ref{sup:classifier} of the Supplementary Material. 

\section{Data availability}
The dataset with synthetic networks and their static and dynamic features: 
DOI 10.4121/0b40e329-7b33-4b8f-a289-ab6d1893b437 available at \url{https://data.4tu.nl/datasets/0b40e329-7b33-4b8f-a289-ab6d1893b437}. 
\section{Code availability}

Code repository for network simulation and feature extraction:\\ https://gitfront.io/r/user-6239985/R9WcT8Msr46T/DynamicNetworkSimulation/

\paragraph{\bf Acknowledgements.}
The work of RvdH and NL was supported in part by the Netherlands Organisation for Scientific Research (NWO) through Gravitation-grant {\sc NETWORKS}-024.002.003. The Web of Science data set used in this research results from a collaboration with the Centre for Science and Technology Studies (CWTS) at Leiden University. We thank Ludo Waltman for preparing the data set. Part of the work by NL was performed at the University of Twente.

\paragraph{\bf Author contributions.}
LT contributed to the writing, contributed to the conception of the machine learning features, contributed to the writing, and performed the machine learning experiments. DB contributed to, and guided, the machine learning experiments. AG contributed to the initial conception of the machine learning features, the dynamic network models, and initial experiments. RvdH and NL contributed to the conception of the project, the dynamic network models, the conception of the machine learning features, and the writing. 

\paragraph{\bf Competing interests.} None.

\printbibliography

\newpage
\onecolumn

\appendix
\input{supplementary}

\end{document}

%% file: supplementary.tex
\maketitle
\appendix
\section{Supplementary Material}
\setcounter{page}{2}

\renewcommand\thefigure{\thesection.\arabic{figure}}  
\renewcommand\thetable{\thesection.\arabic{table}}
\renewcommand\theequation{\thesection.\arabic{equation}}

\subsection{Empirical properties of the evolution of citation networks}
\label{sec:citation}

In citation networks, vertices are scientific publications, and directed edges are citations from one publication to another. The structure of citation networks has received tremendous attention, ever since the groundbreaking work of De Solla Price \cite{Pri65,Pric86}. Recent studies of citation networks include \cite{WanSonBar13, WanYuYu08, WanMeiHic14, GarHofWoe17}. Here we discuss some of the properties of citation networks that form the basis of the network models we consider in this paper.

\begin{figure}[ht]
\centering
\scalebox{0.5}{\input{supp-figures/number_publications.tikz}}
\caption[Number of publications per year]{ Number of publications per year in the fields  (logarithmic vertical axis).} 		
\label{fig-numberpublic-tailtogether}
\end{figure}
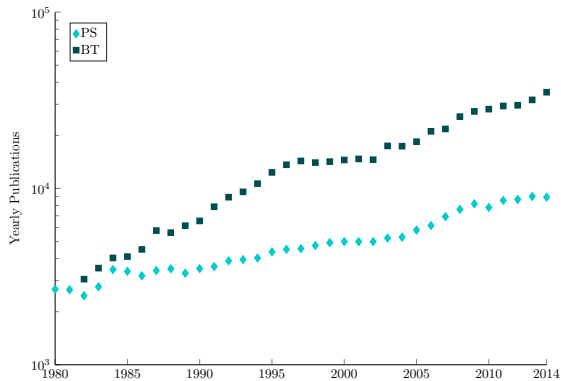

\paragraph{Citation networks grow exponentially in calendar time.}  We illustrate this with  examples from the fields: Biotechnology (BT), Probability and statistics (PS). These examples were generated by us for the new book \cite{RGCN-2}, we include them here to make the current paper self-contained, and refer the reader to \cite{RGCN-2} for more details. 
Figure \ref{fig-numberpublic-tailtogether} displays the exponential growth in citation networks in the fields BT and PS.

\begin{figure}[ht]
	\centering
	\scalebox{0.33}{\input{supp-figures/PS_random_sample.tikz}}
	\scalebox{0.33}{\input{supp-figures/BT_random_sample.tikz}}
	\caption[Time evolution for citations of 20 randomly chosen papers]{Time evolution for citations of 20 randomly chosen papers from 1980 for PS, and from 1982 for BT.}
	\label{fig-randomsample}
\end{figure}
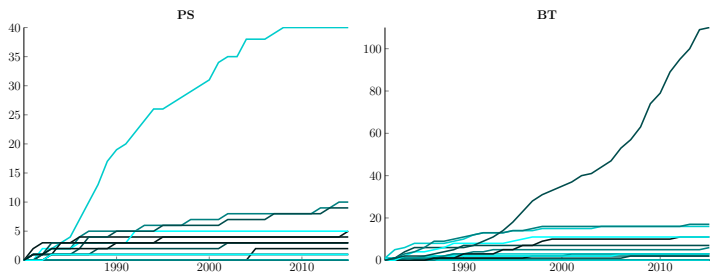

\begin{figure*}[th]
\centering
\begin{subfigure}{0.3\textwidth}
  \scalebox{0.33}{\input{supp-figures/BT_dynamic_PL_data.tikz}}
  \caption{}
\end{subfigure}
\begin{subfigure}{0.3\textwidth}
  \scalebox{0.33}{\input{supp-figures/BT_degree_increment.tikz}}
  \caption{}
\end{subfigure}
\begin{subfigure}{0.3\textwidth}
  \scalebox{0.33}{\input{supp-figures/BT_age_cited.tikz}}
  \caption{}
\end{subfigure}
\caption[Empirical properties of citation networks]{Empirical properties of BT citation network~\cite{RGCN-2}.  (a). {Degree distribution for papers from 1984 over time. Each log-log plot shows the citation distribution in a given later year. As time progresses, the number of citations naturally increases in a way that the tail becomes significantly heavier, stabilizing after a long time. This is modeled with dynamic power laws: the number of citations follows a power law, and its exponent is decreasing with time.} (b). Average citation increment over a 20-years time window for papers published in different years show aging effect. (c). Distribution of the age of cited papers for different citing years. We have used a 20-year time window in order to compare different citing years. The shape of the distribution for different citing years in each field is very similar and resembles a log-normal density.}
\label{fig-empirical-properties-main}
\end{figure*}
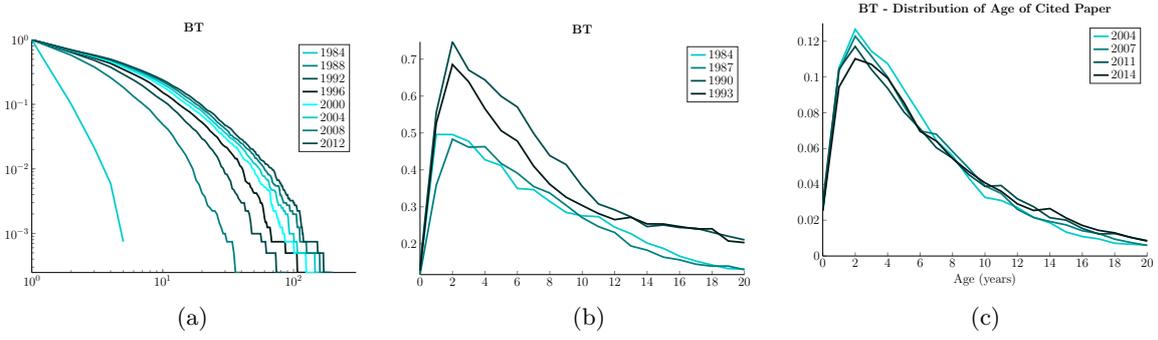

\paragraph{Citation networks have highly variable degrees.} 
Figure \ref{fig-randomsample} shows the citation evolution of 20 random papers from 1980. While this is an anecdotal example, we do see  that some papers keep on attracting citations, while others stop being cited quite quickly. Again, this is a sign of the highly-variable degree evolution. 
{Motivated by the obviously high variability of citation counts, a vast body of literature starting with~\textcite{Pri65} has suggested power-law in-degree distributions in citation networks. The question whether power-laws in networks are `rare' or `omnipresent' has caused considerable scientific debate \cite{ArtSmoVinWit20, BroCla19, Holm19, VoiHooHofKri19}.}
For our data, the log-log plots in Figure \ref{fig-numberpublic-tailtogether}(b) suggests that the in-degree distribution could be a power-law, even though the tails might also be slightly thinner. As time proceeds, important papers attract more citations making power laws more visible. 
Figure \ref{fig-empirical-properties-main}(a) shows the {\em dynamics} of in-degree distribution in BT from a sample of papers published in 1984. We see a strong time dependence, in that the degree distribution of a cohort of papers from a given time period (in this case the year 1984) becomes more and more heavy-tailed. 

\paragraph{Citation increments depend on the age of a paper.} In citation networks, the majority of papers stop receiving citations after some time, while few others keep being cited for longer times (see e.g., Figure \ref{fig-randomsample} for the citation patterns of a few papers). Typically, the average increment of citations increases first, and then decreases over time (see Figure~\ref{fig-empirical-properties-main}(b). This is a manifestation of {\em aging}. Moreover, in Figure \ref{fig-empirical-properties-main}(c), we plot, as an example, the distribution of the age of cited papers, by papers published in a given year. The plot suggests a log-normal distribution of the age of a cited paper. Notably, the shape of this log-normal distribution is rather stable over time, and is similar for different scientific fields.

\subsection{Empirical out-degree distribution used in the synthetic networks}
\label{sup:ecdf}

\begin{figure}[H]
    \centering
    \scalebox{0.46}{\includegraphics{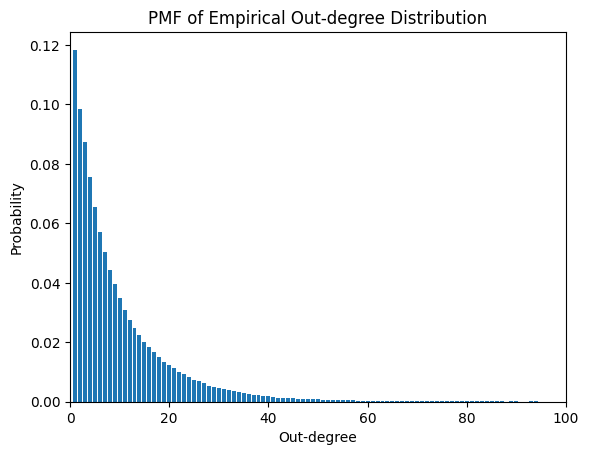}}
    \scalebox{0.46}{\includegraphics{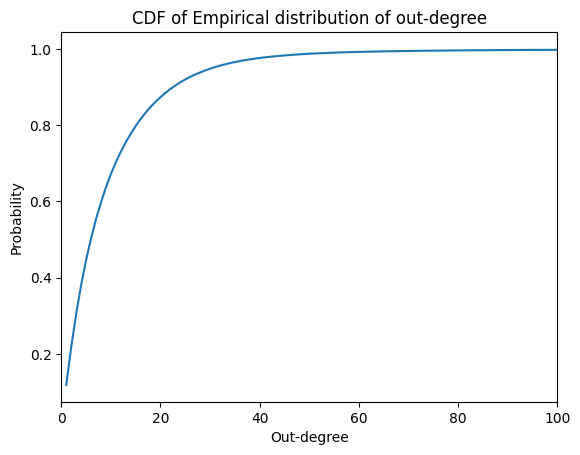}}
    \caption{Empirical Out-degree Distribution}
    \label{fig:ecdf}
\end{figure}

Recall that we denote by $M$ a random variable that represents the out-degree of a vertex in synthetic networks. To model this random variable, we use the empirical distribution of the number of references in a paper in the Web of Science OC network. The OC network was chosen as its out-degree distribution is representative for the other out-degree distributions of the other citation networks. 
Importantly, $M$ counts references only to the papers in the dataset, so it is smaller than the total number of references in a paper. Figure~\ref{fig:ecdf} shows the empirical mass function and cumulative distribution function of $M$.
Table~\ref{tab:empirical_stats} provides the statistical summaries of this empirical distribution. The low mode (equal 1) is due to the fact that we count only references within the dataset.

\begin{table}[H]
    \centering
    \begin{tabular}{c|l}
        $\mathbb{E}[M]$ & $10.29$ \\
        $\text{Var}[M]$ & $181.189$ \\
        Median & $6$ \\
        Mode & $1$ \\
        Min & $1$ \\
        Max & $800$ \\
    \end{tabular}
    \caption{Descriptive statistics for empirical out-degree $M$}
    \label{tab:empirical_stats}
\end{table}

\subsection{Supercriticality of standard CTBP}
\label{sup:supercritical}

 In this section we will derive the supercriticality conditions for all combinations of mechanisms in the standard CTBP with rate given by \eqref{eq:CTBP-rate}.   Let $\xi(t)$, $t>0$, be the arrival process of offspring of a vertex in such CTBP. A supercritical CTBP grows approximately as $e^{\alpha t}$, where $\alpha>0$ is the so-called {\it Malthusian} parameter. Formally, a CTBP is supercritical if there exists $\alpha>0$ such that \begin{equation}
\label{eq:supercritical}
    \int_0^{\infty}e^{-\alpha t}\E{\xi(dt)}=1.
\end{equation}
The point process $\xi(t)$ counts the number of offspring of the root vertex. We rewrite the left-hand side of \eqref{eq:supercritical} using integration by parts, as follows 
\[\int_0^{\infty}e^{-\alpha t}\E{\xi(dt)}=\int_0^{\infty}\alpha e^{-\alpha t}\E{\xi(t)}\,dt=\E{\xi(X)},\]
where $X$ is an exponential random variable with parameter $\alpha$. It follows that $\alpha$ in \eqref{eq:supercritical} exists if and only if $\E{\xi(\infty)}>1$. The supercriticality condition is then 
\begin{equation}
    \label{eq:supercritical-general}
    \E{\xi(\infty)}>1.
\end{equation}
We will now apply \eqref{eq:supercritical-general} to our nine models: 

\begin{itemize}
\item[[U]] The rate of producing offspring at age $t>0$ is $1$, therefore $\E{\xi(\infty)}=\infty$, the CTBP is supercritical.
\item[[P]] The rate of producing offspring at age $t>0$ is $a \xi(t)+b$.
We have that $\E{\xi(\infty)}=\infty$ if $b>0$ and $\xi(\infty)=0$ otherwise. The supercriticality condition is therefore: $b>0$.
\item[[\textsc{F\textsubscript{pl}, F\textsubscript{exp}}]] 
The rate of producing offspring is $\eta$. Therefore, both exponential and power law fitness result in $\E{\xi(\infty)}=\infty$, the corresponding CTBP's are supercritical. 

\item[[\textsc{F\textsubscript{pl}A}]] The rate of producing offspring is $\eta h(t)$, and the total number of offspring is $\eta H(\infty) =\eta$, where $H(t)$ is the cumulative aging function given by $H(t) = \int_{0}^{t}h(s)ds$. Therefore, the supercriticality condition is \[\E{\eta} = \frac{(\tau-1)x_{\rm min}}{\tau-2} >1.\]

\item[[\textsc{F\textsubscript{exp}A}]] Similarly as in \textsc{F\textsubscript{pl}A}, the supercriticality condition is
\[\E{\eta} = \lambda^{-1} >1.\]

\item[[\textsc{F\textsubscript{unif}P}]] The rate of producing offspring is $\eta(a\xi(t)+b)$. The supercriticality condition is:
\[b>0, \quad P(\eta>0)>0.\]
\item[[AP]] The rate of producing offspring at age $t>0$ is $h(t)(a\xi(t)+b)$. By solving the differential equation $y'(t)=h(t)(ay(t)+b)$ with initial conditions $y(0)=0$ and $y'(0)=h(0)b$, we obtain that the average number of offspring produced by time $t$ is $\frac{b}{a}e^{a H(t)}-\frac{b}{a}$. Since $H(\infty)=1$, we obtain the supercriticality condition
\[\frac{b}{a}e^{a}-\frac{b}{a}>1.\]

\item[[\textsc{F\textsubscript{exp}AP}]]  The rate of producing offspring is $\eta h(t)(a\xi(t)+b)$. By solving the differential equation $y'(t)=\eta h(t)(a y(t)+ b)$ with initial conditions $y(0)=0$ and $y'(0)=\eta h(0) b$, we obtain that the average number of offspring produced by time $t$ is $\frac{b}{a}e^{a \eta H(t)}-\frac{b}{a}$. Using again that $H(\infty)=1$, we obtain the supercriticality conditions:
\[\frac{b}{a}\E{e^{\eta a }}-\frac{b}{a} > 1.\]
Recall that $\eta$ has exponential distribution with parameter $\lambda$. If $\lambda<a$, then the left-hand side is infinite, so the process is supercritical. If $\lambda>a$, then the condition becomes
\[\frac{b}{\lambda-a} >1.\]
\item[[A]] The model with only aging is excluded from our analysis because in our chosen setting this model is not supercritical. Indeed,  our assumption that $h$ is the log-normal density, we have that $\E{\xi(\infty)}=1$.

\end{itemize}  

\subsection{Supercriticality of collapsed CTBP}
\label{sup:supercritical-collapsed}

The collapsing procedure that we propose in this paper is motivated by applications in citation networks, and does not permit easy derivation of supercriticality conditions. When $M=m$ is a constant, and the birth times of the vertices in CTBP are not adjusted, the supercriticality condition becomes 
\begin{equation}
    \label{eq:supercritical-general-m}
    \frac{1}{m}\E{\xi(\infty)}>1.
\end{equation}

When $M$ is random, a batch of $M$ vertices born one after another receives factor $1/M$ to their rate of producing offspring. Therefore, the probability that a randomly sampled vertex in CTBP has factor $1/m$, is $\frac{mP(M=m)}{\E{M}}$ as in the size-biased distribution of $M$. Hence, on average, this factor is 
\[\sum_m\frac{1}{m}\cdot \frac{mP(M=m)}{\E{M}}=\frac{1}{\E{M}}.\] 
Yet, we cannot simply replace constant $m$ by $\E{M}$ in \eqref{eq:supercritical-general-m} for two reasons. First,  the factor $1/M$ is assigned to a batch of consecutive vertices of CTBP, and not independently to each vertex. The difference is considerable because, if the factor $1/M$ was sampled independently from the size-biased distribution, then large values of $M$ had a high chance to appear quickly, slowing down the growth of CTBP due to the factor $1/M$, so this makes extinction more likely. On the other hand, in our procedure, we sample from the distribution of $M$ for each batch, so large values of $M$ are likely to appear later, and before they appear, the CTBP might already generate many offspring to ensure survival. At this point, we cannot exactly quantify the effect of large values of $M$ appearing independently for each vertex in CTBP or in batches. Second, and maybe even more importantly, we assign earlier birth times to the vertices in CTBP that create a collapsed vertex $v$. Then, many vertices of CTBP start generating offspring earlier than they would originally, which increases the chance of survival. Again, there are no analytical results that evaluate the effect of earlier birth times on the survival of the CTBP.

\begin{figure}
    \centering
    \includegraphics[width=0.6\textwidth]{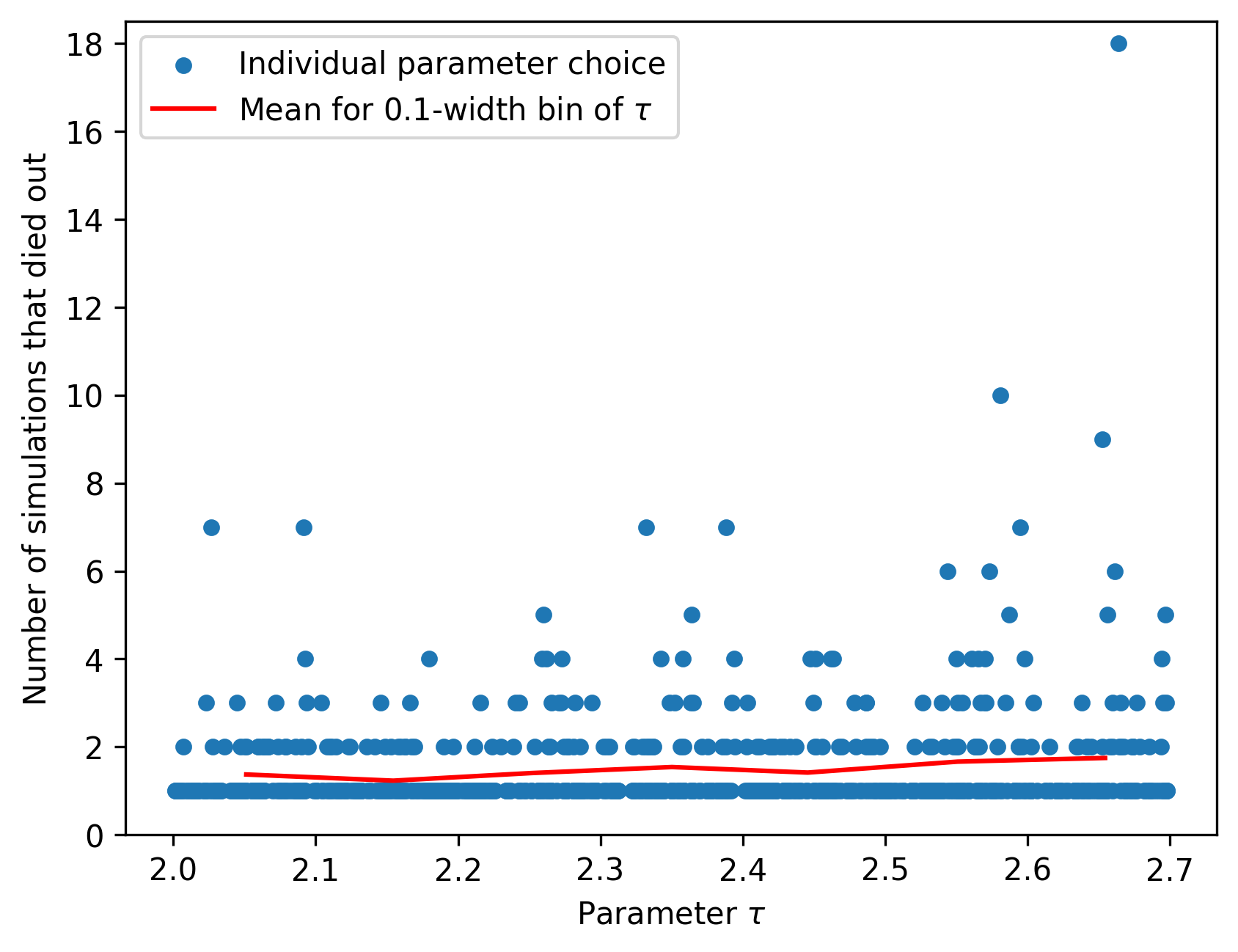}
    \caption{Number of attempts for simulation of class \textsc{F\textsubscript{pl}A} that died out.}
    \label{fig:class4_tau}
\end{figure}

In our experiments we always chose parameters that satisfy \eqref{eq:supercritical-general}, because the collapsed process can survive only if the CTBP with $M=1$ survives. Further, we found that the collapsed CTBP process often survives even when \eqref{eq:supercritical-general-m} is violated, rendering requirement \eqref{eq:supercritical-general-m} conservative. As an example, Figure~\ref{fig:class4_tau} shows survival and extinction of the collapsed process for the model \textsc{F\textsubscript{pl}A}. The figure shows the number of simulations that died out for $\tau$ between $2$ and $2.7$, with $x_{\rm min}$ sampled randomly from the interval $(0.5,1)$ as  stated  in Table~\ref{tab:dynamic_models}. When $\tau=2.7$, the value $\frac{(\tau-1)x_{\rm min}}{\tau-2}$ computed in Section~\ref{sup:supercritical}, ranges from $1.2143$ to $2.4286$, which is much smaller than $\E{M}=10.29$. In Figure~\ref{fig:class4_tau}, when $\tau$ grows,  we see a slight increase in the number of outliers, but all choices of $\tau$ still succeed in under 20 attempts, with the vast majority succeeding in less than 3.

As a final note, the model \textsc{F\textsubscript{exp}AP} has many parameters, so we use \eqref{eq:supercritical-general-m} to define the range of $\lambda$.

\subsection{Power-law degree distribution in F and FA models}
\label{sup:power-law-proof}

\begin{proposition}\label{prop:fitness-only} 
Consider a supercritical CTBP without preferential attachment, and let $(p_k)$ be its limiting degree distribution. Suppose that the fitness of each vertex is an independent copy of random variable $\eta$ with $\E{\eta}<\infty$, and let $\gamma$ be an aging random variable with density $h$. Then the following statements hold: 
\begin{itemize}
  \item[(i)] If $\eta$ has a power-law distribution, then $p_k$ decays as a power law with the same exponent.
  \item[(ii)] If $\eta$ has an exponential distribution, then $p_k$ decays faster than any power law.
\end{itemize}
\end{proposition}

{\it Proof.} Let $X$ be an exponentially distributed random variable with parameter $\alpha$ that is the Malthusian parameter of the supercritical CTBP. We will use the fact that $p_k$ equals the probability mass function of the number of offspring of the root vertex of the CTBP at time $X$~\cite[Definition 3.12]{GarHofWoe17}. Since the root produces children according to a Poisson process, we have that
\begin{equation}
\label{eq:fitness-only}
  p_k = \E{e^{-\gamma\eta X}\frac{(\gamma\eta X)^k}{k!}}.
\end{equation}
In \eqref{eq:fitness-only}, the main contribution comes from $\gamma\eta X=k(1+o(1))$ as $k\to\infty$. Thus, the decay of $\sum_{s=k}^{\infty} p_s$ is the same as that of $\P(\gamma\eta X \ge k)$. Since $X$ and $\gamma$ have all finite moments, and $\gamma$, $\eta$, $X$ are independent, the power law of $\eta$ implies the power law of $\gamma\eta X$ with the same exponent as $\eta$ (see e.g. \cite[Lemma 2.3]{DavisMikosch2008extreme}). This proves (i). Next, if $\eta$ has exponential distribution, then $\gamma\eta X$ has all finite moments and thus $\P(\gamma\eta X \ge k)$ has no power-law decay, which proves (ii). \hfill $\Box$

\subsection{Size and number of networks in the synthetic dataset}
\label{secsup:learning-curves}

\begin{figure}
    \centering
        \begin{subfigure}{0.49\textwidth}
    \includegraphics[width=0.9\textwidth]{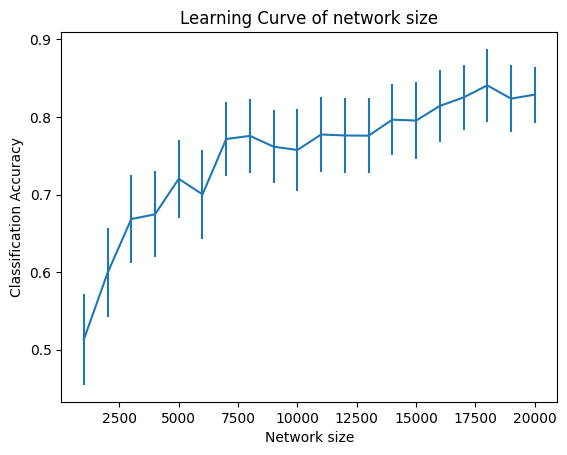}
    \caption{}
    \label{fig:learning_curve_networksize}
    \end{subfigure}
    \begin{subfigure}{0.49\textwidth}
    \includegraphics[width=0.9\textwidth]{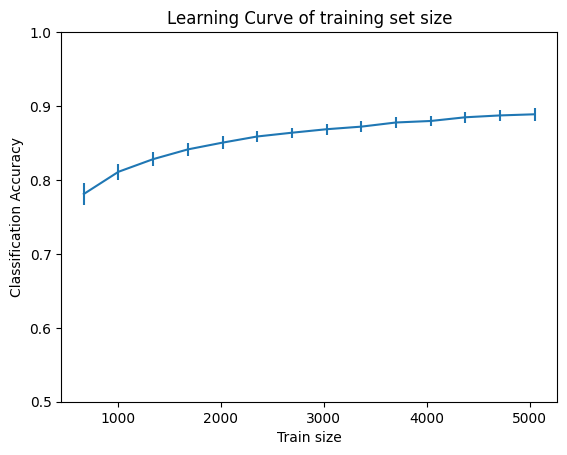}
    \caption{}
    \label{fig:learning_curve_trainingsize}
    \end{subfigure}
    \caption{Learning curves over: (a) the size of a network in the synthetic dataset; (b) the number of synthetic networks in the training set.}
\end{figure}

\paragraph{Choosing the size of synthetic networks.} We aim at synthetic networks' size of the order $10^4$ because our real-life WoS dataset has networks  ranging from $80,000$  to $400,000$ papers.   Generating the networks, especially with aging, is computationally expensive, therefore, the networks' size shouldn't be larger than necessary. Our choice for the size $20,000$ is justified as follows. 
First, we generate a small dataset, 40 networks for each category, with $20,000$ vertices. 
Then, we calculate the time-cohort feature matrix for this dataset at various sizes of the network: \{$1\cdot 10^{3}, \ldots, 20 \cdot 10^{3}$ \}. 
We then plot the \textit{learning curve}, see Figure~\ref{fig:learning_curve_networksize}. The plot shows the mean accuracy of 100 decision trees with random $80$-$20$ train-test splits, as a function of the networks' size.
We use a decision tree as a baseline for the amount of signal in the features.  
The classification accuracy continues to improve as the network grows, but levels off after the networks' size $10,000$. From this, we conclude that the network size $20,000$ in the synthetic dataset is sufficient for our  classification task. 

\paragraph{Choosing the number of synthetic networks.} Our final synthetic dataset consists of $6733$ networks of size $20,000$,  with $750$ or slightly less networks  for each of the nine models (17 parameter combinations in models with aging repeatedly resulted in extinction). To determine whether the dataset is large enough, we plot a learning curve over the training set size, see Figure~\ref{fig:learning_curve_trainingsize}.
Specifically, we vary the size of the training set, and use the rest of synthetic networks in the test set.  
For each training set size, we train $100$ decision trees with $100$ different train-test splits, and consider the classification accuracy.
As before, the decision tree classifier was chosen merely as a metric for the amount of signal in the data.
We observe that the learning curve flattens well before the largest training set size.
We conclude that we have sufficient synthetic data.

\paragraph{Computational resources.} The collapsed CTBP simulation was implemented in Python. The generation of $6733$ networks took 20 hours, using all cores on an Intel i7-7700HQ.\footnote{https://gitfront.io/r/user-6239985/R9WcT8Msr46T/DynamicNetworkSimulation/}

\subsection{Choosing a classifier}
\label{sup:classifier}

We have classified synthetic dynamic networks in one of the nine categories (each category corresponding to one generative model), using state-of-the-art static features, our novel dynamic features, and the combination thereof. We considered a broad variety of classifiers with their scikit-learn implementation~\cite{scikit}: support vector classifiers, tree-based models (decision tree), Naive Bayes (Gausian, Multinomial),  and ensemble models (Random Forest, Histogram-based Gradient Boosting Classification Tree). Performance is evaluated in terms of total classification accuracy and a confusion matrix. We take the standard 20-80 stratified test-train split, to ensure that both the test and train set are balanced for each category. We use $5$-fold cross validation on the training set to select the classification model and tune hyperparameters. We obtain the best performance in the cross validation with a state-of-the-art Histogram-based Gradient Boosting Classification Tree with the logistic loss function. The final accuracy and confusion matrices for the classification of the synthetic data in Section~\ref{ssec:classifying-synthetic}, are on the test set.  For the classification of citation networks in Section~\ref{sec-classifying-citation}, we train the classifier on the full synthetic data set. 

\subsection{Feature importance analysis} \label{sup:feature-importance}
\begin{figure}
\centering
\begin{subfigure}{0.49\textwidth}
\includegraphics[width=0.9\textwidth]{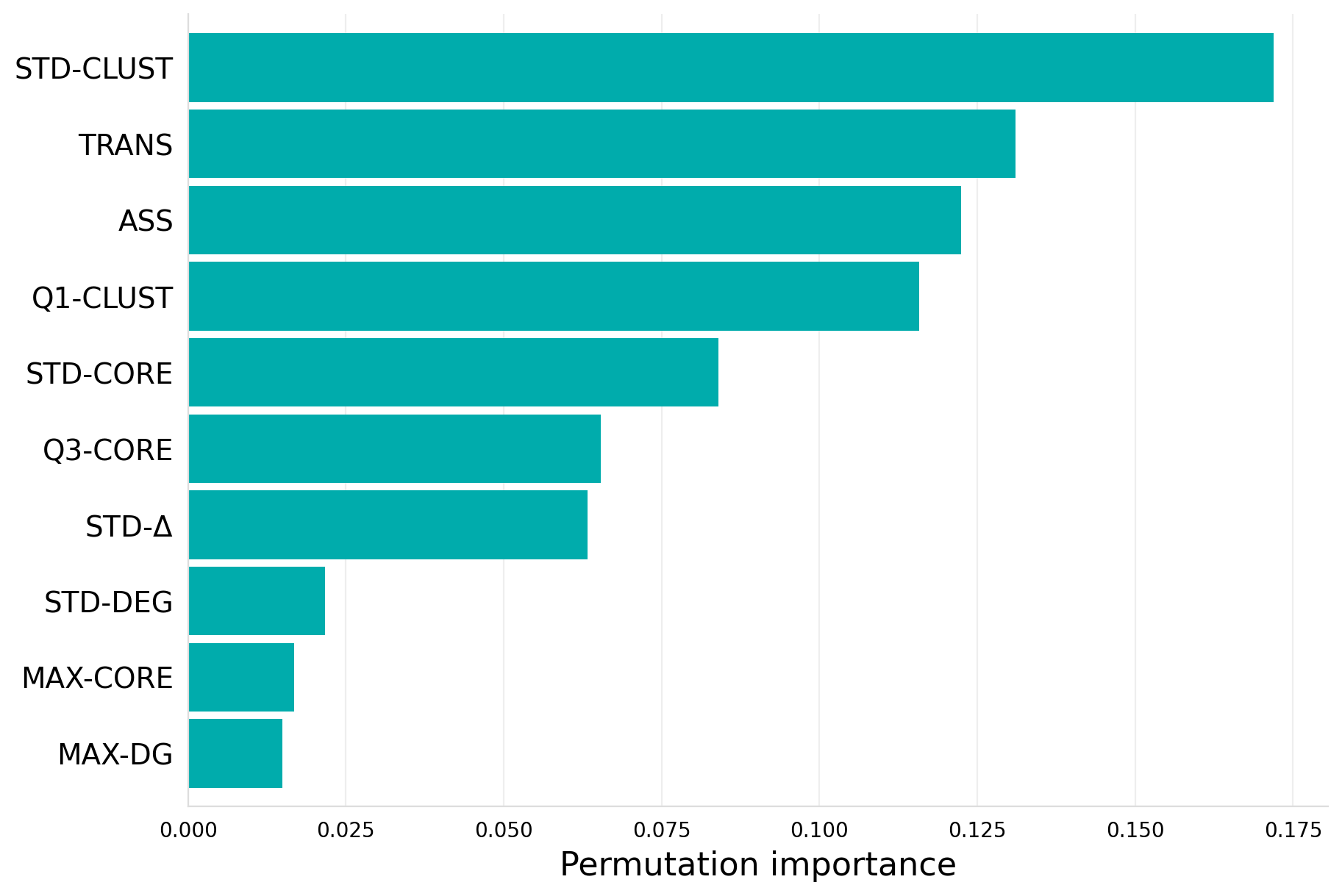}
\caption{}
\end{subfigure}
\begin{subfigure}{0.49\textwidth}
\includegraphics[width=0.9\textwidth]{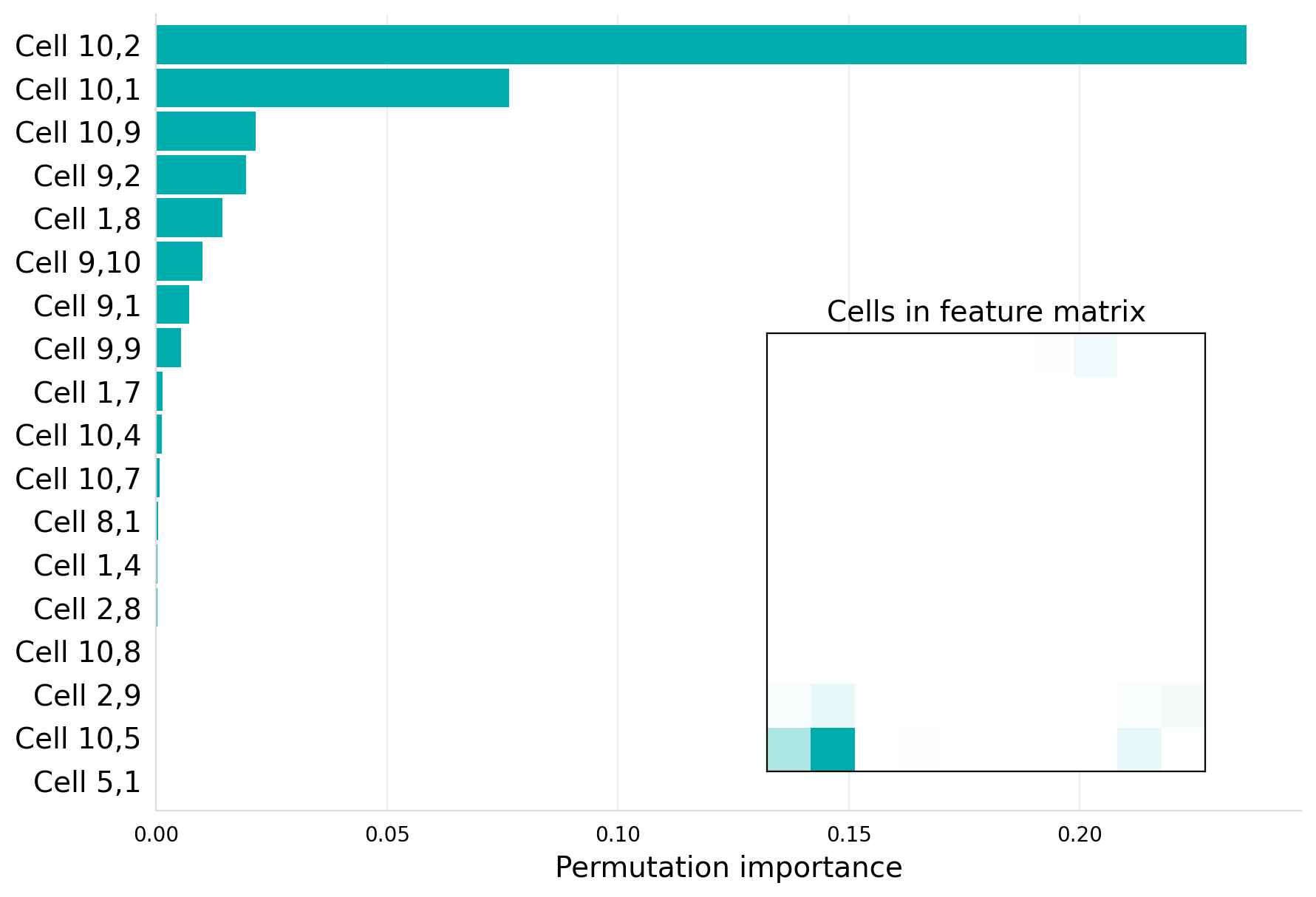}
\caption{}
\end{subfigure}
\begin{subfigure}{0.49\textwidth}
\includegraphics[width=0.9\textwidth]{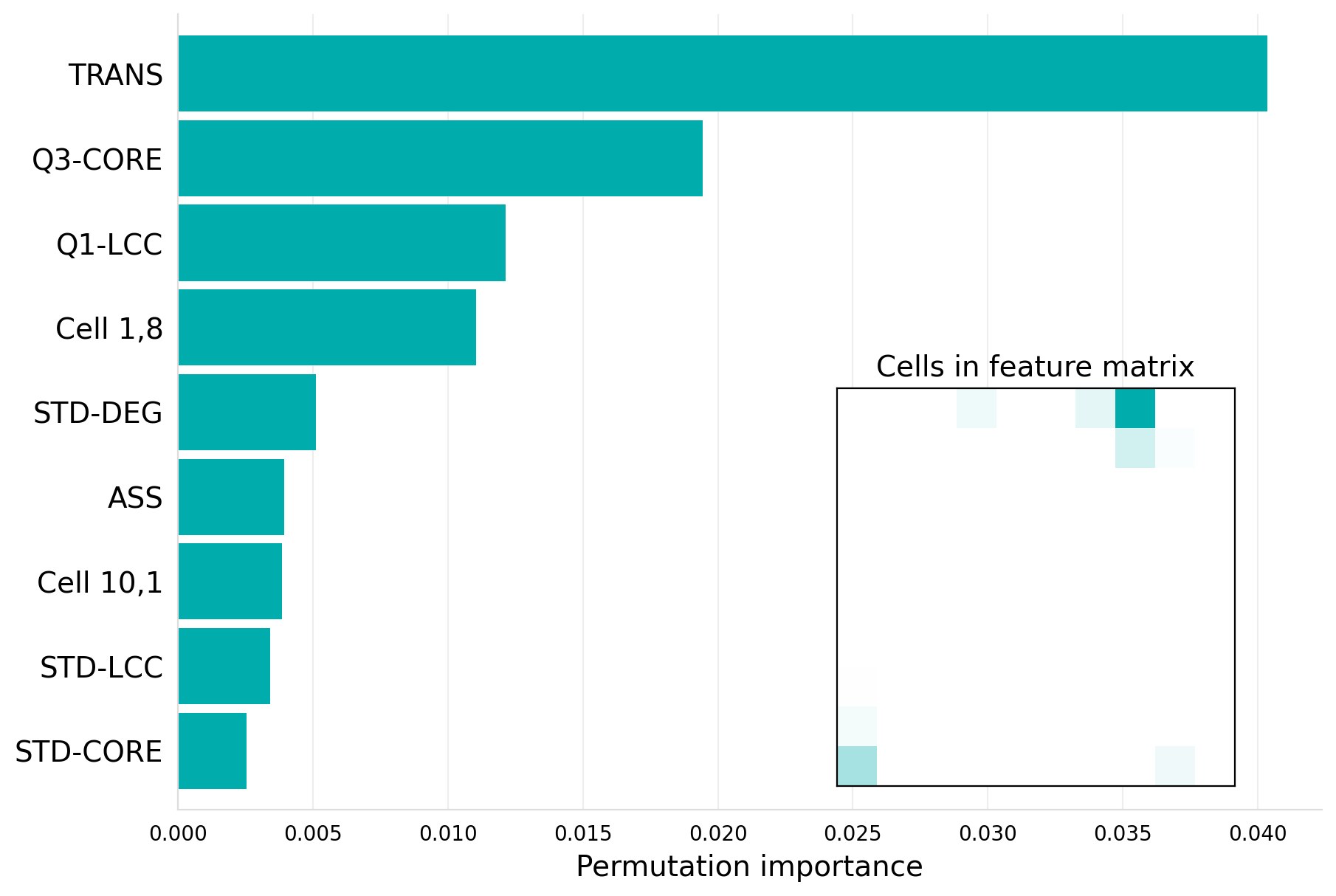}
\caption{}
\end{subfigure}
  \caption{Permutation importance of: (a)  static features, (b) dynamic features with time-cohorts, (c) both dynamic and static features with time-cohorts.  Abbreviations for the static features are as follows. STD-CLUST: standard deviation of clustering coefficient; TRANS: transitivity (in the giant connected component); ASS: Assortativity; Q1-CLUST: first quantile of clustering coefficient; STD-CORE: standard deviation of coreness, Q3-CORE: third quantile of coreness; STD-$\Delta$: standard deviation of triangles; STD-DEG: standard deviation of degree, MAX-CORE: maximum coreness, MAX-DEG: maximum degree; Q1- and STD-LCC: 1st quantile and resp. standard deviation of local clustering coefficient.}
\label{fig:feature-importance}
\end{figure}

Figure~\ref{fig:feature-importance} presents the permutation importance results. Among static features, in Figure~\ref{fig:feature-importance}(a),the standard deviation of the clustering coefficient comes out as most distinctive feature. Assortativity comes in third. In the future, this should be verified on networks of different sizes as assortativity can be size-dependent in power-law networks~\cite{Litvak2013Uncovering}. The low values of the permutation importance suggest that none of the features dominates the performance. We did not study the redundancy among static features any further because we use them merely as a benchmark from the literature.    In Figure~\ref{fig:feature-importance}(b),(c), the total permutation importance of the top features (dynamic features in Figure~\ref{fig:feature-importance}(b), and the combination of static and dynamic features in Figure~\ref{fig:feature-importance}(c)) turns out to be much smaller than 1. This renders permutation importance uninformative. Indeed, the permutation importance of a feature equals to the decrease in performance when this feature is randomly permuted over the dataset. When the sum of permutation importance scores is small, it means that there are correlations between the features: if information carried by one feature is removed due to the random permutation, then other features supply a similar information to the classifier, so the performance decrease is small.

\onecolumn

\subsection{Size-cohort dynamic feature matrix}
\begin{figure}[H]
\centering
\begin{subfigure}{.45\textwidth}
\includegraphics[width=.9\linewidth]{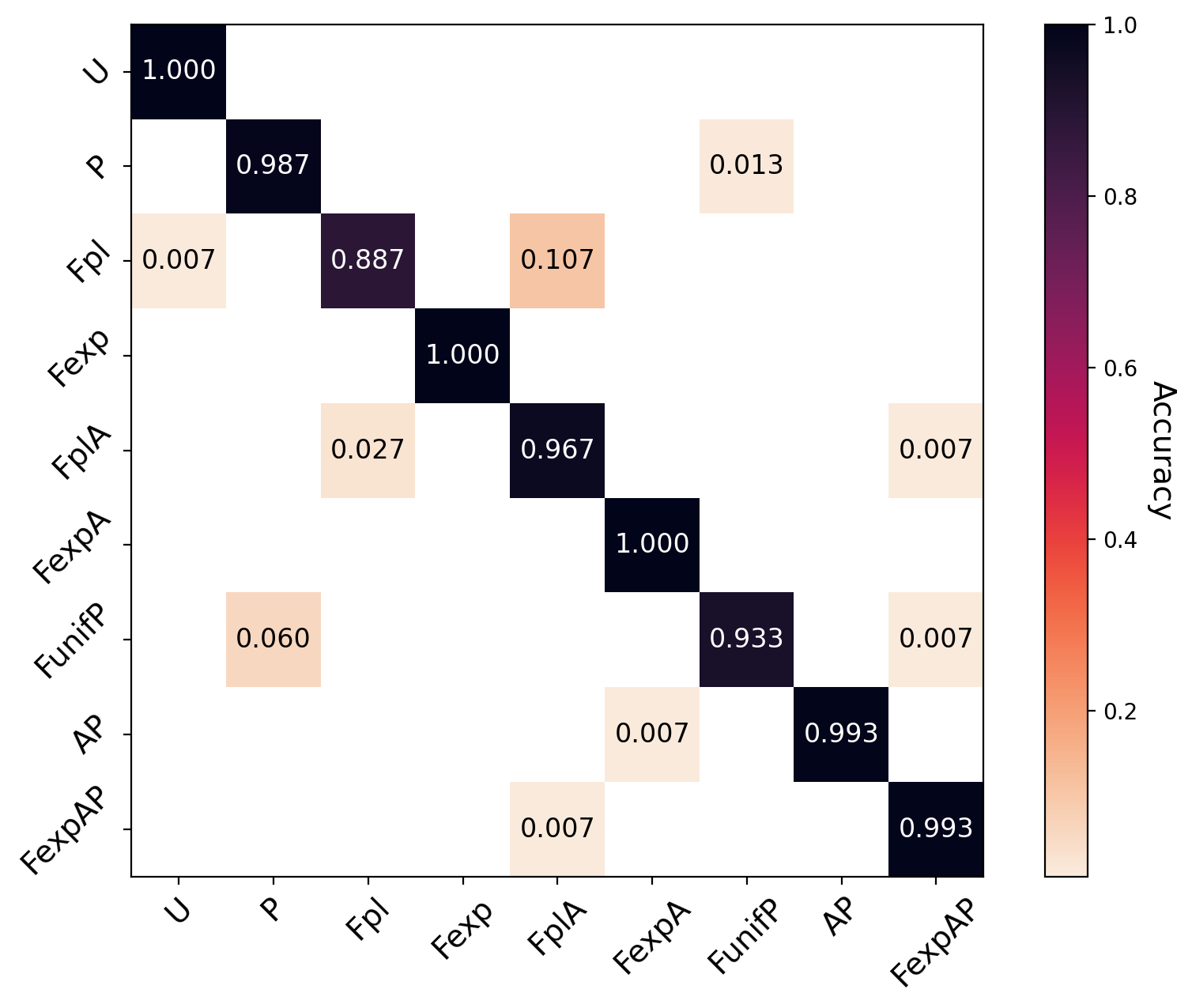}
\caption{}
\end{subfigure}
\begin{subfigure}{.45\textwidth}
\includegraphics[width=.9\linewidth]{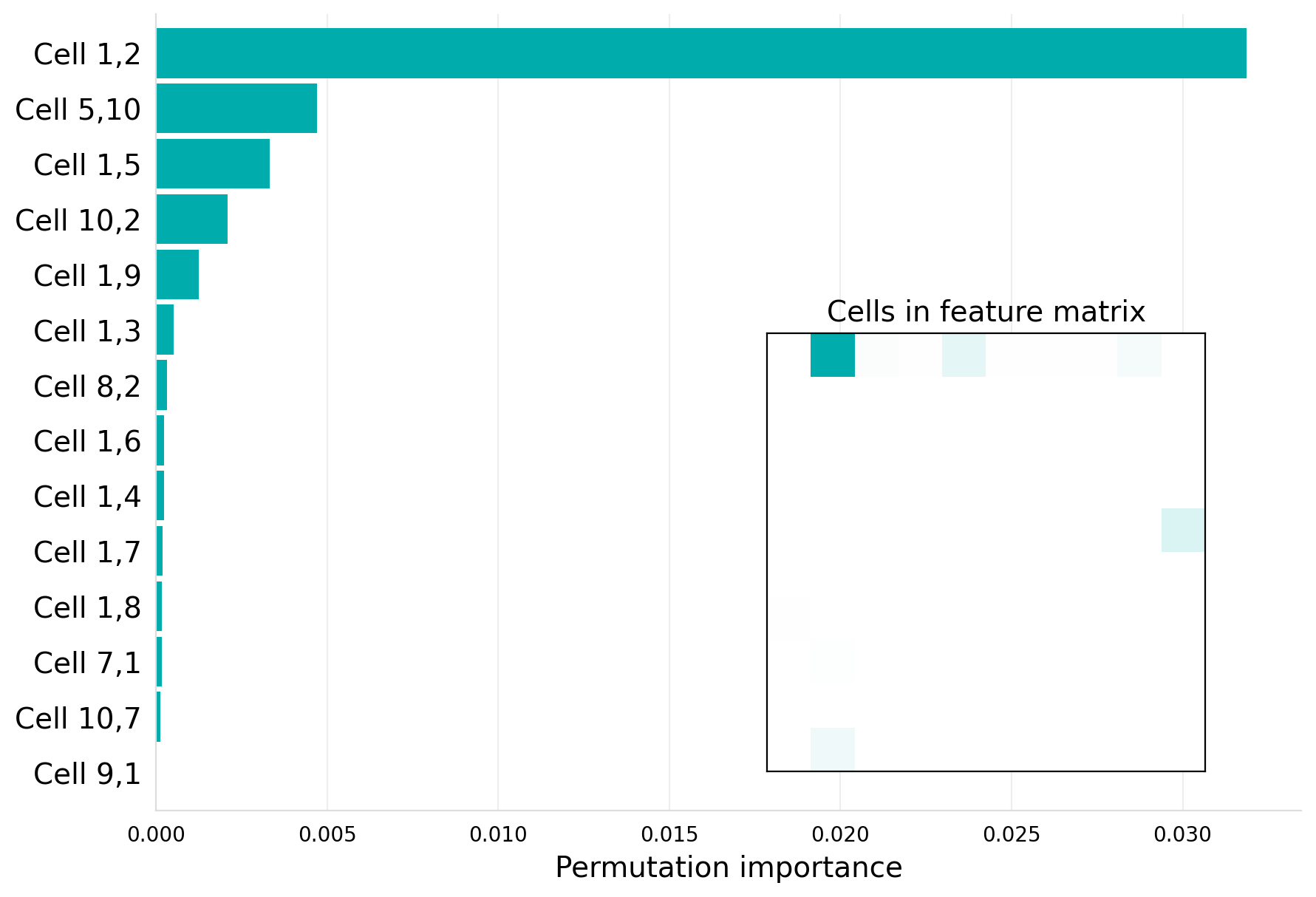}
\caption{}
\end{subfigure}
\begin{subfigure}{.45\textwidth}
\includegraphics[width=.9\textwidth]{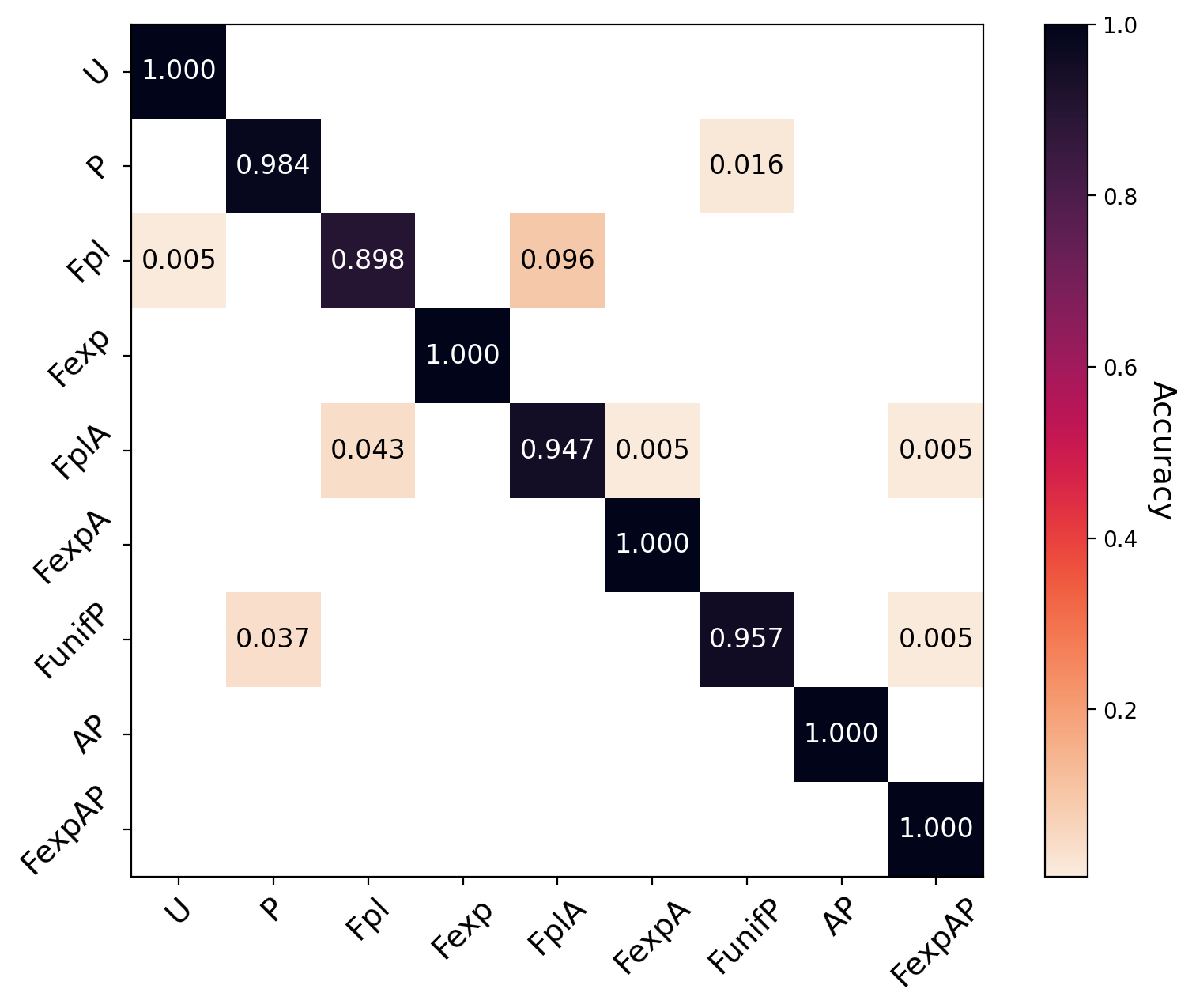}
\caption{}
\end{subfigure}
\begin{subfigure}{.45\textwidth}
\includegraphics[width=.9\linewidth]{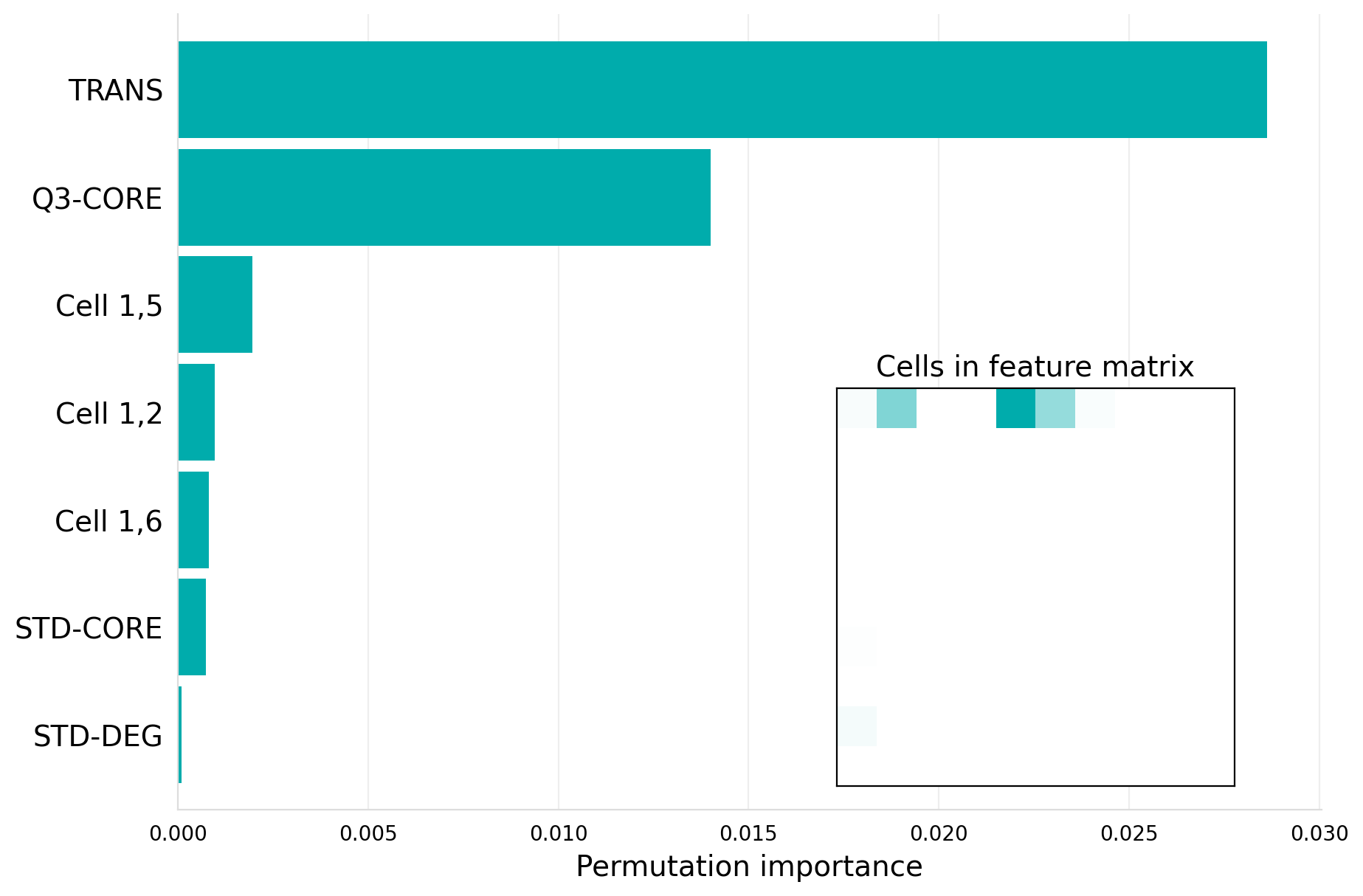}
\caption{}
\end{subfigure}
\caption{(a) Confusion matrix using size-cohort dynamic feature matrix, (b) Feature permutation of the size-cohort dynamic feature matrix, (c) Confusion matrix using both static and size-cohort dynamic features, (d) Feature permutation of both dynamic and static features. TRANS: transitivity (gcc), Q3-CORE: 3rd quantile of Core, STD-CORE: standard deviation of core, STD-DEG: standard deviation of degree.}
\label{fig:size_cohort_feature_importance}
\end{figure}

Here, we report on the performance of the classifier trained on the \textit{size}-cohort feature matrix, i.e. each row  is computed based on the arrival of an equal number of new vertices (see Section~\ref{ssec:ML-methods}). Recall that the main text presents the results for  the feature matrix with \textit{time}-cohorts, where the rows are computed based on the arrivals in equal time intervals. The {size-cohort} dynamic features perform comparable to the {time-cohort} dynamic features, yielding an accuracy of $97.32\%$, see the confusion matrix in Figure~\ref{fig:size_cohort_feature_importance}(a). The feature importance in Figure~\ref{fig:size_cohort_feature_importance}(b) again has sum of the scores much lower than one, therefore, is not informative. 
When training on both static and size-cohort dynamic features, we get the $97.62\%$ accuracy, see the confusion matrix in figure~\ref{fig:size_cohort_feature_importance}(c). 
Permutation importance, presented in Figure~\ref{fig:size_cohort_feature_importance}(d), is again uninformative. 

\subsection{UMAP Visualization of Features and Citation Networks} 
\label{sec:umap}

\review{
In search for an interpretable explanation of why different features may place the same citation network in different classes, we created a two-dimensional embedding of the features using UMAP (Uniform Manifold Approximation and Projection), a method for general non-linear dimension reduction \cite{mcinnes2020umap}. UMAP can be seen as an unsupervised classification method based on only two features derived from the vector of all features.  
Figure~\ref{fig:UMAP} shows the two-dimension UMAP embedding of four feature combinations for the synthetic networks. The same embedding is applied to feature combinations of the citation networks, and they are overlayed in the visualization. We see that UMAP separates some of the classes, thus showing that the features presented in this paper can be used for unsupervised model selection of dynamic networks. However, a striking observation is that there is a significant overlap between classes of synthetic networks. This underlines the complexity of the problem. Indeed, even for synthetic networks, a two-dimensional embedding is not able to identify the differences between some classes. We also see that the placement of citation networks in the embedding space is not consistent with the supervised machine learning classifications (see Table~\ref{tab:overview_wos_results}). 
However, given the large overlap of the classes, this placement is not reliable and only confirms the sensitivity of classification to chosen features. We conclude that there is no simple explanation why different features classify citation networks differently. The most likely explanation is that none of the models accurately describes all aspects of citation networks. Then, different features capture different aspects of real-life networks, and thus may produce a different label. The experiments with UMAP embedding confirm our cautionary tale that applying machine learned model selection in practice requires significant care in justification and validation of feature design.

\begin{figure}[H]
\centering
\begin{subfigure}{.45\textwidth}
\includegraphics[width=.9\linewidth]{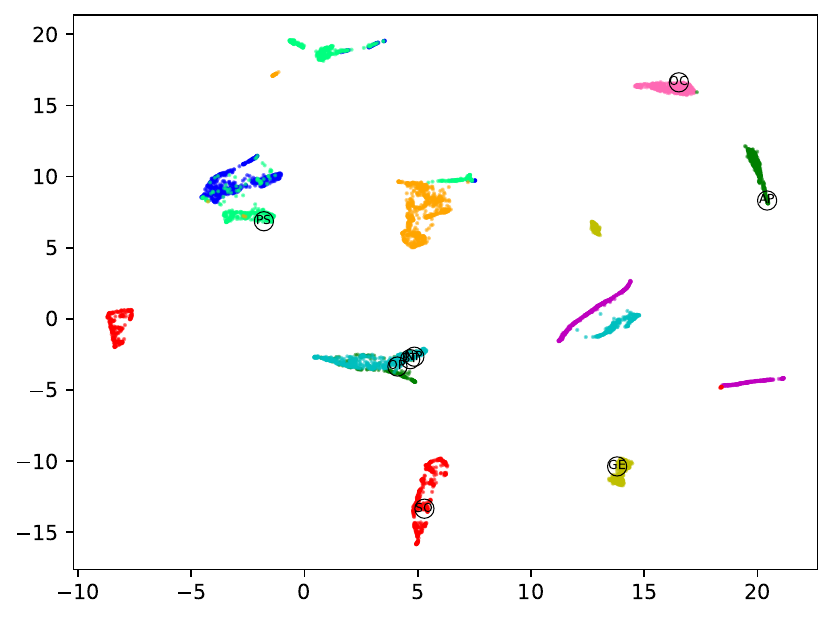}
\caption{Size-cohort matrix}
\end{subfigure}
\begin{subfigure}{.45\textwidth}
\includegraphics[width=.9\linewidth]{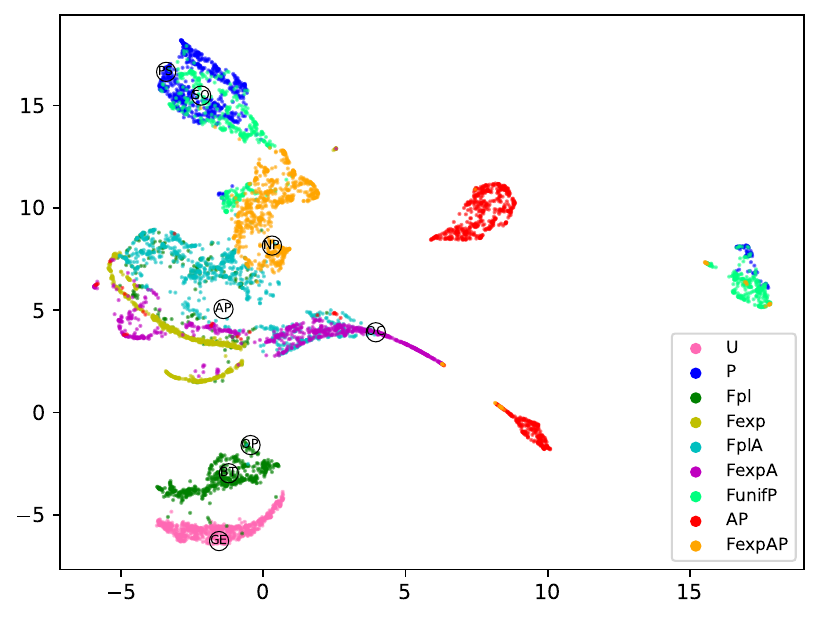}
\caption{Time-cohort matrix}
\end{subfigure}
\begin{subfigure}{.45\textwidth}
\includegraphics[width=.9\textwidth]{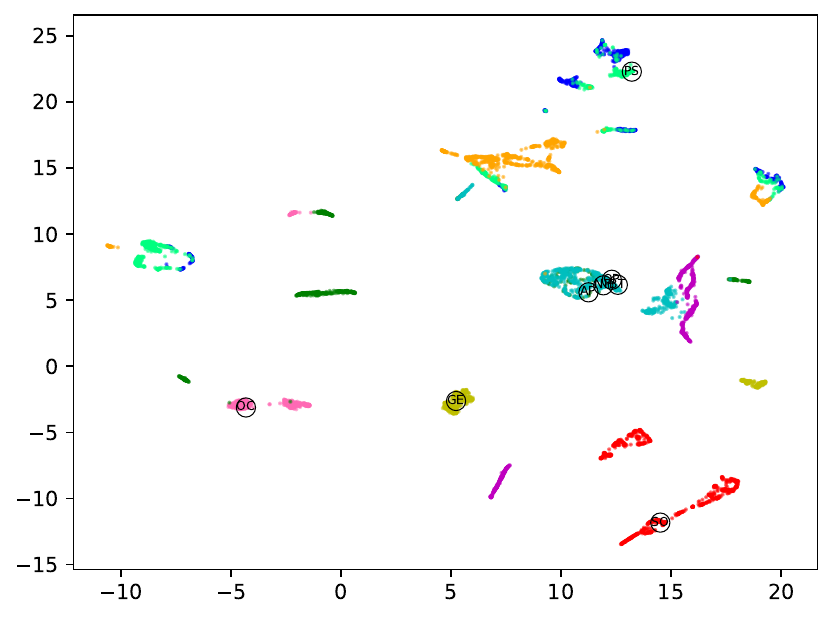}
\caption{Size-cohort matrix + static features}
\end{subfigure}
\begin{subfigure}{.45\textwidth}
\includegraphics[width=.9\linewidth]{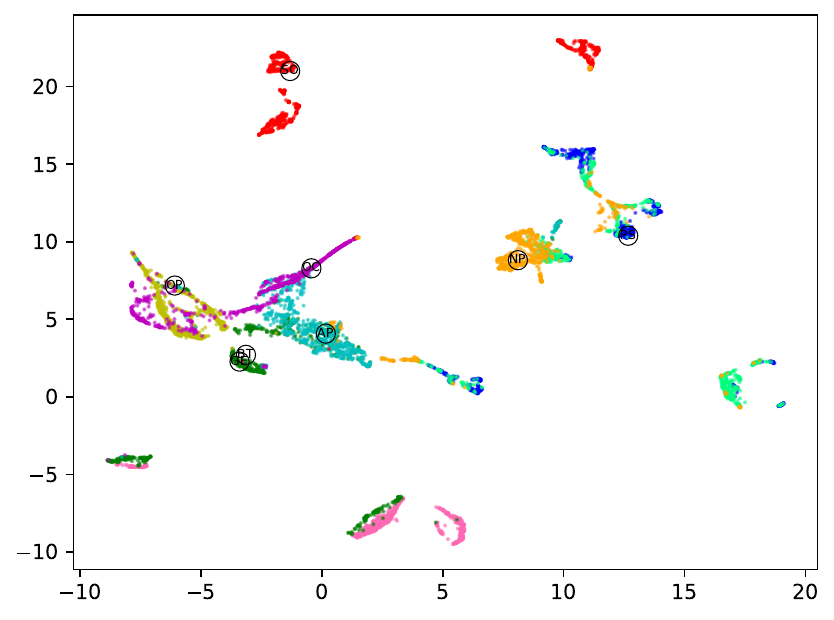}
\caption{Time-cohort matrix + static features}
\end{subfigure}
\caption{UMAP embedding in two dimensions for feature combinations. Citation networks denoted by black circle with field acronym.}
\label{fig:UMAP}
\end{figure}

} 

\subsection{Citation network classifications}
In Section~\ref{sec-classifying-citation} we have reported the most likely model for each real-life network. In Tables~\ref{tab:static_features_class_probabilities}-- \ref{tab:size_cohort_both_features_class_probabilities} below we present the entire distribution over the models for each citation network and all combinations of the features.  The column {\bf `Prediction'} states the most likely model.

\begin{table}[H]
    \centering
    \scalebox{0.7}{
    \begin{tabular}{l|l|lllllllll}
    WoS Field & Prediction &        \textsc{U} &         \textsc{P} &      \textsc{F\textsubscript{pl}} &     \textsc{F\textsubscript{exp}} &     \textsc{F\textsubscript{pl}A} &     \textsc{F\textsubscript{exp}A} &   \textsc{F\textsubscript{unif}P} &        AP &   \textsc{F\textsubscript{exp}AP} \\
    \hline
    AP &      \textsc{F\textsubscript{exp}A} &  10.194\% &   0.0098\% &  0.8629\% &  0.0087\% &  1.7351\% &  \textbf{87.1539\%} &  0.0302\% &   0.0033\% &  0.0022\% \\
    BT &         AP &  0.0054\% &   0.2101\% &  0.0042\% &   0.017\% &  1.6216\% &   7.7802\% &  0.0047\% &  \textbf{83.0959\%} &  7.2608\% \\
    GE &         AP &  0.0004\% &   0.0074\% &  0.0002\% &  0.0007\% &  0.1266\% &   0.1152\% &  0.0002\% &  \textbf{98.8932\%} &  0.8562\% \\
    NP &      \textsc{F\textsubscript{exp}A} &   0.008\% &   1.3284\% &  0.0437\% &   0.022\% &  2.0946\% &  \textbf{91.7197\%} &  0.0034\% &   4.6799\% &  0.1002\% \\
    OC &      \textsc{F\textsubscript{exp}A} &  0.3216\% &  48.8172\% &  0.2431\% &  0.0114\% &  1.0426\% &  \textbf{48.9337\%} &  0.0134\% &    0.544\% &  0.0732\% \\
    OP &         AP &  0.0033\% &    0.353\% &  0.0026\% &  0.0088\% &   0.241\% &  14.5056\% &  0.0015\% &  \textbf{81.4642\%} &  3.4201\% \\
    PS &         AP &   0.005\% &   0.2505\% &  0.0088\% &  0.0196\% &  3.5427\% &  14.1684\% &  0.0066\% &   \textbf{72.324\%} &  9.6744\% \\
    SO &         AP &  0.0007\% &   0.0437\% &  0.0009\% &  0.0034\% &  0.2198\% &   0.5241\% &  0.0009\% &  \textbf{95.7194\%} &  3.4872\% \\
    \end{tabular}
    }
    \caption{Class probabilities for citation networks classified using only static features.}
    \label{tab:static_features_class_probabilities}
\end{table}

\begin{table}[H]
    \centering
    \scalebox{0.7}{
    \begin{tabular}{l|l|lllllllll}
    WoS Field & Prediction &        \textsc{U} &        \textsc{P} &      \textsc{F\textsubscript{pl}} &     \textsc{F\textsubscript{exp}} &     \textsc{F\textsubscript{pl}A} &    \textsc{F\textsubscript{exp}A} &   \textsc{F\textsubscript{unif}P} &        AP &   \textsc{F\textsubscript{exp}AP} \\
    \hline
    AP &         AP &  0.0001\% &  0.0001\% &  0.0002\% &  0.0001\% &  0.0008\% &  0.6312\% &  0.0001\% &  \textbf{99.3624\%} &  0.0048\% \\
    BT &         AP &  0.0001\% &  0.0001\% &  0.0002\% &  0.0001\% &  0.0005\% &  0.1622\% &  0.0001\% &  \textbf{99.7829\%} &   0.054\% \\
    GE &         AP &  0.0002\% &  0.0001\% &  0.0004\% &  0.0002\% &  0.0032\% &  0.5906\% &  0.0004\% &  \textbf{99.4024\%} &  0.0025\% \\
    NP &         AP &  0.0002\% &  0.0002\% &  0.0004\% &  0.0002\% &  0.0012\% &  0.6091\% &  0.0002\% &  \textbf{96.7235\%} &   2.665\% \\
    OC &         AP &  0.0001\% &  0.0001\% &  0.0003\% &  0.0001\% &  0.0008\% &   0.245\% &  0.0001\% &  \textbf{99.5958\% }&  0.1576\% \\
    OP &         AP &  0.0001\% &     0.0\% &  0.0001\% &  0.0001\% &  0.0012\% &  0.1917\% &  0.0001\% &  \textbf{99.7396\%} &  0.0671\% \\
    PS &         AP &  0.0013\% &  0.0017\% &  0.0093\% &  0.0015\% &  0.0082\% &  1.5193\% &  0.0012\% &  \textbf{95.0033\%} &  3.4543\% \\
    SO &         AP &  0.0002\% &  0.0002\% &  0.0012\% &  0.0002\% &   0.001\% &  0.4056\% &  0.0005\% &   \textbf{99.578\%} &  0.0131\% \\
    \end{tabular}
    }
    \caption{Class probabilities for citation networks classified using only time-cohort dynamic features.}
    \label{tab:time_cohort_dynamic_features_class_probabilities}
\end{table}

\begin{table}[H]
    \centering
    \scalebox{0.7}{
    \begin{tabular}{l|l|lllllllll}
    WoS Field & Prediction &         \textsc{U} &         \textsc{P} &      \textsc{F\textsubscript{pl}} &     \textsc{F\textsubscript{exp}} &     \textsc{F\textsubscript{pl}A} &    \textsc{F\textsubscript{exp}A} &    \textsc{F\textsubscript{unif}P} &        AP &    \textsc{F\textsubscript{exp}AP} \\
    \hline
    AP &     \textsc{F\textsubscript{exp}AP} &    0.006\% &   0.0009\% &  0.0041\% &  0.0176\% &  0.0127\% &  0.0026\% &   6.1683\% &   3.3167\% & \textbf{ 90.4711\%} \\
    BT &     \textsc{F\textsubscript{exp}AP} &   0.0203\% &   0.0005\% &  0.0048\% &  0.0064\% &  0.0384\% &  0.0949\% &   7.2838\% &   0.6714\% &  \textbf{91.8795\%} \\
    GE &     \textsc{F\textsubscript{exp}AP} &   0.1216\% &   0.0174\% &  0.0552\% &  0.3496\% &  0.1957\% &  1.4084\% &  14.3512\% &   4.8238\% &  \textbf{78.6771\%} \\
    NP &     \textsc{F\textsubscript{exp}AP} &   0.0078\% &   0.0006\% &  0.0732\% &  0.0069\% &  0.1857\% &  0.0481\% &     0.87\% &   1.1023\% &  \textbf{97.7054\%} \\
    OC &     \textsc{F\textsubscript{exp}AP} &  12.6796\% &   0.0009\% &  0.0018\% &  0.0209\% &  0.0026\% &  0.0027\% &  17.0127\% &   4.1622\% &  \textbf{66.1166\%} \\
    OP &     \textsc{F\textsubscript{exp}AP} &   0.0038\% &   0.0016\% &  0.0035\% &  0.0079\% &  0.0335\% &  0.0862\% &  10.7796\% &   0.6463\% &  \textbf{88.4376\%} \\
    PS &     \textsc{F\textsubscript{exp}AP} &   0.0002\% &   0.0005\% &  0.0006\% &  0.0022\% &  0.0021\% &  0.0176\% &   1.1467\% &   0.0716\% & \textbf{ 98.7585\%} \\
    SO &          \textsc{P} &   0.8341\% &  \textbf{49.6114\%} &  0.0571\% &  1.0126\% &  0.3739\% &  1.2897\% &   8.8134\% &  12.6348\% &   25.373\% \\
    \end{tabular}
    }
    \caption{Class probabilities for citation networks classified using only size-cohort dynamic features.}
    \label{tab:size-cohortdynamic_features_class_probabilities}
\end{table}

\begin{table}[H]
    \centering
    \scalebox{0.7}{
    \begin{tabular}{l|l|lllllllll}
    WoS Field & Prediction &        \textsc{U} &        \textsc{P} &      \textsc{F\textsubscript{pl}} &     \textsc{F\textsubscript{exp}} &     \textsc{F\textsubscript{pl}A} &     \textsc{F\textsubscript{exp}A} &   \textsc{F\textsubscript{unif}P} &        AP &    \textsc{F\textsubscript{exp}AP} \\
    \hline
    AP &      \textsc{F\textsubscript{exp}A} &  0.0295\% &    0.01\% &  0.1138\% &  0.0525\% &  4.6638\% &  \textbf{94.2683\%} &  0.0238\% &   0.6141\% &   0.2241\% \\
    BT &     \textsc{F\textsubscript{exp}AP} &  0.0061\% &  0.0091\% &  0.0372\% &  0.0278\% &    0.08\% &   1.4155\% &  0.0081\% &  36.3496\% &  \textbf{62.0666\%} \\
    GE &         AP &  0.0005\% &  0.0001\% &  0.0002\% &  0.0002\% &  0.0009\% &   0.0167\% &  0.0001\% &  \textbf{99.8106\%} &   0.1707\% \\
    NP &      \textsc{F\textsubscript{exp}A} &  0.0069\% &  0.0086\% &  0.0436\% &  0.0399\% &  0.1774\% &  \textbf{59.7829\%} &  0.0062\% &   8.1347\% &  31.7999\% \\
    OC &      \textsc{F\textsubscript{exp}A} &   0.272\% &   0.053\% &  0.2732\% &  0.0929\% &  0.2399\% &  \textbf{73.6001\%} &  0.0391\% &   11.526\% &  13.9039\% \\
    OP &     \textsc{F\textsubscript{exp}AP} &   0.005\% &  0.0061\% &  0.0274\% &  0.0278\% &  0.0862\% &  10.2059\% &  0.0047\% &  29.6566\% &  \textbf{59.9803\%} \\
    PS &     \textsc{F\textsubscript{exp}AP} &  0.0053\% &  0.0065\% &   0.029\% &  0.0266\% &   0.101\% &   7.5905\% &  0.0048\% &  21.7334\% &  \textbf{70.5028\%} \\
    SO &         AP &  0.1384\% &  0.0011\% &  0.0032\% &  0.0074\% &  0.0191\% &   0.2522\% &  0.0048\% &  \textbf{50.9388\%} &   48.635\% \\
    \end{tabular}
    }
    \caption{Class probabilities for citation networks classified using both time-cohort dynamic features and static features.}
    \label{tab:time_cohort_both_features_class_probabilities}
\end{table}

\begin{table}[H]
    \centering
    \scalebox{0.7}{
    \begin{tabular}{l|l|lllllllll}
    WoS Field & Prediction &         \textsc{U} &        \textsc{P} &      \textsc{F\textsubscript{pl}} &     \textsc{F\textsubscript{exp}} &     \textsc{F\textsubscript{pl}A} &    \textsc{F\textsubscript{exp}A} &    \textsc{F\textsubscript{unif}P} &        AP &    \textsc{F\textsubscript{exp}AP} \\
    \hline
    AP &     \textsc{F\textsubscript{exp}AP} &   8.2686\% &  0.3277\% &  5.3107\% &  9.5221\% &  1.2493\% &  5.9404\% &  20.9093\% &   0.8591\% &  \textbf{47.6128\%} \\
    BT &     \textsc{F\textsubscript{exp}AP} &   0.0667\% &  0.0009\% &  0.0185\% &  0.0503\% &  0.0077\% &  0.0955\% &   0.0426\% &  21.8597\% &   \textbf{77.858\%} \\
    GE &         AP &   0.0011\% &  0.0001\% &  0.0033\% &  0.0048\% &  0.0004\% &  0.0015\% &   0.0034\% &  \textbf{99.7651\% }&   0.2204\% \\
    NP &     \textsc{F\textsubscript{exp}AP} &   0.0127\% &  0.0005\% &  0.0077\% &  0.0082\% &  0.0048\% &  0.1248\% &   0.0097\% &   7.2476\% &   \textbf{92.584\%} \\
    OC &     \textsc{F\textsubscript{exp}AP} &  47.4841\% &  0.0018\% &  0.0324\% &  0.0351\% &  0.0042\% &  0.1073\% &   0.9745\% &   0.7513\% &  \textbf{50.6094\%} \\
    OP &     \textsc{F\textsubscript{exp}AP} &   0.0165\% &   0.003\% &  0.0165\% &  0.0554\% &  0.0118\% &  0.0985\% &   0.0709\% &  23.5339\% &  \textbf{76.1935\%} \\
    PS &     \textsc{F\textsubscript{exp}AP} &   0.0027\% &  0.0026\% &  0.0155\% &  0.1971\% &  0.0217\% &  0.0442\% &   0.0165\% &  10.0808\% &  \textbf{89.6187\%} \\
    SO &         AP &   0.1376\% &  1.2988\% &  0.0141\% &  0.4178\% &  0.0046\% &  4.9587\% &   0.0125\% &  \textbf{88.0893\%} &   5.0665\% \\
    \end{tabular}
    }
    \caption{Class probabilities for citation networks classified using both size-cohort dynamic features and static features.}
    \label{tab:size_cohort_both_features_class_probabilities}
\end{table}

\subsection{Correlations in Dynamic Feature matrix}
\label{ssec:correlations}

\begin{figure}[H]
    \centering
    \begin{subfigure}{0.45\textwidth}
        \centering
        \includegraphics[width=0.9\textwidth]{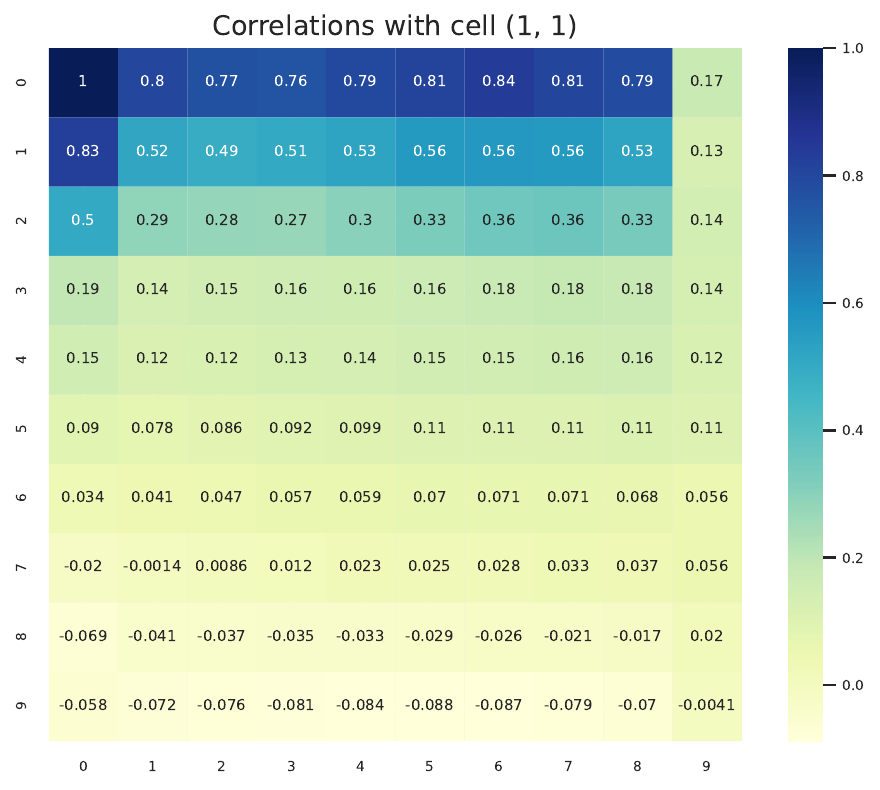}
    \end{subfigure}
    \begin{subfigure}{0.45\textwidth}
        \includegraphics[width=0.9\textwidth]{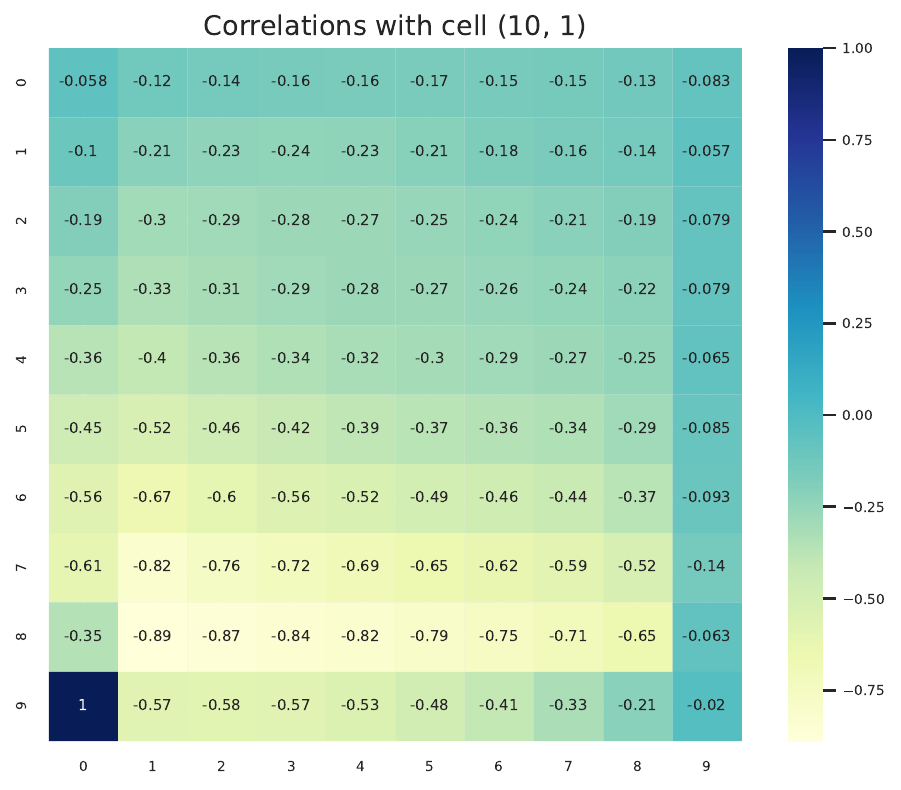}
    \end{subfigure}
    \\
    \begin{subfigure}{0.45\textwidth}
        \includegraphics[width=0.9\textwidth]{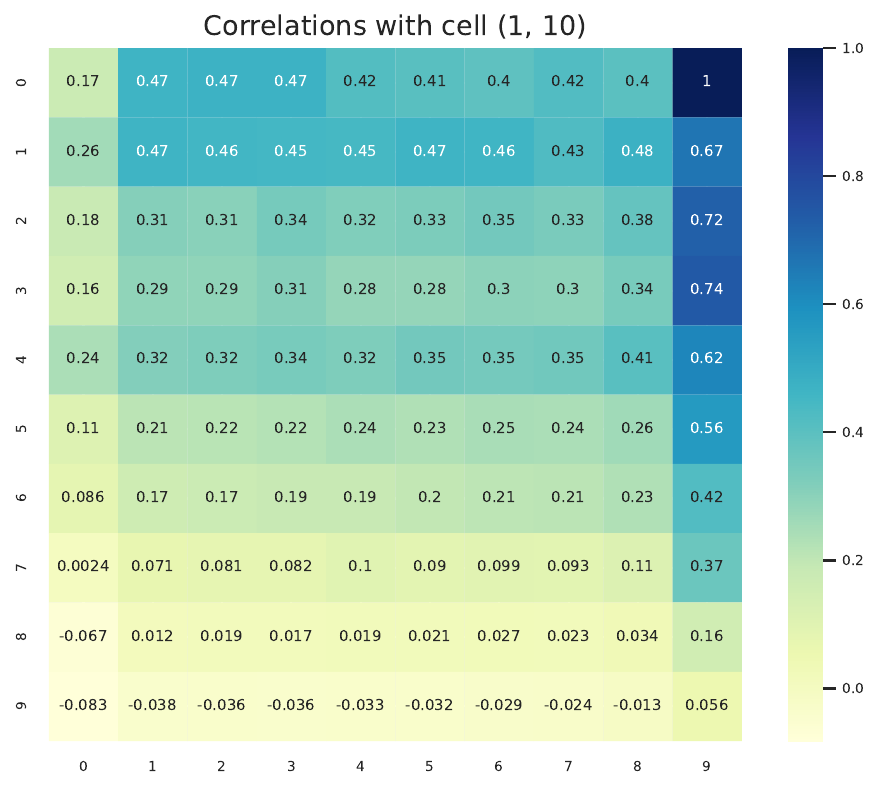}
    \end{subfigure}
    \begin{subfigure}{0.45\textwidth}
        \includegraphics[width=0.9\textwidth]{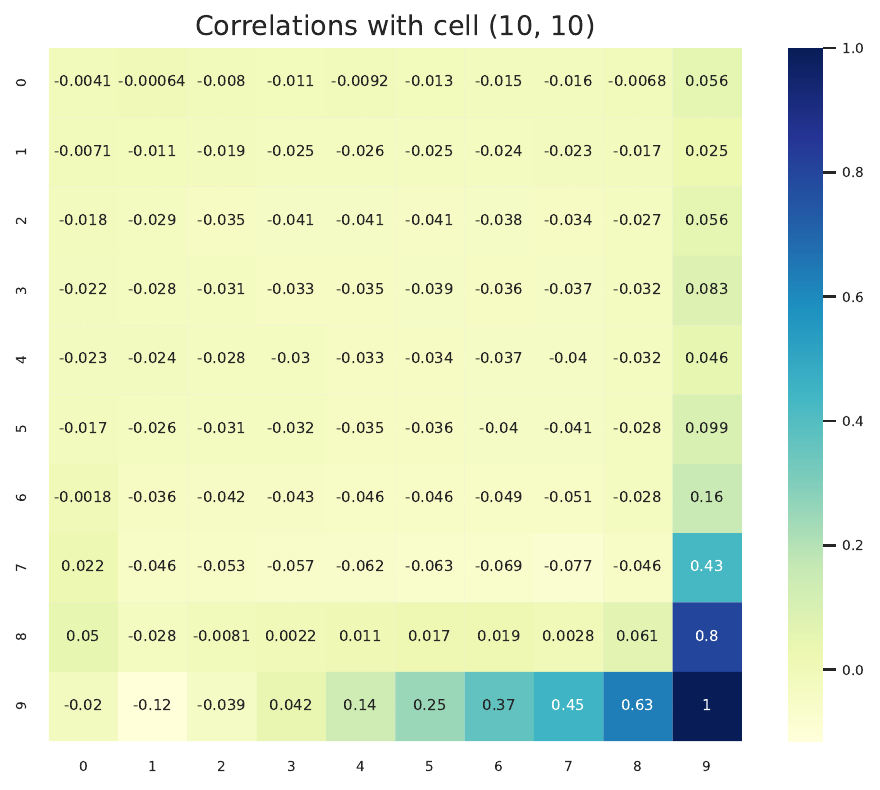}
    \end{subfigure}
    \caption{Correlations in time-cohort dynamic feature matrix}
    \label{fig:correlations-time-cohort}
\end{figure}

\begin{figure}[H]
    \centering
    \begin{subfigure}{0.45\textwidth}
        \centering
        \includegraphics[width=0.9\textwidth]{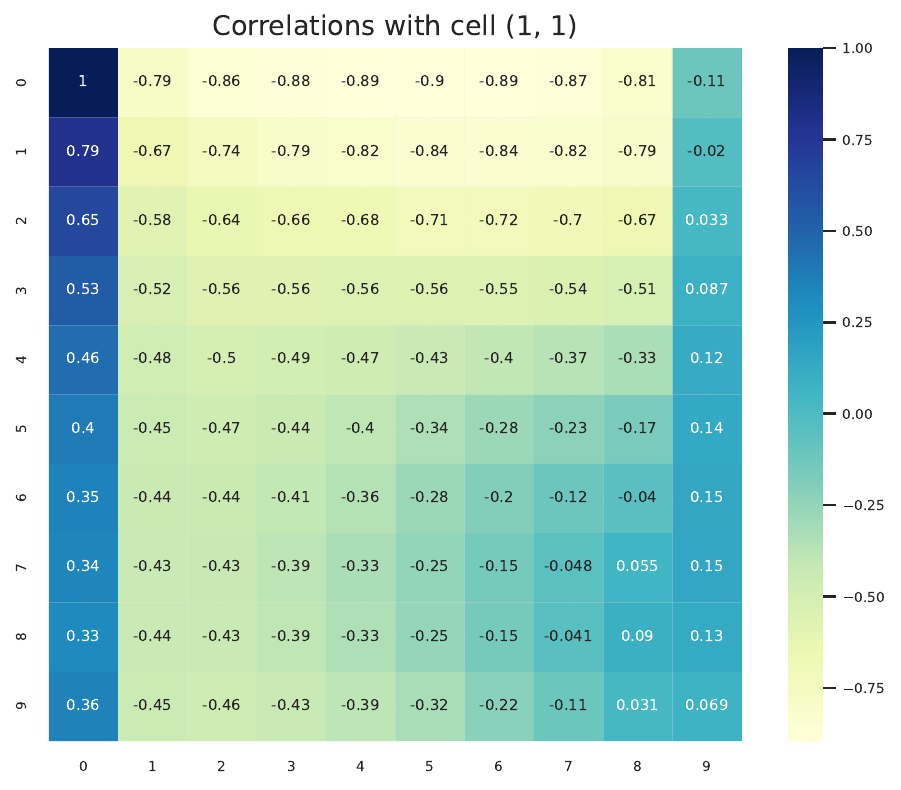}
    \end{subfigure}
    \begin{subfigure}{0.45\textwidth}
        \includegraphics[width=0.9\textwidth]{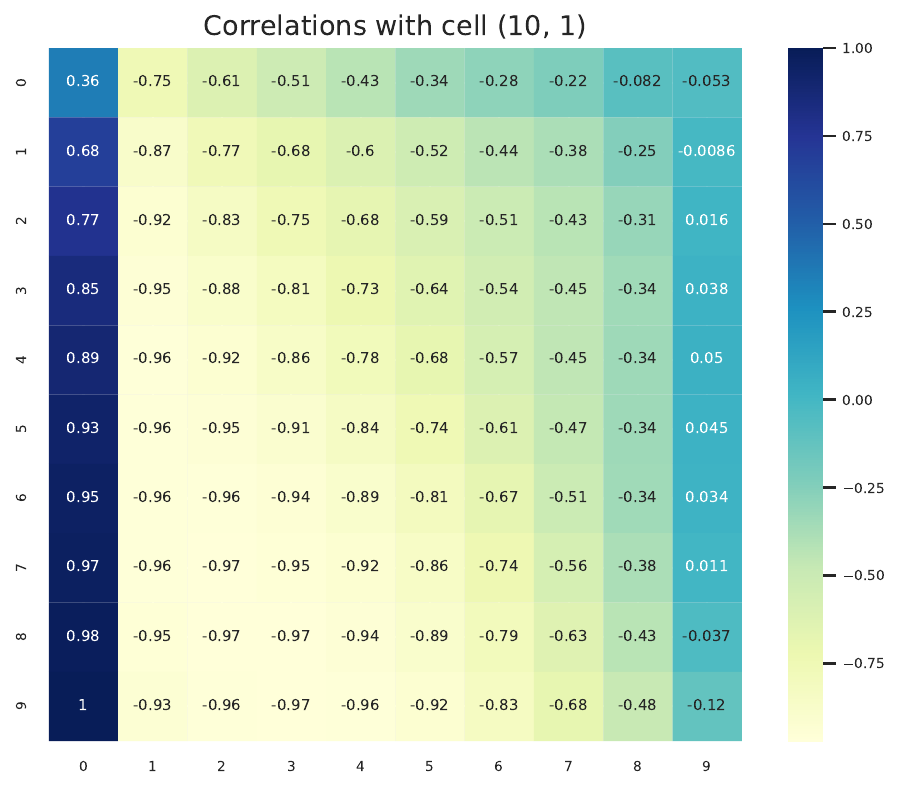}
    \end{subfigure}
    \\
    \begin{subfigure}{0.45\textwidth}
        \includegraphics[width=0.9\textwidth]{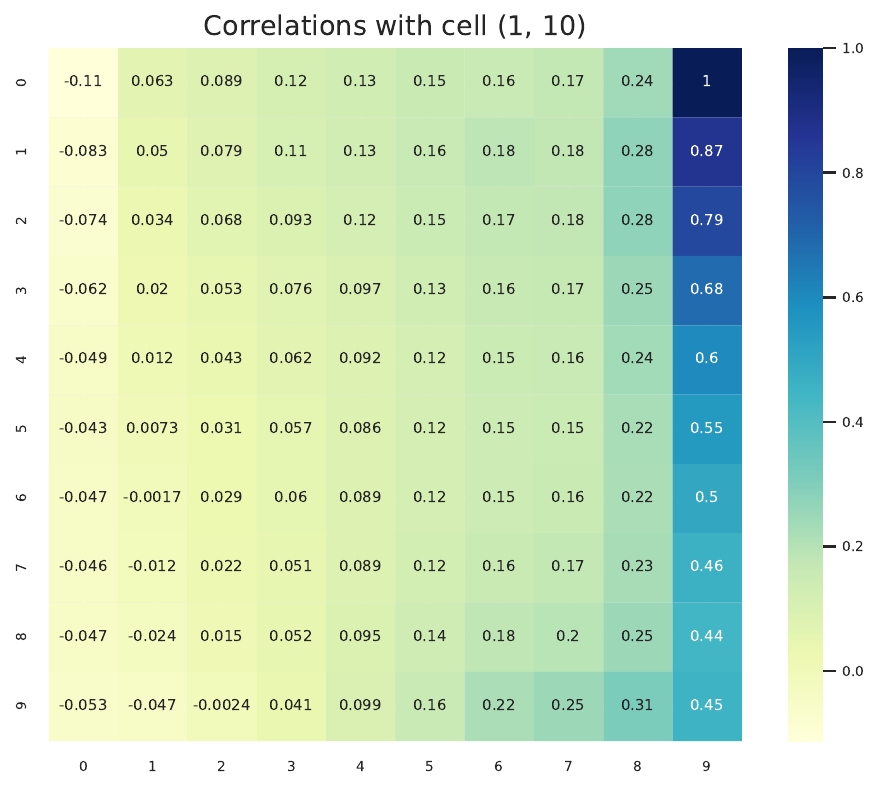}
    \end{subfigure}
    \begin{subfigure}{0.45\textwidth}
        \includegraphics[width=0.9\textwidth]{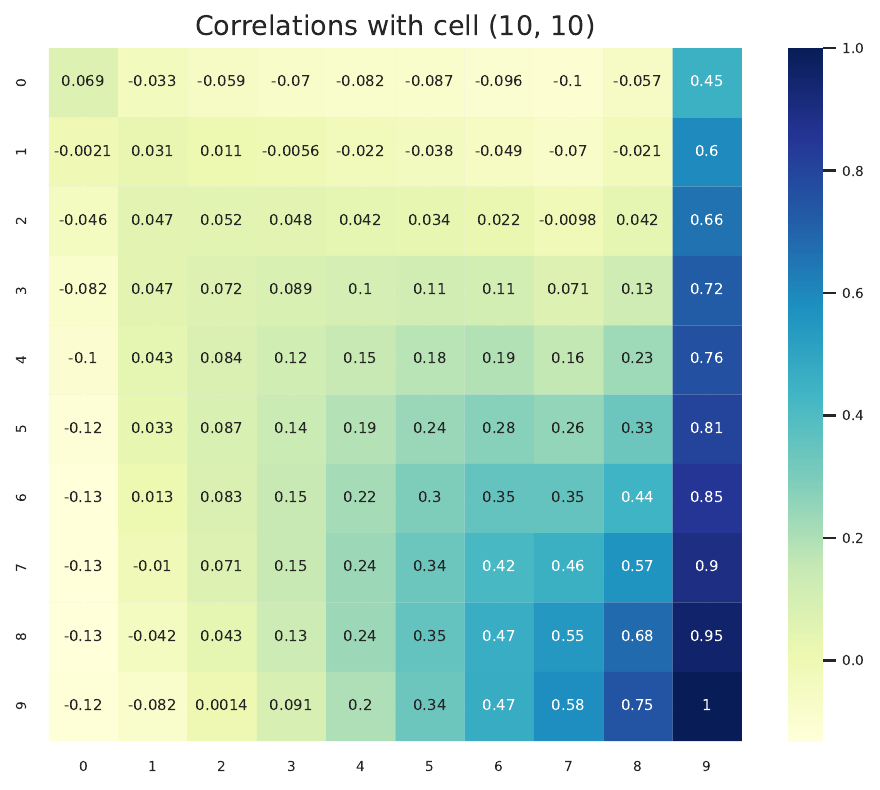}
    \end{subfigure}
    \caption{Correlations in size-cohort dynamic feature matrix}
    \label{fig:correlations-size-cohort}
\end{figure}

Within the dynamic feature matrix there is correlation between cells. In Figure~\ref{fig:correlations-time-cohort} and Figure~\ref{fig:correlations-size-cohort} we show the correlations between the corner-cells and the other cells in the feature matrix, across all networks in the dataset. We see that cells adjacent to the corners are more stronger correlated, especially in  top corners that correspond to the number of edges received from early cohorts.  This suggests redundancy in the features. Nevertheless, in the experiments, excluding some cells from the feature matrix leads to an immediate decrease in classification accuracy. Indeed, Figures~\ref{fig:correlations-time-cohort},~\ref{fig:correlations-size-cohort} don't show any clear pattern that can be helpful for excluding some of the features.    

%% file: supp-figures/number_publications.tikz
\tikzstyle{every node}=[font=\normalsize]

\begin{tikzpicture}
	\begin{axis}[%
		width=0.951\fwidth,
		height=\fheight,
		at={(0\fwidth,0\fheight)},
		scale only axis,
		xmin=1980,
		xmax=2014,
		xminorticks=true,
		yminorticks=true,
		axis x line*=bottom,
		axis y line*=left,
		ylabel={Yearly Publications},
		xtick ={1980,1985,1990,...,2015, 2014},
		x tick label style = {/pgf/number format/.cd,%
														scaled x ticks = false,
														set thousands separator={},
														fixed},
		ymode=log,
		ymin=1000,
		ymax=100000,
		ytick = {1000.0, 10000.0,  100000.0},
		axis background/.style={fill=white},
		legend style={legend pos=north west, legend cell align=left, align=left, draw=white!15!black}
		]
		\addplot [color=mycolor1, line width = 2pt,mark size=2pt, mark=diamond*, only marks, mark options={solid, mycolor1}]
		  table[row sep=crcr]{%
			1980	2677\\
			1981	2664\\
			1982	2461\\
			1983	2767\\
			1984	3464\\
			1985	3385\\
			1986	3197\\
			1987	3421\\
			1988	3502\\
			1989	3305\\
			1990	3510\\
			1991	3604\\
			1992	3885\\
			1993	3948\\
			1994	4032\\
			1995	4368\\
			1996	4511\\
			1997	4556\\
			1998	4743\\
			1999	4928\\
			2000	4997\\
			2001	4992\\
			2002	4991\\
			2003	5229\\
			2004	5295\\
			2005	5804\\
			2006	6159\\
			2007	6920\\
			2008	7613\\
			2009	8189\\
			2010	7809\\
			2011	8557\\
			2012	8657\\
			2013	9002\\
			2014	8923\\
		};
		\addlegendentry{PS}

		\addplot [color=mycolor3, line width = 2pt,mark size=1.5pt, mark=square*, only marks, mark options={solid, mycolor3}]
		  table[row sep=crcr]{%
			1982	3058\\
			1983	3533\\
			1984	4035\\
			1985	4107\\
			1986	4511\\
			1987	5766\\
			1988	5613\\
			1989	6150\\
			1990	6545\\
			1991	7878\\
			1992	8923\\
			1993	9565\\
			1994	10647\\
			1995	12328\\
			1996	13627\\
			1997	14325\\
			1998	14017\\
			1999	14193\\
			2000	14512\\
			2001	14710\\
			2002	14558\\
			2003	17434\\
			2004	17364\\
			2005	18419\\
			2006	21092\\
			2007	21751\\
			2008	25546\\
			2009	27362\\
			2010	28178\\
			2011	29360\\
			2012	29636\\
			2013	31794\\
			2014	35156\\
		};
		\addlegendentry{BT}
	
	\end{axis}
\end{tikzpicture}

%% file: supp-figures/PS_random_sample.tikz
\tikzstyle{every node}=[font=\Large]

\begin{tikzpicture}
	\begin{axis}[%
		width=0.951\fwidth,
		height=\fheight,
		at={(0\fwidth,0\fheight)},
		scale only axis,
		xmin=1980,
		xmax=2015,
		axis x line*=bottom,
		axis y line*=left,
		xtick ={1990,2000,2010},
		x tick label style = {/pgf/number format/.cd,%
														scaled x ticks = false,
														set thousands separator={},
														fixed},
		ymin=0,
		ymax=40,
		axis background/.style={fill=white},
		title style={font=\bfseries\Large},
		title={PS}
		]
		\addplot [color=mycolor1, line width=1.8pt, forget plot]
		  table[row sep=crcr]{%
		1980	0\\
		1981	0\\
		1982	0\\
		1983	0\\
		1984	0\\
		1985	0\\
		1986	0\\
		1987	0\\
		1988	0\\
		1989	0\\
		1990	0\\
		1991	0\\
		1992	0\\
		1993	0\\
		1994	0\\
		1995	0\\
		1996	0\\
		1997	0\\
		1998	0\\
		1999	0\\
		2000	0\\
		2001	0\\
		2002	0\\
		2003	0\\
		2004	0\\
		2005	0\\
		2006	0\\
		2007	0\\
		2008	0\\
		2009	0\\
		2010	0\\
		2011	0\\
		2012	0\\
		2013	0\\
		2014	0\\
		2015	0\\
		};
		\addplot [color=mycolor2, line width=1.8pt, forget plot]
		  table[row sep=crcr]{%
		1980	0\\
		1981	1\\
		1982	1\\
		1983	2\\
		1984	2\\
		1985	2\\
		1986	4\\
		1987	5\\
		1988	5\\
		1989	5\\
		1990	5\\
		1991	5\\
		1992	5\\
		1993	6\\
		1994	6\\
		1995	6\\
		1996	6\\
		1997	6\\
		1998	7\\
		1999	7\\
		2000	7\\
		2001	7\\
		2002	8\\
		2003	8\\
		2004	8\\
		2005	8\\
		2006	8\\
		2007	8\\
		2008	8\\
		2009	8\\
		2010	8\\
		2011	8\\
		2012	9\\
		2013	9\\
		2014	10\\
		2015	10\\
		};
		\addplot [color=mycolor3, line width=1.8pt, forget plot]
		  table[row sep=crcr]{%
		1980	0\\
		1981	1\\
		1982	1\\
		1983	1\\
		1984	1\\
		1985	1\\
		1986	1\\
		1987	1\\
		1988	2\\
		1989	2\\
		1990	2\\
		1991	2\\
		1992	2\\
		1993	2\\
		1994	2\\
		1995	2\\
		1996	2\\
		1997	2\\
		1998	2\\
		1999	2\\
		2000	2\\
		2001	2\\
		2002	3\\
		2003	3\\
		2004	3\\
		2005	3\\
		2006	3\\
		2007	3\\
		2008	3\\
		2009	3\\
		2010	3\\
		2011	3\\
		2012	3\\
		2013	3\\
		2014	3\\
		2015	3\\
		};
		\addplot [color=mycolor4, line width=1.8pt, forget plot]
		  table[row sep=crcr]{%
		1980	0\\
		1981	2\\
		1982	3\\
		1983	3\\
		1984	3\\
		1985	3\\
		1986	4\\
		1987	4\\
		1988	4\\
		1989	4\\
		1990	4\\
		1991	4\\
		1992	4\\
		1993	4\\
		1994	4\\
		1995	4\\
		1996	4\\
		1997	4\\
		1998	4\\
		1999	4\\
		2000	4\\
		2001	4\\
		2002	4\\
		2003	4\\
		2004	4\\
		2005	4\\
		2006	4\\
		2007	4\\
		2008	4\\
		2009	4\\
		2010	4\\
		2011	4\\
		2012	4\\
		2013	4\\
		2014	4\\
		2015	5\\
		};
		\addplot [color=mycolor5, line width=1.8pt, forget plot]
		  table[row sep=crcr]{%
		1980	0\\
		1981	0\\
		1982	2\\
		1983	2\\
		1984	2\\
		1985	2\\
		1986	3\\
		1987	3\\
		1988	3\\
		1989	3\\
		1990	3\\
		1991	3\\
		1992	5\\
		1993	5\\
		1994	5\\
		1995	5\\
		1996	5\\
		1997	5\\
		1998	5\\
		1999	5\\
		2000	5\\
		2001	5\\
		2002	5\\
		2003	5\\
		2004	5\\
		2005	5\\
		2006	5\\
		2007	5\\
		2008	5\\
		2009	5\\
		2010	5\\
		2011	5\\
		2012	5\\
		2013	5\\
		2014	5\\
		2015	5\\
		};
		\addplot [color=mycolor1, line width=1.8pt, forget plot]
		  table[row sep=crcr]{%
		1980	0\\
		1981	1\\
		1982	1\\
		1983	2\\
		1984	3\\
		1985	4\\
		1986	7\\
		1987	10\\
		1988	13\\
		1989	17\\
		1990	19\\
		1991	20\\
		1992	22\\
		1993	24\\
		1994	26\\
		1995	26\\
		1996	27\\
		1997	28\\
		1998	29\\
		1999	30\\
		2000	31\\
		2001	34\\
		2002	35\\
		2003	35\\
		2004	38\\
		2005	38\\
		2006	38\\
		2007	39\\
		2008	40\\
		2009	40\\
		2010	40\\
		2011	40\\
		2012	40\\
		2013	40\\
		2014	40\\
		2015	40\\
		};
		\addplot [color=mycolor2, line width=1.8pt, forget plot]
		  table[row sep=crcr]{%
		1980	0\\
		1981	0\\
		1982	0\\
		1983	1\\
		1984	1\\
		1985	1\\
		1986	1\\
		1987	1\\
		1988	1\\
		1989	1\\
		1990	1\\
		1991	1\\
		1992	1\\
		1993	1\\
		1994	1\\
		1995	1\\
		1996	1\\
		1997	1\\
		1998	1\\
		1999	1\\
		2000	1\\
		2001	1\\
		2002	1\\
		2003	1\\
		2004	1\\
		2005	1\\
		2006	1\\
		2007	1\\
		2008	1\\
		2009	1\\
		2010	1\\
		2011	1\\
		2012	1\\
		2013	1\\
		2014	1\\
		2015	1\\
		};
		\addplot [color=mycolor1, line width=1.8pt, forget plot]
		  table[row sep=crcr]{%
		1980	0\\
		1981	0\\
		1982	0\\
		1983	0\\
		1984	0\\
		1985	0\\
		1986	0\\
		1987	0\\
		1988	0\\
		1989	0\\
		1990	0\\
		1991	0\\
		1992	0\\
		1993	0\\
		1994	0\\
		1995	0\\
		1996	0\\
		1997	0\\
		1998	0\\
		1999	0\\
		2000	0\\
		2001	0\\
		2002	0\\
		2003	0\\
		2004	0\\
		2005	0\\
		2006	0\\
		2007	0\\
		2008	0\\
		2009	0\\
		2010	0\\
		2011	0\\
		2012	0\\
		2013	0\\
		2014	0\\
		2015	0\\
		};
		\addplot [color=mycolor2, line width=1.8pt, forget plot]
		  table[row sep=crcr]{%
		1980	0\\
		1981	0\\
		1982	0\\
		1983	0\\
		1984	0\\
		1985	0\\
		1986	0\\
		1987	0\\
		1988	0\\
		1989	0\\
		1990	0\\
		1991	0\\
		1992	0\\
		1993	0\\
		1994	0\\
		1995	0\\
		1996	0\\
		1997	0\\
		1998	0\\
		1999	0\\
		2000	0\\
		2001	0\\
		2002	0\\
		2003	0\\
		2004	0\\
		2005	0\\
		2006	0\\
		2007	0\\
		2008	0\\
		2009	0\\
		2010	0\\
		2011	0\\
		2012	0\\
		2013	0\\
		2014	0\\
		2015	0\\
		};
		\addplot [color=mycolor3, line width=1.8pt, forget plot]
		  table[row sep=crcr]{%
		1980	0\\
		1981	1\\
		1982	1\\
		1983	1\\
		1984	1\\
		1985	1\\
		1986	2\\
		1987	4\\
		1988	4\\
		1989	4\\
		1990	5\\
		1991	5\\
		1992	5\\
		1993	5\\
		1994	5\\
		1995	6\\
		1996	6\\
		1997	6\\
		1998	6\\
		1999	6\\
		2000	6\\
		2001	6\\
		2002	7\\
		2003	7\\
		2004	7\\
		2005	7\\
		2006	7\\
		2007	8\\
		2008	8\\
		2009	8\\
		2010	8\\
		2011	8\\
		2012	8\\
		2013	9\\
		2014	9\\
		2015	9\\
		};
		\addplot [color=mycolor4, line width=1.8pt, forget plot]
		  table[row sep=crcr]{%
		1980	0\\
		1981	0\\
		1982	0\\
		1983	0\\
		1984	0\\
		1985	0\\
		1986	0\\
		1987	0\\
		1988	0\\
		1989	0\\
		1990	0\\
		1991	0\\
		1992	0\\
		1993	0\\
		1994	0\\
		1995	0\\
		1996	0\\
		1997	0\\
		1998	0\\
		1999	0\\
		2000	0\\
		2001	0\\
		2002	0\\
		2003	0\\
		2004	0\\
		2005	0\\
		2006	0\\
		2007	0\\
		2008	0\\
		2009	0\\
		2010	0\\
		2011	0\\
		2012	0\\
		2013	0\\
		2014	0\\
		2015	0\\
		};
		\addplot [color=mycolor5, line width=1.8pt, forget plot]
		  table[row sep=crcr]{%
		1980	0\\
		1981	0\\
		1982	0\\
		1983	0\\
		1984	0\\
		1985	0\\
		1986	0\\
		1987	0\\
		1988	0\\
		1989	0\\
		1990	0\\
		1991	0\\
		1992	0\\
		1993	0\\
		1994	0\\
		1995	0\\
		1996	0\\
		1997	0\\
		1998	0\\
		1999	0\\
		2000	0\\
		2001	0\\
		2002	0\\
		2003	0\\
		2004	0\\
		2005	0\\
		2006	0\\
		2007	0\\
		2008	0\\
		2009	0\\
		2010	0\\
		2011	0\\
		2012	0\\
		2013	0\\
		2014	0\\
		2015	0\\
		};
		\addplot [color=mycolor3, line width=1.8pt, forget plot]
		  table[row sep=crcr]{%
		1980	0\\
		1981	0\\
		1982	1\\
		1983	1\\
		1984	1\\
		1985	1\\
		1986	1\\
		1987	1\\
		1988	1\\
		1989	1\\
		1990	1\\
		1991	1\\
		1992	1\\
		1993	1\\
		1994	1\\
		1995	1\\
		1996	1\\
		1997	1\\
		1998	1\\
		1999	1\\
		2000	1\\
		2001	1\\
		2002	1\\
		2003	1\\
		2004	1\\
		2005	1\\
		2006	1\\
		2007	1\\
		2008	1\\
		2009	1\\
		2010	1\\
		2011	1\\
		2012	1\\
		2013	1\\
		2014	1\\
		2015	1\\
		};
		\addplot [color=mycolor4, line width=1.8pt, forget plot]
		  table[row sep=crcr]{%
		1980	0\\
		1981	0\\
		1982	0\\
		1983	0\\
		1984	0\\
		1985	0\\
		1986	0\\
		1987	0\\
		1988	0\\
		1989	0\\
		1990	0\\
		1991	0\\
		1992	0\\
		1993	0\\
		1994	0\\
		1995	0\\
		1996	0\\
		1997	0\\
		1998	0\\
		1999	0\\
		2000	0\\
		2001	0\\
		2002	0\\
		2003	0\\
		2004	0\\
		2005	2\\
		2006	2\\
		2007	2\\
		2008	2\\
		2009	2\\
		2010	2\\
		2011	2\\
		2012	2\\
		2013	2\\
		2014	2\\
		2015	2\\
		};
		\addplot [color=mycolor1, line width=1.8pt, forget plot]
		  table[row sep=crcr]{%
		1980	0\\
		1981	0\\
		1982	0\\
		1983	0\\
		1984	0\\
		1985	0\\
		1986	0\\
		1987	0\\
		1988	0\\
		1989	0\\
		1990	0\\
		1991	0\\
		1992	0\\
		1993	0\\
		1994	0\\
		1995	0\\
		1996	0\\
		1997	0\\
		1998	0\\
		1999	0\\
		2000	0\\
		2001	0\\
		2002	0\\
		2003	0\\
		2004	0\\
		2005	0\\
		2006	0\\
		2007	0\\
		2008	0\\
		2009	0\\
		2010	0\\
		2011	0\\
		2012	0\\
		2013	0\\
		2014	0\\
		2015	0\\
		};
		\addplot [color=mycolor2, line width=1.8pt, forget plot]
		  table[row sep=crcr]{%
		1980	0\\
		1981	0\\
		1982	0\\
		1983	0\\
		1984	0\\
		1985	0\\
		1986	0\\
		1987	0\\
		1988	0\\
		1989	0\\
		1990	0\\
		1991	0\\
		1992	0\\
		1993	0\\
		1994	0\\
		1995	0\\
		1996	0\\
		1997	0\\
		1998	0\\
		1999	0\\
		2000	0\\
		2001	0\\
		2002	0\\
		2003	0\\
		2004	0\\
		2005	0\\
		2006	0\\
		2007	0\\
		2008	0\\
		2009	0\\
		2010	0\\
		2011	0\\
		2012	0\\
		2013	0\\
		2014	0\\
		2015	0\\
		};
		\addplot [color=mycolor3, line width=1.8pt, forget plot]
		  table[row sep=crcr]{%
		1980	0\\
		1981	0\\
		1982	1\\
		1983	3\\
		1984	3\\
		1985	3\\
		1986	3\\
		1987	3\\
		1988	3\\
		1989	3\\
		1990	3\\
		1991	3\\
		1992	3\\
		1993	3\\
		1994	3\\
		1995	3\\
		1996	3\\
		1997	3\\
		1998	3\\
		1999	3\\
		2000	3\\
		2001	3\\
		2002	3\\
		2003	3\\
		2004	3\\
		2005	3\\
		2006	3\\
		2007	3\\
		2008	3\\
		2009	3\\
		2010	3\\
		2011	3\\
		2012	3\\
		2013	3\\
		2014	3\\
		2015	3\\
		};
		\addplot [color=mycolor4, line width=1.8pt, forget plot]
		  table[row sep=crcr]{%
		1980	0\\
		1981	0\\
		1982	0\\
		1983	1\\
		1984	3\\
		1985	3\\
		1986	3\\
		1987	3\\
		1988	3\\
		1989	3\\
		1990	3\\
		1991	3\\
		1992	3\\
		1993	3\\
		1994	3\\
		1995	3\\
		1996	3\\
		1997	3\\
		1998	3\\
		1999	3\\
		2000	3\\
		2001	3\\
		2002	3\\
		2003	3\\
		2004	3\\
		2005	3\\
		2006	3\\
		2007	3\\
		2008	3\\
		2009	3\\
		2010	3\\
		2011	3\\
		2012	3\\
		2013	3\\
		2014	3\\
		2015	3\\
		};
		\addplot [color=mycolor5, line width=1.8pt, forget plot]
		  table[row sep=crcr]{%
		1980	0\\
		1981	0\\
		1982	0\\
		1983	1\\
		1984	1\\
		1985	1\\
		1986	1\\
		1987	1\\
		1988	1\\
		1989	1\\
		1990	1\\
		1991	1\\
		1992	1\\
		1993	1\\
		1994	1\\
		1995	1\\
		1996	1\\
		1997	1\\
		1998	1\\
		1999	1\\
		2000	1\\
		2001	1\\
		2002	1\\
		2003	1\\
		2004	1\\
		2005	1\\
		2006	1\\
		2007	1\\
		2008	1\\
		2009	1\\
		2010	1\\
		2011	1\\
		2012	1\\
		2013	1\\
		2014	1\\
		2015	1\\
		};
		\addplot [color=mycolor4, line width=1.8pt, forget plot]
		  table[row sep=crcr]{%
		1980	0\\
		1981	1\\
		1982	1\\
		1983	2\\
		1984	2\\
		1985	2\\
		1986	2\\
		1987	2\\
		1988	2\\
		1989	2\\
		1990	3\\
		1991	3\\
		1992	3\\
		1993	3\\
		1994	3\\
		1995	4\\
		1996	4\\
		1997	4\\
		1998	4\\
		1999	4\\
		2000	4\\
		2001	4\\
		2002	4\\
		2003	4\\
		2004	4\\
		2005	4\\
		2006	4\\
		2007	4\\
		2008	4\\
		2009	4\\
		2010	4\\
		2011	4\\
		2012	4\\
		2013	4\\
		2014	4\\
		2015	4\\
		};
	\end{axis}
\end{tikzpicture}%

%% file: supp-figures/BT_random_sample.tikz
\tikzstyle{every node}=[font=\Large]

\begin{tikzpicture}
	\begin{axis}[%
		width=0.951\fwidth,
		height=\fheight,
		at={(0\fwidth,0\fheight)},
		scale only axis,
		xmin=1982,
		xmax=2015,
		ymin=0,
		ymax=110,
		xminorticks=true,
		yminorticks=true,
		axis x line*=bottom,
		axis y line*=left,
		xtick ={1980,1990,2000,2010},
		x tick label style = {/pgf/number format/.cd,%
														scaled x ticks = false,
														set thousands separator={},
														fixed},
		axis background/.style={fill=white},
		title style={font=\bfseries\Large},
		title={BT}
		]
		\addplot [color=mycolor2, line width=1.8pt, forget plot]
		 table[row sep=crcr]{%
			1982	0\\
			1983	0\\
			1984	0\\
			1985	0\\
			1986	0\\
			1987	0\\
			1988	0\\
			1989	0\\
			1990	0\\
			1991	0\\
			1992	0\\
			1993	0\\
			1994	0\\
			1995	0\\
			1996	0\\
			1997	0\\
			1998	0\\
			1999	0\\
			2000	0\\
			2001	0\\
			2002	0\\
			2003	0\\
			2004	0\\
			2005	0\\
			2006	0\\
			2007	0\\
			2008	0\\
			2009	0\\
			2010	0\\
			2011	0\\
			2012	0\\
			2013	0\\
			2014	0\\
			2015	0\\
		};
				\addplot [color=mycolor3, line width=1.8pt, forget plot]
		table[row sep=crcr]{%
			1982	0\\
			1983	0\\
			1984	0\\
			1985	0\\
			1986	0\\
			1987	0\\
			1988	0\\
			1989	0\\
			1990	0\\
			1991	0\\
			1992	0\\
			1993	0\\
			1994	0\\
			1995	0\\
			1996	0\\
			1997	0\\
			1998	0\\
			1999	0\\
			2000	0\\
			2001	0\\
			2002	0\\
			2003	0\\
			2004	0\\
			2005	0\\
			2006	0\\
			2007	0\\
			2008	0\\
			2009	0\\
			2010	0\\
			2011	0\\
			2012	0\\
			2013	0\\
			2014	0\\
			2015	0\\
		};
		\addplot [color=mycolor4, line width=1.8pt, forget plot]
		table[row sep=crcr]{%
			1982	0\\
			1983	0\\
			1984	0\\
			1985	0\\
			1986	0\\
			1987	0\\
			1988	0\\
			1989	0\\
			1990	0\\
			1991	0\\
			1992	0\\
			1993	0\\
			1994	0\\
			1995	0\\
			1996	0\\
			1997	0\\
			1998	0\\
			1999	0\\
			2000	0\\
			2001	0\\
			2002	0\\
			2003	0\\
			2004	0\\
			2005	0\\
			2006	0\\
			2007	0\\
			2008	0\\
			2009	0\\
			2010	0\\
			2011	0\\
			2012	0\\
			2013	0\\
			2014	0\\
			2015	0\\
		};
		\addplot [color=mycolor2, line width=1.8pt, forget plot]
		table[row sep=crcr]{%
			1982	0\\
			1983	0\\
			1984	0\\
			1985	0\\
			1986	0\\
			1987	0\\
			1988	0\\
			1989	0\\
			1990	0\\
			1991	0\\
			1992	0\\
			1993	0\\
			1994	0\\
			1995	0\\
			1996	0\\
			1997	0\\
			1998	0\\
			1999	0\\
			2000	0\\
			2001	0\\
			2002	0\\
			2003	0\\
			2004	0\\
			2005	0\\
			2006	0\\
			2007	0\\
			2008	0\\
			2009	0\\
			2010	0\\
			2011	0\\
			2012	0\\
			2013	0\\
			2014	0\\
			2015	0\\
		};
		\addplot [color=mycolor1, line width=1.8pt, forget plot]
		table[row sep=crcr]{%
			1982	0\\
			1983	0\\
			1984	0\\
			1985	0\\
			1986	0\\
			1987	0\\
			1988	0\\
			1989	0\\
			1990	0\\
			1991	0\\
			1992	0\\
			1993	0\\
			1994	0\\
			1995	0\\
			1996	0\\
			1997	0\\
			1998	0\\
			1999	0\\
			2000	0\\
			2001	0\\
			2002	0\\
			2003	0\\
			2004	0\\
			2005	0\\
			2006	0\\
			2007	0\\
			2008	0\\
			2009	0\\
			2010	0\\
			2011	0\\
			2012	0\\
			2013	0\\
			2014	0\\
			2015	0\\
		};
	
				\addplot [color=mycolor1, line width=1.8pt, forget plot]
		table[row sep=crcr]{%
			1982	0\\
			1983	0\\
			1984	0\\
			1985	0\\
			1986	0\\
			1987	0\\
			1988	0\\
			1989	0\\
			1990	0\\
			1991	0\\
			1992	0\\
			1993	0\\
			1994	0\\
			1995	0\\
			1996	0\\
			1997	0\\
			1998	0\\
			1999	0\\
			2000	0\\
			2001	0\\
			2002	0\\
			2003	0\\
			2004	0\\
			2005	0\\
			2006	0\\
			2007	0\\
			2008	0\\
			2009	0\\
			2010	0\\
			2011	0\\
			2012	0\\
			2013	0\\
			2014	0\\
			2015	0\\
		};
		
		\addplot [color=mycolor3, line width=1.8pt, forget plot]
		table[row sep=crcr]{%
			1982	0\\
			1983	0\\
			1984	0\\
			1985	0\\
			1986	0\\
			1987	0\\
			1988	0\\
			1989	0\\
			1990	0\\
			1991	0\\
			1992	0\\
			1993	0\\
			1994	0\\
			1995	0\\
			1996	0\\
			1997	0\\
			1998	0\\
			1999	0\\
			2000	0\\
			2001	0\\
			2002	0\\
			2003	0\\
			2004	0\\
			2005	0\\
			2006	0\\
			2007	0\\
			2008	0\\
			2009	0\\
			2010	0\\
			2011	0\\
			2012	0\\
			2013	0\\
			2014	0\\
			2015	0\\
		};
	
				\addplot [color=mycolor1, line width=1.8pt, forget plot]
		table[row sep=crcr]{%
			1982	0\\
			1983	0\\
			1984	0\\
			1985	0\\
			1986	0\\
			1987	0\\
			1988	0\\
			1989	0\\
			1990	0\\
			1991	0\\
			1992	0\\
			1993	0\\
			1994	0\\
			1995	0\\
			1996	0\\
			1997	0\\
			1998	0\\
			1999	0\\
			2000	0\\
			2001	0\\
			2002	0\\
			2003	0\\
			2004	0\\
			2005	0\\
			2006	0\\
			2007	0\\
			2008	0\\
			2009	0\\
			2010	0\\
			2011	0\\
			2012	0\\
			2013	0\\
			2014	0\\
			2015	0\\
		};
	
		\addplot [color=mycolor2, line width=1.8pt, forget plot]
		table[row sep=crcr]{%
			1982	0\\
			1983	0\\
			1984	0\\
			1985	0\\
			1986	0\\
			1987	0\\
			1988	2\\
			1989	2\\
			1990	2\\
			1991	2\\
			1992	2\\
			1993	2\\
			1994	2\\
			1995	2\\
			1996	2\\
			1997	2\\
			1998	2\\
			1999	2\\
			2000	2\\
			2001	2\\
			2002	2\\
			2003	2\\
			2004	2\\
			2005	2\\
			2006	2\\
			2007	2\\
			2008	2\\
			2009	2\\
			2010	2\\
			2011	2\\
			2012	2\\
			2013	2\\
			2014	2\\
			2015	2\\
		};
	
		\addplot [color=mycolor3, line width=1.8pt, forget plot]
		table[row sep=crcr]{%
			1982	0\\
			1983	0\\
			1984	0\\
			1985	0\\
			1986	0\\
			1987	1\\
			1988	1\\
			1989	1\\
			1990	1\\
			1991	1\\
			1992	1\\
			1993	1\\
			1994	1\\
			1995	1\\
			1996	1\\
			1997	1\\
			1998	1\\
			1999	1\\
			2000	1\\
			2001	1\\
			2002	1\\
			2003	1\\
			2004	1\\
			2005	1\\
			2006	1\\
			2007	2\\
			2008	2\\
			2009	2\\
			2010	2\\
			2011	2\\
			2012	2\\
			2013	2\\
			2014	2\\
			2015	2\\
		};
	
		\addplot [color=mycolor4, line width=1.8pt, forget plot]
		table[row sep=crcr]{%
			1982	1\\
			1983	1\\
			1984	1\\
			1985	1\\
			1986	1\\
			1987	1\\
			1988	1\\
			1989	1\\
			1990	1\\
			1991	1\\
			1992	1\\
			1993	1\\
			1994	1\\
			1995	1\\
			1996	1\\
			1997	2\\
			1998	2\\
			1999	2\\
			2000	2\\
			2001	3\\
			2002	3\\
			2003	3\\
			2004	3\\
			2005	3\\
			2006	3\\
			2007	3\\
			2008	3\\
			2009	3\\
			2010	3\\
			2011	3\\
			2012	3\\
			2013	3\\
			2014	3\\
			2015	3\\
		};
	
			\addplot [color=mycolor5, line width=1.8pt, forget plot]
	table[row sep=crcr]{%
		1982	1\\
		1983	1\\
		1984	1\\
		1985	1\\
		1986	1\\
		1987	2\\
		1988	2\\
		1989	2\\
		1990	2\\
		1991	2\\
		1992	2\\
		1993	2\\
		1994	2\\
		1995	2\\
		1996	2\\
		1997	2\\
		1998	2\\
		1999	2\\
		2000	2\\
		2001	2\\
		2002	2\\
		2003	2\\
		2004	2\\
		2005	2\\
		2006	3\\
		2007	3\\
		2008	3\\
		2009	3\\
		2010	3\\
		2011	3\\
		2012	3\\
		2013	3\\
		2014	3\\
		2015	3\\
	};
	
		\addplot [color=mycolor1, line width=1.8pt, forget plot]
		table[row sep=crcr]{%
			1982	0\\
			1983	0\\
			1984	0\\
			1985	1\\
			1986	2\\
			1987	2\\
			1988	2\\
			1989	3\\
			1990	3\\
			1991	3\\
			1992	3\\
			1993	3\\
			1994	3\\
			1995	3\\
			1996	3\\
			1997	3\\
			1998	3\\
			1999	3\\
			2000	3\\
			2001	3\\
			2002	3\\
			2003	3\\
			2004	3\\
			2005	3\\
			2006	3\\
			2007	3\\
			2008	3\\
			2009	3\\
			2010	3\\
			2011	3\\
			2012	3\\
			2013	3\\
			2014	3\\
			2015	3\\
		};

		\addplot [color=mycolor2, line width=1.8pt, forget plot]
		table[row sep=crcr]{%
			1982	0\\
			1983	1\\
			1984	1\\
			1985	1\\
			1986	1\\
			1987	1\\
			1988	2\\
			1989	3\\
			1990	3\\
			1991	4\\
			1992	4\\
			1993	5\\
			1994	5\\
			1995	5\\
			1996	5\\
			1997	5\\
			1998	5\\
			1999	5\\
			2000	5\\
			2001	5\\
			2002	5\\
			2003	5\\
			2004	5\\
			2005	5\\
			2006	5\\
			2007	5\\
			2008	5\\
			2009	5\\
			2010	5\\
			2011	5\\
			2012	5\\
			2013	5\\
			2014	5\\
			2015	6\\
		};

		\addplot [color=mycolor3, line width=1.8pt, forget plot]
		  table[row sep=crcr]{%
			1982	0\\
			1983	1\\
			1984	4\\
			1985	6\\
			1986	6\\
			1987	6\\
			1988	6\\
			1989	6\\
			1990	6\\
			1991	7\\
			1992	7\\
			1993	7\\
			1994	7\\
			1995	7\\
			1996	7\\
			1997	7\\
			1998	7\\
			1999	7\\
			2000	7\\
			2001	7\\
			2002	7\\
			2003	7\\
			2004	7\\
			2005	7\\
			2006	7\\
			2007	7\\
			2008	7\\
			2009	7\\
			2010	7\\
			2011	7\\
			2012	7\\
			2013	7\\
			2014	7\\
			2015	7\\
		};
	
		\addplot [color=mycolor4, line width=1.8pt, forget plot]
		table[row sep=crcr]{%
			1982	0\\
			1983	0\\
			1984	0\\
			1985	0\\
			1986	0\\
			1987	1\\
			1988	1\\
			1989	1\\
			1990	3\\
			1991	3\\
			1992	3\\
			1993	3\\
			1994	5\\
			1995	5\\
			1996	7\\
			1997	7\\
			1998	9\\
			1999	10\\
			2000	10\\
			2001	10\\
			2002	10\\
			2003	10\\
			2004	10\\
			2005	10\\
			2006	10\\
			2007	10\\
			2008	10\\
			2009	10\\
			2010	10\\
			2011	10\\
			2012	11\\
			2013	11\\
			2014	11\\
			2015	11\\
		};

		\addplot [color=mycolor5, line width=1.8pt, forget plot]
		table[row sep=crcr]{%
			1982	0\\
			1983	0\\
			1984	2\\
			1985	4\\
			1986	4\\
			1987	5\\
			1988	6\\
			1989	8\\
			1990	8\\
			1991	8\\
			1992	8\\
			1993	8\\
			1994	8\\
			1995	9\\
			1996	10\\
			1997	11\\
			1998	11\\
			1999	11\\
			2000	11\\
			2001	11\\
			2002	11\\
			2003	11\\
			2004	11\\
			2005	11\\
			2006	11\\
			2007	11\\
			2008	11\\
			2009	11\\
			2010	11\\
			2011	11\\
			2012	11\\
			2013	11\\
			2014	11\\
			2015	11\\
		};

		\addplot [color=mycolor1, line width=1.8pt, forget plot]
		table[row sep=crcr]{%
			1982	1\\
			1983	5\\
			1984	6\\
			1985	8\\
			1986	8\\
			1987	8\\
			1988	8\\
			1989	9\\
			1990	10\\
			1991	12\\
			1992	13\\
			1993	13\\
			1994	13\\
			1995	14\\
			1996	14\\
			1997	14\\
			1998	15\\
			1999	15\\
			2000	15\\
			2001	15\\
			2002	15\\
			2003	15\\
			2004	16\\
			2005	16\\
			2006	16\\
			2007	16\\
			2008	16\\
			2009	16\\
			2010	16\\
			2011	16\\
			2012	16\\
			2013	16\\
			2014	16\\
			2015	16\\
		};

		\addplot [color=mycolor2, line width=1.8pt, forget plot]
		  table[row sep=crcr]{%
			1982	0\\
			1983	1\\
			1984	3\\
			1985	4\\
			1986	6\\
			1987	9\\
			1988	9\\
			1989	10\\
			1990	11\\
			1991	12\\
			1992	13\\
			1993	13\\
			1994	13\\
			1995	14\\
			1996	14\\
			1997	15\\
			1998	16\\
			1999	16\\
			2000	16\\
			2001	16\\
			2002	16\\
			2003	16\\
			2004	16\\
			2005	16\\
			2006	16\\
			2007	16\\
			2008	16\\
			2009	16\\
			2010	16\\
			2011	16\\
			2012	16\\
			2013	17\\
			2014	17\\
			2015	17\\
		};

		\addplot [color=mycolor3, line width=1.8pt, forget plot]
		  table[row sep=crcr]{%
			1982	0\\
			1983	1\\
			1984	2\\
			1985	2\\
			1986	2\\
			1987	3\\
			1988	4\\
			1989	5\\
			1990	7\\
			1991	7\\
			1992	9\\
			1993	11\\
			1994	14\\
			1995	18\\
			1996	23\\
			1997	28\\
			1998	31\\
			1999	33\\
			2000	35\\
			2001	37\\
			2002	40\\
			2003	41\\
			2004	44\\
			2005	47\\
			2006	53\\
			2007	57\\
			2008	63\\
			2009	74\\
			2010	79\\
			2011	89\\
			2012	95\\
			2013	100\\
			2014	109\\
			2015	110\\
		};

	\end{axis}
\end{tikzpicture}%

%% file: supp-figures/BT_dynamic_PL_data.tikz
\tikzstyle{every node}=[font=\Large]

\begin{tikzpicture}

	\begin{axis}[%
		width=0.951\fwidth,
		height=\fheight,
		at={(0\fwidth,0\fheight)},
		scale only axis,
		axis x line*=bottom,
		axis y line*=left,
		xmode=log,
		xmin=1,
		xmax=301,
		xminorticks=true,
		ymode=log,
		ymin=0.000247892910262766,
		ymax=1.00024789291026,
		yminorticks=true,
		axis background/.style={fill=white},
		title style={font=\bfseries\Large},
		title={BT},
		legend style={legend cell align=left, align=left, draw=white!15!black}
		]
		\addplot [color=mycolor1,line width = 2pt]
		  table[row sep=crcr]{%
		1	1.00024789291026\\
		2	0.10039662865642\\
		3	0.0213187902825979\\
		4	0.0059494298463064\\
		5	0.000743678730788299\\
		};
		\addlegendentry{1984}
		
		\addplot [color=mycolor2,line width = 2pt]
		  table[row sep=crcr]{%
		1	1.00024789291026\\
		2	0.577342588001983\\
		3	0.379771938522558\\
		4	0.259791769955379\\
		5	0.18368864650471\\
		6	0.132126921170055\\
		7	0.102875557759048\\
		8	0.0815567674764502\\
		9	0.0614774417451661\\
		10	0.0490827962320278\\
		11	0.0406544372830937\\
		12	0.0322260783341596\\
		13	0.0245413981160139\\
		14	0.0200793257312841\\
		15	0.0163609320773426\\
		16	0.0126425384234011\\
		17	0.00991571641051066\\
		18	0.0066931085770947\\
		19	0.00570153693604363\\
		20	0.00421417947446703\\
		21	0.0037183936539415\\
		22	0.00272682201289043\\
		23	0.00247892910262767\\
		24	0.0022310361923649\\
		25	0.0022310361923649\\
		26	0.00198314328210213\\
		27	0.0014873574615766\\
		28	0.00123946455131383\\
		29	0.000991571641051066\\
		30	0.000991571641051066\\
		31	0.000743678730788299\\
		32	0.000743678730788299\\
		33	0.000743678730788299\\
		34	0.000743678730788299\\
		35	0.000495785820525533\\
		36	0.000247892910262766\\
		37	0.000247892910262766\\
		38	0.000247892910262766\\
		39	0.000247892910262766\\
		};
		\addlegendentry{1988}
		
		\addplot [color=mycolor3,line width = 2pt]
		  table[row sep=crcr]{%
		1	1.00024789291026\\
		2	0.638324243926624\\
		3	0.48116013882003\\
		4	0.376549330689142\\
		5	0.292017848289539\\
		6	0.23748140803173\\
		7	0.195091720376797\\
		8	0.158899355478433\\
		9	0.132126921170055\\
		10	0.114526524541398\\
		11	0.0999008428358949\\
		12	0.0842835894893406\\
		13	0.0733763014377789\\
		14	0.0624690133862172\\
		15	0.0535448686167576\\
		16	0.0478433316807139\\
		17	0.0421417947446703\\
		18	0.0386712940009916\\
		19	0.0337134357957362\\
		20	0.0292513634110064\\
		21	0.0275161130391671\\
		22	0.0242935052057511\\
		23	0.0213187902825979\\
		24	0.0195835399107586\\
		25	0.0173525037183937\\
		26	0.0156172533465543\\
		27	0.0133862171541894\\
		28	0.0109072880515617\\
		29	0.0101636093207734\\
		30	0.00842835894893406\\
		31	0.00818046603867129\\
		32	0.00718889439762023\\
		33	0.00644521566683193\\
		34	0.00495785820525533\\
		35	0.00470996529499256\\
		36	0.00421417947446703\\
		37	0.00396628656420426\\
		38	0.00347050074367873\\
		39	0.00322260783341596\\
		40	0.0029747149231532\\
		41	0.0029747149231532\\
		42	0.0029747149231532\\
		43	0.0022310361923649\\
		44	0.0022310361923649\\
		45	0.0022310361923649\\
		46	0.00198314328210213\\
		47	0.0014873574615766\\
		48	0.000991571641051066\\
		49	0.000991571641051066\\
		50	0.000991571641051066\\
		51	0.000991571641051066\\
		52	0.000991571641051066\\
		53	0.000991571641051066\\
		54	0.000991571641051066\\
		55	0.000991571641051066\\
		56	0.000991571641051066\\
		57	0.000743678730788299\\
		58	0.000743678730788299\\
		59	0.000743678730788299\\
		60	0.000743678730788299\\
		61	0.000743678730788299\\
		62	0.000495785820525533\\
		63	0.000495785820525533\\
		64	0.000495785820525533\\
		65	0.000495785820525533\\
		66	0.000495785820525533\\
		67	0.000495785820525533\\
		68	0.000495785820525533\\
		69	0.000495785820525533\\
		70	0.000495785820525533\\
		71	0.000495785820525533\\
		72	0.000495785820525533\\
		73	0.000495785820525533\\
		74	0.000247892910262766\\
		75	0.000247892910262766\\
		76	0.000247892910262766\\
		77	0.000247892910262766\\
		78	0.000247892910262766\\
		79	0.000247892910262766\\
		80	0.000247892910262766\\
		81	0.000247892910262766\\
		82	0.000247892910262766\\
		83	0.000247892910262766\\
		84	0.000247892910262766\\
		85	0.000247892910262766\\
		86	0.000247892910262766\\
		87	0.000247892910262766\\
		88	0.000247892910262766\\
		89	0.000247892910262766\\
		};
		\addlegendentry{1992}
		
		\addplot [color=mycolor4,line width = 2pt]
		  table[row sep=crcr]{%
		1	1.00024789291026\\
		2	0.664600892414477\\
		3	0.522310361923649\\
		4	0.431333663857214\\
		5	0.348289538919187\\
		6	0.292513634110064\\
		7	0.25086762518592\\
		8	0.215171046108081\\
		9	0.181209717402082\\
		10	0.152702032721864\\
		11	0.135349529003471\\
		12	0.118740704015865\\
		13	0.103867129400099\\
		14	0.0924640555280119\\
		15	0.0837878036688151\\
		16	0.0768468021814576\\
		17	0.0694100148735746\\
		18	0.0627169062964799\\
		19	0.0567674764501735\\
		20	0.0525532969757065\\
		21	0.0478433316807139\\
		22	0.0436291522062469\\
		23	0.0389191869112543\\
		24	0.0359444719881011\\
		25	0.0332176499752107\\
		26	0.0312345066931086\\
		27	0.0280118988596926\\
		28	0.0250371839365394\\
		29	0.0233019335647001\\
		30	0.0220624690133862\\
		31	0.0213187902825979\\
		32	0.0195835399107586\\
		33	0.018096182449182\\
		34	0.0168567178978681\\
		35	0.0156172533465543\\
		36	0.0143777887952405\\
		37	0.0143777887952405\\
		38	0.0126425384234011\\
		39	0.0111551809618245\\
		40	0.010659395141299\\
		41	0.00966782350024789\\
		42	0.00793257312840853\\
		43	0.00644521566683193\\
		44	0.00619732275656916\\
		45	0.0059494298463064\\
		46	0.0059494298463064\\
		47	0.0052057511155181\\
		48	0.00495785820525533\\
		49	0.0044620723847298\\
		50	0.00421417947446703\\
		51	0.00421417947446703\\
		52	0.00421417947446703\\
		53	0.00396628656420426\\
		54	0.0037183936539415\\
		55	0.00347050074367873\\
		56	0.0029747149231532\\
		57	0.0029747149231532\\
		58	0.00272682201289043\\
		59	0.0022310361923649\\
		60	0.00173525037183937\\
		61	0.0014873574615766\\
		62	0.0014873574615766\\
		63	0.00123946455131383\\
		64	0.00123946455131383\\
		65	0.00123946455131383\\
		66	0.00123946455131383\\
		67	0.000991571641051066\\
		68	0.000743678730788299\\
		69	0.000743678730788299\\
		70	0.000743678730788299\\
		71	0.000743678730788299\\
		72	0.000743678730788299\\
		73	0.000743678730788299\\
		74	0.000743678730788299\\
		75	0.000743678730788299\\
		76	0.000743678730788299\\
		77	0.000743678730788299\\
		78	0.000743678730788299\\
		79	0.000743678730788299\\
		80	0.000743678730788299\\
		81	0.000743678730788299\\
		82	0.000743678730788299\\
		83	0.000743678730788299\\
		84	0.000743678730788299\\
		85	0.000743678730788299\\
		86	0.000743678730788299\\
		87	0.000495785820525533\\
		88	0.000495785820525533\\
		89	0.000495785820525533\\
		90	0.000495785820525533\\
		91	0.000495785820525533\\
		92	0.000495785820525533\\
		93	0.000495785820525533\\
		94	0.000495785820525533\\
		95	0.000495785820525533\\
		96	0.000495785820525533\\
		97	0.000495785820525533\\
		98	0.000495785820525533\\
		99	0.000495785820525533\\
		100	0.000495785820525533\\
		101	0.000495785820525533\\
		102	0.000495785820525533\\
		103	0.000495785820525533\\
		104	0.000495785820525533\\
		105	0.000495785820525533\\
		106	0.000495785820525533\\
		107	0.000247892910262766\\
		108	0.000247892910262766\\
		109	0.000247892910262766\\
		110	0.000247892910262766\\
		111	0.000247892910262766\\
		112	0.000247892910262766\\
		113	0.000247892910262766\\
		114	0.000247892910262766\\
		115	0.000247892910262766\\
		116	0.000247892910262766\\
		117	0.000247892910262766\\
		118	0.000247892910262766\\
		119	0.000247892910262766\\
		120	0.000247892910262766\\
		121	0.000247892910262766\\
		122	0.000247892910262766\\
		123	0.000247892910262766\\
		124	0.000247892910262766\\
		125	0.000247892910262766\\
		126	0.000247892910262766\\
		127	0.000247892910262766\\
		128	0.000247892910262766\\
		};
		\addlegendentry{1996}
		
		\addplot [color=mycolor5,line width = 2pt]
		  table[row sep=crcr]{%
		1	1.00024789291026\\
		2	0.676995537927615\\
		3	0.545364402578086\\
		4	0.459841348537432\\
		5	0.380267724343084\\
		6	0.323252354982647\\
		7	0.282350024789291\\
		8	0.247149231531978\\
		9	0.213187902825979\\
		10	0.187902825979177\\
		11	0.166088249876054\\
		12	0.146008924144769\\
		13	0.126425384234011\\
		14	0.114030738720873\\
		15	0.102627664848785\\
		16	0.0941993058998513\\
		17	0.0852751611303917\\
		18	0.0780862667327714\\
		19	0.0728805156172534\\
		20	0.0654437283093704\\
		21	0.0614774417451661\\
		22	0.0557759048091225\\
		23	0.0528011898859693\\
		24	0.0473475458601884\\
		25	0.0443728309370352\\
		26	0.0394149727317799\\
		27	0.0364402578086267\\
		28	0.0352007932573128\\
		29	0.0327218641546852\\
		30	0.030738720872583\\
		31	0.0285076846802181\\
		32	0.0267724343083788\\
		33	0.0252850768468022\\
		34	0.0245413981160139\\
		35	0.0225582548339117\\
		36	0.0220624690133862\\
		37	0.0210708973723352\\
		38	0.0193356470004958\\
		39	0.0178482895389192\\
		40	0.0163609320773426\\
		41	0.0151214675260288\\
		42	0.0138820029747149\\
		43	0.0136341100644522\\
		44	0.0131383242439266\\
		45	0.0126425384234011\\
		46	0.0121467526028756\\
		47	0.01165096678235\\
		48	0.0104115022310362\\
		49	0.00991571641051066\\
		50	0.00966782350024789\\
		51	0.00842835894893406\\
		52	0.00793257312840853\\
		53	0.00768468021814576\\
		54	0.00694100148735746\\
		55	0.00644521566683193\\
		56	0.0059494298463064\\
		57	0.00570153693604363\\
		58	0.00570153693604363\\
		59	0.00545364402578086\\
		60	0.0052057511155181\\
		61	0.0052057511155181\\
		62	0.00495785820525533\\
		63	0.00470996529499256\\
		64	0.00470996529499256\\
		65	0.00470996529499256\\
		66	0.0044620723847298\\
		67	0.00322260783341596\\
		68	0.00272682201289043\\
		69	0.00272682201289043\\
		70	0.0022310361923649\\
		71	0.0022310361923649\\
		72	0.0022310361923649\\
		73	0.0022310361923649\\
		74	0.00198314328210213\\
		75	0.00173525037183937\\
		76	0.00173525037183937\\
		77	0.0014873574615766\\
		78	0.0014873574615766\\
		79	0.00123946455131383\\
		80	0.00123946455131383\\
		81	0.00123946455131383\\
		82	0.00123946455131383\\
		83	0.000991571641051066\\
		84	0.000991571641051066\\
		85	0.000991571641051066\\
		86	0.000743678730788299\\
		87	0.000743678730788299\\
		88	0.000743678730788299\\
		89	0.000743678730788299\\
		90	0.000743678730788299\\
		91	0.000743678730788299\\
		92	0.000743678730788299\\
		93	0.000743678730788299\\
		94	0.000743678730788299\\
		95	0.000743678730788299\\
		96	0.000743678730788299\\
		97	0.000743678730788299\\
		98	0.000743678730788299\\
		99	0.000743678730788299\\
		100	0.000743678730788299\\
		101	0.000495785820525533\\
		102	0.000495785820525533\\
		103	0.000495785820525533\\
		104	0.000495785820525533\\
		105	0.000495785820525533\\
		106	0.000495785820525533\\
		107	0.000495785820525533\\
		108	0.000495785820525533\\
		109	0.000495785820525533\\
		110	0.000495785820525533\\
		111	0.000495785820525533\\
		112	0.000495785820525533\\
		113	0.000495785820525533\\
		114	0.000495785820525533\\
		115	0.000495785820525533\\
		116	0.000495785820525533\\
		117	0.000495785820525533\\
		118	0.000495785820525533\\
		119	0.000495785820525533\\
		120	0.000495785820525533\\
		121	0.000495785820525533\\
		122	0.000495785820525533\\
		123	0.000495785820525533\\
		124	0.000495785820525533\\
		125	0.000247892910262766\\
		126	0.000247892910262766\\
		127	0.000247892910262766\\
		128	0.000247892910262766\\
		129	0.000247892910262766\\
		130	0.000247892910262766\\
		131	0.000247892910262766\\
		132	0.000247892910262766\\
		133	0.000247892910262766\\
		134	0.000247892910262766\\
		135	0.000247892910262766\\
		136	0.000247892910262766\\
		137	0.000247892910262766\\
		138	0.000247892910262766\\
		139	0.000247892910262766\\
		140	0.000247892910262766\\
		141	0.000247892910262766\\
		142	0.000247892910262766\\
		143	0.000247892910262766\\
		144	0.000247892910262766\\
		145	0.000247892910262766\\
		146	0.000247892910262766\\
		147	0.000247892910262766\\
		148	0.000247892910262766\\
		149	0.000247892910262766\\
		150	0.000247892910262766\\
		151	0.000247892910262766\\
		152	0.000247892910262766\\
		153	0.000247892910262766\\
		154	0.000247892910262766\\
		155	0.000247892910262766\\
		156	0.000247892910262766\\
		157	0.000247892910262766\\
		158	0.000247892910262766\\
		};
		\addlegendentry{2000}
		
		\addplot [color=mycolor1,line width = 2pt]
		  table[row sep=crcr]{%
		1	1.00024789291026\\
		2	0.687159147248389\\
		3	0.557015369360436\\
		4	0.476945959345563\\
		5	0.399851264253842\\
		6	0.343083787803669\\
		7	0.300941993058998\\
		8	0.264501735250372\\
		9	0.233019335647001\\
		10	0.20699058006941\\
		11	0.184928111056024\\
		12	0.165344571145265\\
		13	0.146008924144769\\
		14	0.130887456618741\\
		15	0.117501239464551\\
		16	0.107089737233515\\
		17	0.0989092711948438\\
		18	0.090233019335647\\
		19	0.0855230540406544\\
		20	0.0780862667327714\\
		21	0.071145265245414\\
		22	0.0654437283093704\\
		23	0.0592464055528012\\
		24	0.0552801189885969\\
		25	0.0508180466038671\\
		26	0.0478433316807139\\
		27	0.0433812592959841\\
		28	0.0418939018344075\\
		29	0.0386712940009916\\
		30	0.0359444719881011\\
		31	0.0347050074367873\\
		32	0.0317302925136341\\
		33	0.0299950421417947\\
		34	0.0290034705007437\\
		35	0.0280118988596926\\
		36	0.026524541398116\\
		37	0.0250371839365394\\
		38	0.0240456122954884\\
		39	0.0228061477441745\\
		40	0.0215666831928607\\
		41	0.0203272186415469\\
		42	0.0193356470004958\\
		43	0.0183440753594447\\
		44	0.0166088249876054\\
		45	0.0158651462568171\\
		46	0.0151214675260288\\
		47	0.0143777887952405\\
		48	0.0136341100644522\\
		49	0.0126425384234011\\
		50	0.0121467526028756\\
		51	0.0118988596926128\\
		52	0.0114030738720873\\
		53	0.0109072880515617\\
		54	0.0104115022310362\\
		55	0.00991571641051066\\
		56	0.00941993058998513\\
		57	0.00842835894893406\\
		58	0.00793257312840853\\
		59	0.00768468021814576\\
		60	0.00768468021814576\\
		61	0.00718889439762023\\
		62	0.00694100148735746\\
		63	0.0066931085770947\\
		64	0.0066931085770947\\
		65	0.0059494298463064\\
		66	0.0052057511155181\\
		67	0.00495785820525533\\
		68	0.00470996529499256\\
		69	0.00470996529499256\\
		70	0.00470996529499256\\
		71	0.00470996529499256\\
		72	0.0044620723847298\\
		73	0.0044620723847298\\
		74	0.0037183936539415\\
		75	0.00322260783341596\\
		76	0.0029747149231532\\
		77	0.00272682201289043\\
		78	0.00272682201289043\\
		79	0.00272682201289043\\
		80	0.00272682201289043\\
		81	0.00247892910262767\\
		82	0.00247892910262767\\
		83	0.00247892910262767\\
		84	0.0022310361923649\\
		85	0.0022310361923649\\
		86	0.0022310361923649\\
		87	0.00198314328210213\\
		88	0.00198314328210213\\
		89	0.00198314328210213\\
		90	0.00198314328210213\\
		91	0.00173525037183937\\
		92	0.00123946455131383\\
		93	0.000991571641051066\\
		94	0.000991571641051066\\
		95	0.000991571641051066\\
		96	0.000991571641051066\\
		97	0.000991571641051066\\
		98	0.000991571641051066\\
		99	0.000991571641051066\\
		100	0.000991571641051066\\
		101	0.000743678730788299\\
		102	0.000743678730788299\\
		103	0.000743678730788299\\
		104	0.000743678730788299\\
		105	0.000743678730788299\\
		106	0.000743678730788299\\
		107	0.000495785820525533\\
		108	0.000495785820525533\\
		109	0.000495785820525533\\
		110	0.000495785820525533\\
		111	0.000495785820525533\\
		112	0.000495785820525533\\
		113	0.000495785820525533\\
		114	0.000495785820525533\\
		115	0.000495785820525533\\
		116	0.000495785820525533\\
		117	0.000495785820525533\\
		118	0.000495785820525533\\
		119	0.000495785820525533\\
		120	0.000495785820525533\\
		121	0.000495785820525533\\
		122	0.000495785820525533\\
		123	0.000495785820525533\\
		124	0.000495785820525533\\
		125	0.000495785820525533\\
		126	0.000495785820525533\\
		127	0.000495785820525533\\
		128	0.000495785820525533\\
		129	0.000495785820525533\\
		130	0.000495785820525533\\
		131	0.000495785820525533\\
		132	0.000495785820525533\\
		133	0.000495785820525533\\
		134	0.000495785820525533\\
		135	0.000495785820525533\\
		136	0.000495785820525533\\
		137	0.000495785820525533\\
		138	0.000495785820525533\\
		139	0.000495785820525533\\
		140	0.000495785820525533\\
		141	0.000495785820525533\\
		142	0.000495785820525533\\
		143	0.000495785820525533\\
		144	0.000495785820525533\\
		145	0.000495785820525533\\
		146	0.000247892910262766\\
		147	0.000247892910262766\\
		148	0.000247892910262766\\
		149	0.000247892910262766\\
		150	0.000247892910262766\\
		151	0.000247892910262766\\
		152	0.000247892910262766\\
		153	0.000247892910262766\\
		154	0.000247892910262766\\
		155	0.000247892910262766\\
		156	0.000247892910262766\\
		157	0.000247892910262766\\
		158	0.000247892910262766\\
		159	0.000247892910262766\\
		160	0.000247892910262766\\
		161	0.000247892910262766\\
		162	0.000247892910262766\\
		163	0.000247892910262766\\
		164	0.000247892910262766\\
		165	0.000247892910262766\\
		166	0.000247892910262766\\
		167	0.000247892910262766\\
		168	0.000247892910262766\\
		169	0.000247892910262766\\
		170	0.000247892910262766\\
		171	0.000247892910262766\\
		172	0.000247892910262766\\
		173	0.000247892910262766\\
		174	0.000247892910262766\\
		175	0.000247892910262766\\
		176	0.000247892910262766\\
		177	0.000247892910262766\\
		178	0.000247892910262766\\
		179	0.000247892910262766\\
		180	0.000247892910262766\\
		181	0.000247892910262766\\
		182	0.000247892910262766\\
		183	0.000247892910262766\\
		};
		\addlegendentry{2004}
		
		\addplot [color=mycolor2,line width = 2pt]
		  table[row sep=crcr]{%
		1	1.00024789291026\\
		2	0.693604362915221\\
		3	0.568666336142786\\
		4	0.489092711948438\\
		5	0.416212196331185\\
		6	0.359196826970749\\
		7	0.317055032226078\\
		8	0.280118988596926\\
		9	0.250619732275657\\
		10	0.224838869608329\\
		11	0.201289043133366\\
		12	0.180466038671294\\
		13	0.162865642042638\\
		14	0.146008924144769\\
		15	0.131631135349529\\
		16	0.121715418939018\\
		17	0.110064452156668\\
		18	0.100892414476946\\
		19	0.0949429846306396\\
		20	0.0862667327714427\\
		21	0.0790778383738225\\
		22	0.0741199801685672\\
		23	0.0689142290530491\\
		24	0.0634605850272682\\
		25	0.0587506197322757\\
		26	0.0532969757064948\\
		27	0.0505701536936044\\
		28	0.0470996529499256\\
		29	0.0448686167575607\\
		30	0.0418939018344075\\
		31	0.0399107585523054\\
		32	0.037183936539415\\
		33	0.0332176499752107\\
		34	0.0327218641546852\\
		35	0.0304908279623203\\
		36	0.029747149231532\\
		37	0.0287555775904809\\
		38	0.0275161130391671\\
		39	0.0257808626673277\\
		40	0.0247892910262766\\
		41	0.0242935052057511\\
		42	0.0228061477441745\\
		43	0.0215666831928607\\
		44	0.0205751115518096\\
		45	0.0200793257312841\\
		46	0.0193356470004958\\
		47	0.0185919682697075\\
		48	0.0173525037183937\\
		49	0.0171046108081309\\
		50	0.0158651462568171\\
		51	0.0146256817055032\\
		52	0.0146256817055032\\
		53	0.0138820029747149\\
		54	0.0131383242439266\\
		55	0.0126425384234011\\
		56	0.0121467526028756\\
		57	0.01165096678235\\
		58	0.0111551809618245\\
		59	0.010659395141299\\
		60	0.00991571641051066\\
		61	0.00941993058998513\\
		62	0.00917203767972236\\
		63	0.00867625185919683\\
		64	0.00818046603867129\\
		65	0.00818046603867129\\
		66	0.00793257312840853\\
		67	0.00768468021814576\\
		68	0.00718889439762023\\
		69	0.00619732275656916\\
		70	0.0059494298463064\\
		71	0.00570153693604363\\
		72	0.00570153693604363\\
		73	0.00545364402578086\\
		74	0.00545364402578086\\
		75	0.00545364402578086\\
		76	0.00545364402578086\\
		77	0.00545364402578086\\
		78	0.00495785820525533\\
		79	0.00396628656420426\\
		80	0.00396628656420426\\
		81	0.0037183936539415\\
		82	0.00347050074367873\\
		83	0.00322260783341596\\
		84	0.0029747149231532\\
		85	0.0029747149231532\\
		86	0.0029747149231532\\
		87	0.0029747149231532\\
		88	0.0029747149231532\\
		89	0.00272682201289043\\
		90	0.00272682201289043\\
		91	0.00272682201289043\\
		92	0.00272682201289043\\
		93	0.00272682201289043\\
		94	0.0022310361923649\\
		95	0.0022310361923649\\
		96	0.0022310361923649\\
		97	0.0022310361923649\\
		98	0.0022310361923649\\
		99	0.00198314328210213\\
		100	0.00198314328210213\\
		101	0.00198314328210213\\
		102	0.00198314328210213\\
		103	0.00198314328210213\\
		104	0.00198314328210213\\
		105	0.00198314328210213\\
		106	0.0014873574615766\\
		107	0.00123946455131383\\
		108	0.00123946455131383\\
		109	0.00123946455131383\\
		110	0.00123946455131383\\
		111	0.00123946455131383\\
		112	0.000743678730788299\\
		113	0.000743678730788299\\
		114	0.000743678730788299\\
		115	0.000743678730788299\\
		116	0.000743678730788299\\
		117	0.000743678730788299\\
		118	0.000743678730788299\\
		119	0.000743678730788299\\
		120	0.000743678730788299\\
		121	0.000743678730788299\\
		122	0.000743678730788299\\
		123	0.000743678730788299\\
		124	0.000743678730788299\\
		125	0.000743678730788299\\
		126	0.000495785820525533\\
		127	0.000495785820525533\\
		128	0.000495785820525533\\
		129	0.000495785820525533\\
		130	0.000495785820525533\\
		131	0.000495785820525533\\
		132	0.000495785820525533\\
		133	0.000495785820525533\\
		134	0.000495785820525533\\
		135	0.000495785820525533\\
		136	0.000495785820525533\\
		137	0.000495785820525533\\
		138	0.000495785820525533\\
		139	0.000495785820525533\\
		140	0.000495785820525533\\
		141	0.000495785820525533\\
		142	0.000495785820525533\\
		143	0.000495785820525533\\
		144	0.000495785820525533\\
		145	0.000495785820525533\\
		146	0.000495785820525533\\
		147	0.000495785820525533\\
		148	0.000495785820525533\\
		149	0.000495785820525533\\
		150	0.000495785820525533\\
		151	0.000495785820525533\\
		152	0.000495785820525533\\
		153	0.000495785820525533\\
		154	0.000495785820525533\\
		155	0.000495785820525533\\
		156	0.000495785820525533\\
		157	0.000495785820525533\\
		158	0.000495785820525533\\
		159	0.000247892910262766\\
		160	0.000247892910262766\\
		161	0.000247892910262766\\
		162	0.000247892910262766\\
		163	0.000247892910262766\\
		164	0.000247892910262766\\
		165	0.000247892910262766\\
		166	0.000247892910262766\\
		167	0.000247892910262766\\
		168	0.000247892910262766\\
		169	0.000247892910262766\\
		170	0.000247892910262766\\
		171	0.000247892910262766\\
		172	0.000247892910262766\\
		173	0.000247892910262766\\
		174	0.000247892910262766\\
		175	0.000247892910262766\\
		176	0.000247892910262766\\
		177	0.000247892910262766\\
		178	0.000247892910262766\\
		179	0.000247892910262766\\
		180	0.000247892910262766\\
		181	0.000247892910262766\\
		182	0.000247892910262766\\
		183	0.000247892910262766\\
		184	0.000247892910262766\\
		185	0.000247892910262766\\
		186	0.000247892910262766\\
		187	0.000247892910262766\\
		188	0.000247892910262766\\
		189	0.000247892910262766\\
		190	0.000247892910262766\\
		191	0.000247892910262766\\
		192	0.000247892910262766\\
		193	0.000247892910262766\\
		194	0.000247892910262766\\
		};
		\addlegendentry{2008}
		
		\addplot [color=mycolor3,line width = 2pt]
		  table[row sep=crcr]{%
		1	1.00024789291026\\
		2	0.697818542389688\\
		3	0.578582052553297\\
		4	0.500495785820526\\
		5	0.428606841844323\\
		6	0.372335151214675\\
		7	0.330441249380268\\
		8	0.294000991571641\\
		9	0.264501735250372\\
		10	0.238472979672781\\
		11	0.215171046108081\\
		12	0.192117005453644\\
		13	0.177243430837878\\
		14	0.161626177491324\\
		15	0.146008924144769\\
		16	0.135597421913733\\
		17	0.123450669310858\\
		18	0.11204759543877\\
		19	0.104610808130887\\
		20	0.0959345562716906\\
		21	0.0872583044124938\\
		22	0.0805651958353991\\
		23	0.0775904809122459\\
		24	0.0728805156172534\\
		25	0.0681705503222608\\
		26	0.0622211204759544\\
		27	0.0572632622706991\\
		28	0.0542885473475459\\
		29	0.0508180466038671\\
		30	0.0473475458601884\\
		31	0.0461080813088746\\
		32	0.0438770451165097\\
		33	0.0423896876549331\\
		34	0.0396628656420426\\
		35	0.0369360436291522\\
		36	0.0344571145265245\\
		37	0.0327218641546852\\
		38	0.0312345066931086\\
		39	0.0294992563212692\\
		40	0.0287555775904809\\
		41	0.0277640059494298\\
		42	0.0260287555775905\\
		43	0.0255329697570649\\
		44	0.0240456122954884\\
		45	0.0230540406544373\\
		46	0.0218145761031234\\
		47	0.0205751115518096\\
		48	0.0203272186415469\\
		49	0.0200793257312841\\
		50	0.0188398611799703\\
		51	0.0183440753594447\\
		52	0.0166088249876054\\
		53	0.0163609320773426\\
		54	0.0161130391670798\\
		55	0.014873574615766\\
		56	0.0138820029747149\\
		57	0.0136341100644522\\
		58	0.0133862171541894\\
		59	0.0131383242439266\\
		60	0.0126425384234011\\
		61	0.0118988596926128\\
		62	0.0118988596926128\\
		63	0.01165096678235\\
		64	0.0114030738720873\\
		65	0.010659395141299\\
		66	0.0104115022310362\\
		67	0.0104115022310362\\
		68	0.0101636093207734\\
		69	0.00966782350024789\\
		70	0.00966782350024789\\
		71	0.00917203767972236\\
		72	0.00867625185919683\\
		73	0.00768468021814576\\
		74	0.00718889439762023\\
		75	0.0066931085770947\\
		76	0.0066931085770947\\
		77	0.0066931085770947\\
		78	0.00644521566683193\\
		79	0.0059494298463064\\
		80	0.0059494298463064\\
		81	0.00545364402578086\\
		82	0.00545364402578086\\
		83	0.00495785820525533\\
		84	0.00470996529499256\\
		85	0.00470996529499256\\
		86	0.0044620723847298\\
		87	0.00421417947446703\\
		88	0.00396628656420426\\
		89	0.00396628656420426\\
		90	0.00396628656420426\\
		91	0.00347050074367873\\
		92	0.00347050074367873\\
		93	0.00347050074367873\\
		94	0.00322260783341596\\
		95	0.00322260783341596\\
		96	0.00322260783341596\\
		97	0.00322260783341596\\
		98	0.00322260783341596\\
		99	0.00322260783341596\\
		100	0.0029747149231532\\
		101	0.0029747149231532\\
		102	0.00272682201289043\\
		103	0.00272682201289043\\
		104	0.00272682201289043\\
		105	0.00272682201289043\\
		106	0.00247892910262767\\
		107	0.0022310361923649\\
		108	0.0022310361923649\\
		109	0.0022310361923649\\
		110	0.0022310361923649\\
		111	0.0022310361923649\\
		112	0.00198314328210213\\
		113	0.00173525037183937\\
		114	0.0014873574615766\\
		115	0.00123946455131383\\
		116	0.00123946455131383\\
		117	0.000991571641051066\\
		118	0.000991571641051066\\
		119	0.000743678730788299\\
		120	0.000743678730788299\\
		121	0.000743678730788299\\
		122	0.000743678730788299\\
		123	0.000743678730788299\\
		124	0.000743678730788299\\
		125	0.000743678730788299\\
		126	0.000743678730788299\\
		127	0.000743678730788299\\
		128	0.000743678730788299\\
		129	0.000743678730788299\\
		130	0.000743678730788299\\
		131	0.000743678730788299\\
		132	0.000743678730788299\\
		133	0.000743678730788299\\
		134	0.000743678730788299\\
		135	0.000743678730788299\\
		136	0.000743678730788299\\
		137	0.000743678730788299\\
		138	0.000743678730788299\\
		139	0.000743678730788299\\
		140	0.000743678730788299\\
		141	0.000743678730788299\\
		142	0.000743678730788299\\
		143	0.000743678730788299\\
		144	0.000743678730788299\\
		145	0.000743678730788299\\
		146	0.000743678730788299\\
		147	0.000743678730788299\\
		148	0.000743678730788299\\
		149	0.000743678730788299\\
		150	0.000743678730788299\\
		151	0.000743678730788299\\
		152	0.000743678730788299\\
		153	0.000495785820525533\\
		154	0.000495785820525533\\
		155	0.000495785820525533\\
		156	0.000495785820525533\\
		157	0.000495785820525533\\
		158	0.000495785820525533\\
		159	0.000495785820525533\\
		160	0.000495785820525533\\
		161	0.000495785820525533\\
		162	0.000495785820525533\\
		163	0.000495785820525533\\
		164	0.000495785820525533\\
		165	0.000495785820525533\\
		166	0.000495785820525533\\
		167	0.000495785820525533\\
		168	0.000495785820525533\\
		169	0.000495785820525533\\
		170	0.000247892910262766\\
		171	0.000247892910262766\\
		172	0.000247892910262766\\
		173	0.000247892910262766\\
		174	0.000247892910262766\\
		175	0.000247892910262766\\
		176	0.000247892910262766\\
		177	0.000247892910262766\\
		178	0.000247892910262766\\
		179	0.000247892910262766\\
		180	0.000247892910262766\\
		181	0.000247892910262766\\
		182	0.000247892910262766\\
		183	0.000247892910262766\\
		184	0.000247892910262766\\
		185	0.000247892910262766\\
		186	0.000247892910262766\\
		187	0.000247892910262766\\
		188	0.000247892910262766\\
		189	0.000247892910262766\\
		190	0.000247892910262766\\
		191	0.000247892910262766\\
		192	0.000247892910262766\\
		193	0.000247892910262766\\
		194	0.000247892910262766\\
		195	0.000247892910262766\\
		196	0.000247892910262766\\
		197	0.000247892910262766\\
		198	0.000247892910262766\\
		199	0.000247892910262766\\
		200	0.000247892910262766\\
		201	0.000247892910262766\\
		};
		\addlegendentry{2012}
	
	\end{axis}
\end{tikzpicture}%

%% file: supp-figures/BT_degree_increment.tikz
\tikzstyle{every node}=[font=\Large]

\begin{tikzpicture}
	\begin{axis}[%
		width=0.951\fwidth,
		height=\fheight,
		at={(0\fwidth,0\fheight)},
		scale only axis,
		axis x line*=bottom,
		axis y line*=left,
		xmin=0,
		xmax=20,
		ymin=0.116371834894207,
		ymax=0.746676852559205,
		axis background/.style={fill=white},
		title style={font=\bfseries\Large},
		title={BT},
		legend style={legend cell align=left, align=left, draw=white!15!black}
		]
		\addplot [color=mycolor1, line width = 2pt]
		  table[row sep=crcr]{%
		0	0.128376703841388\\
		1	0.495910780669145\\
		2	0.495662949194548\\
		3	0.477323420074349\\
		4	0.427509293680297\\
		5	0.411648079306072\\
		6	0.349690210656753\\
		7	0.346468401486989\\
		8	0.314498141263941\\
		9	0.285006195786865\\
		10	0.275588599752169\\
		11	0.273358116480793\\
		12	0.244857496902107\\
		13	0.2272614622057\\
		14	0.201734820322181\\
		15	0.187360594795539\\
		16	0.166294919454771\\
		17	0.153159851301115\\
		18	0.142503097893432\\
		19	0.132589838909542\\
		20	0.131350681536555\\
		};
		\addlegendentry{1984}
		
		\addplot [color=mycolor2, line width = 2pt]
		  table[row sep=crcr]{%
		0	0.116371834894207\\
		1	0.358480749219563\\
		2	0.483003815469997\\
		3	0.46149843912591\\
		4	0.462885882761013\\
		5	0.417273673257024\\
		6	0.391259105098855\\
		7	0.355185570586195\\
		8	0.337148803329865\\
		9	0.304370447450572\\
		10	0.270551508844953\\
		11	0.24644467568505\\
		12	0.230489073881374\\
		13	0.19389524800555\\
		14	0.182622268470343\\
		15	0.163718348942074\\
		16	0.156434269857787\\
		17	0.144640998959417\\
		18	0.139264654873396\\
		19	0.139611515782171\\
		20	0.129725979882067\\
		};
		\addlegendentry{1987}
		
		\addplot [color=mycolor3, line width = 2pt]
		  table[row sep=crcr]{%
		0	0.138884644766998\\
		1	0.555538579067991\\
		2	0.746676852559205\\
		3	0.670282658517953\\
		4	0.643697478991597\\
		5	0.59938884644767\\
		6	0.570359052711994\\
		7	0.499006875477464\\
		8	0.438502673796791\\
		9	0.414514896867838\\
		10	0.355538579067991\\
		11	0.307715813598167\\
		12	0.291825821237586\\
		13	0.272727272727273\\
		14	0.246447669977082\\
		15	0.250725744843392\\
		16	0.243850267379679\\
		17	0.242169595110772\\
		18	0.231016042780749\\
		19	0.220320855614973\\
		20	0.210542398777693\\
		};
		\addlegendentry{1990}
		
		\addplot [color=mycolor4, line width = 2pt]
		  table[row sep=crcr]{%
		0	0.12723470987977\\
		1	0.526607422895975\\
		2	0.685415577626764\\
		3	0.638473601672765\\
		4	0.565708311552535\\
		5	0.506220595922635\\
		6	0.477992681651856\\
		7	0.410350235232619\\
		8	0.361003659174072\\
		9	0.326084683742812\\
		10	0.303502352326189\\
		11	0.281756403554626\\
		12	0.265237846314689\\
		13	0.271615263983272\\
		14	0.25363303711448\\
		15	0.253423941453215\\
		16	0.246523784631469\\
		17	0.23993727130162\\
		18	0.240460010454783\\
		19	0.207527443805541\\
		20	0.203031887088343\\
		};
		\addlegendentry{1993}
	
	\end{axis}
\end{tikzpicture}

%% file: supp-figures/BT_age_cited.tikz
\tikzstyle{every node}=[font=\Large]

\begin{tikzpicture}

	\begin{axis}[%
		width=0.951\fwidth,
		height=\fheight,
		at={(0\fwidth,0\fheight)},
		scale only axis,
		xmin=0,
		xmax=20,
		ymin=0,
		ymax=0.13,
		xminorticks=true,
		yminorticks=true,
		axis x line*=bottom,
		axis y line*=left,
		axis background/.style={fill=white},
		yticklabel style={
			/pgf/number format/fixed,
			/pgf/number format/precision=5
		},
		scaled y ticks=false,
		xlabel={Age (years)},
		title style={font=\bfseries\Large},
		title={BT - Distribution of Age of Cited Paper},
		legend style={legend cell align=left, align=left, draw=white!15!black}
		]
		\addplot [color=mycolor1, line width = 2pt]
		  table[row sep=crcr]{%
			0	0.0255029231689914\\
			1	0.105095981218064\\
			2	0.126697509552088\\
			3	0.114429406619712\\
			4	0.107466740321318\\
			5	0.0931271003084289\\
			6	0.0789715969249183\\
			7	0.064528380058003\\
			8	0.0559545182525434\\
			9	0.0435828384661419\\
			10	0.0327417944114533\\
			11	0.0310155135110252\\
			12	0.0270795930580491\\
			13	0.0215900197946877\\
			14	0.0185632739492704\\
			15	0.0134074483266584\\
			16	0.0108755696726971\\
			17	0.00959812180638033\\
			18	0.00711227730976385\\
			19	0.00655986742162685\\
			20	0.00609952584817935\\
		};
		\addlegendentry{2004}
		
		\addplot [color=mycolor2,line width = 2pt]
		  table[row sep=crcr]{%
			0	0.0253005932552901\\
			1	0.103327940103423\\
			2	0.122783223882491\\
			3	0.111592271818787\\
			4	0.0999809650709051\\
			5	0.0838885187652676\\
			6	0.0698423273373307\\
			7	0.0680340090733162\\
			8	0.0584293011008534\\
			9	0.0490308048602519\\
			10	0.0391247105104533\\
			11	0.0349608197709464\\
			12	0.0260540591986295\\
			13	0.0214856762158561\\
			14	0.0192411408267504\\
			15	0.0174010976809111\\
			16	0.0140937787506742\\
			17	0.012570984423083\\
			18	0.00940642746105771\\
			19	0.0075187969924812\\
			20	0.00593255290124044\\
		};
		\addlegendentry{2007}
		
		\addplot [color=mycolor3,line width = 2pt]
		  table[row sep=crcr]{%
			0	0.0300296043485135\\
			1	0.104362143507488\\
			2	0.117087106570203\\
			3	0.104193131573065\\
			4	0.0936816796515083\\
			5	0.080520556758024\\
			6	0.0710558884303153\\
			7	0.0605171765193355\\
			8	0.0546726347870177\\
			9	0.0454751143556556\\
			10	0.0389109089025673\\
			11	0.0393743287227605\\
			12	0.031926899612363\\
			13	0.0275434933131246\\
			14	0.0214154476908063\\
			15	0.0200524482196501\\
			16	0.0154018940240651\\
			17	0.0122997072277136\\
			18	0.0125395951346371\\
			19	0.0103696999765564\\
			20	0.00857054067463022\\
		};
		\addlegendentry{2011}
		
		\addplot [color=mycolor4,line width = 2pt]
		  table[row sep=crcr]{%
			0	0.0252174545102125\\
			1	0.0943861604438737\\
			2	0.110217333365639\\
			3	0.107038499745596\\
			4	0.099396700021806\\
			5	0.0858042788263514\\
			6	0.0696629757953141\\
			7	0.0642114699682601\\
			8	0.0545247498364548\\
			9	0.047144622392363\\
			10	0.0408887165943837\\
			11	0.0361446950791074\\
			12	0.0291328471397766\\
			13	0.0254112858285077\\
			14	0.0264870496450464\\
			15	0.0212826787488188\\
			16	0.017091076490684\\
			17	0.0143483633368061\\
			18	0.0129043200155065\\
			19	0.0103457466140092\\
			20	0.00835897560148281\\
		};
		\addlegendentry{2014}
	
	\end{axis}
\end{tikzpicture}%